\documentclass{aa}  

\usepackage{graphicx}
\usepackage{txfonts}
\usepackage[colorlinks,citecolor=blue,linkcolor=blue,urlcolor=blue]{hyperref}

\usepackage{xcolor}
\usepackage{gensymb}
\usepackage{array}
\usepackage{subcaption}
\usepackage[labelformat = empty,position=top]{subcaption}
\usepackage[export]{adjustbox}
\usepackage{makeidx}
\usepackage{multirow}
\usepackage{siunitx}

\usepackage{etoolbox}
\makeatletter
\patchcmd\@combinedblfloats{\box\@outputbox}{\unvbox\@outputbox}{}{%
  \errmessage{\noexpand\@combinedblfloats could not be patched}%
}%
\makeatother

\begin{document}

   \title{MAGNUM survey: A MUSE-Chandra resolved view on ionized outflows and photoionization in the Seyfert galaxy NGC 1365\thanks{Based on observations made with ESO Telescopes at the La Silla Paranal Observatory under program ID 094.B-0321(A).}}

   \subtitle{}

   \author{G. Venturi\inst{1}\fnmsep\inst{2}\fnmsep\inst{3}\fnmsep\inst{4},
          E. Nardini\inst{2},
          A. Marconi\inst{1}\fnmsep\inst{2},
          S. Carniani\inst{3}\fnmsep\inst{4},
          M. Mingozzi\inst{5}\fnmsep\inst{2},
          G. Cresci\inst{2},
          F. Mannucci\inst{2},
          G. Risaliti\inst{1}\fnmsep\inst{2},
          R. Maiolino\inst{3}\fnmsep\inst{4},
          B. Balmaverde\inst{6},
          A. Bongiorno\inst{7},
          M. Brusa\inst{5}\fnmsep\inst{8},
          A. Capetti\inst{9},
          C. Cicone\inst{6},
          S. Ciroi\inst{10}\fnmsep\inst{11},
          C. Feruglio\inst{12},
          F. Fiore\inst{12}\fnmsep\inst{7},
          A. Gallazzi\inst{2},
          F. La Franca\inst{13},
          V. Mainieri\inst{14},
          K. Matsuoka\inst{1}\fnmsep\inst{2},
          T. Nagao\inst{15},
          M. Perna\inst{2},
          E. Piconcelli\inst{7},
          E. Sani\inst{16},
          P. Tozzi\inst{2}
          \and
          S. Zibetti\inst{2}
          }

   \institute{Dipartimento di Fisica e Astronomia, Università degli Studi di Firenze, Via G. Sansone 1, 50019 Sesto Fiorentino, Firenze, Italy\\
              \email{giacomo.venturi@unifi.it}
         \and
             INAF -- Osservatorio Astrofisco di Arcetri, Largo E. Fermi 5, I-50125 Firenze, Italy\\
             \email{gventuri@arcetri.astro.it}
          \and
             Cavendish Laboratory, University of Cambridge, 19 J. J. Thomson Ave., Cambridge CB3 0HE, UK
          \and
             Kavli Institute for Cosmology, University of Cambridge, Madingley Road, Cambridge CB3 0HA, UK         
          \and
             Dipartimento di Fisica e Astronomia, Università degli Studi di Bologna, Via Piero Gobetti 93/2, I-40129 Bologna, Italy
          \and
             INAF -- Osservatorio Astronomico di Brera, Via Brera 28, I-20121 Milano, Italy
          \and
             INAF -- Osservatorio Astronomico di Roma, Via Frascati 33, I-00044 Monte Porzio Catone, Italy
          \and
             INAF -- Osservatorio di Astrofisica e Scienza dello Spazio di Bologna, Via Piero Gobetti 93/3, I-40129 Bologna, Italy
          \and
             INAF -- Osservatorio Astrofisico di Torino, Via Osservatorio 20, 10025 Pino Torinese, Italy
          \and
             Dipartimento di Fisica e Astronomia ``G. Galilei'', Università di Padova, Vicolo dell’Osservatorio 3, 35122 Padova, Italy
          \and
             INAF -- Osservatorio Astronomico di Padova, Vicolo dell’Osservatorio 5, 35122 Padova, Italy
          \and
             INAF -- Osservatorio Astronomico di Trieste, Via G.B. Tiepolo 11, I-34143 Trieste, Italy
          \and
             Dipartimento di Matematica e Fisica, Università Roma Tre, Via della Vasca Navale 84, I-00146 Roma, Italy
          \and
             European Southern Observatory, Karl-Schwarzschild-Str. 2, 85748 Garching bei München, Germany
          \and
             Research Center for Space and Cosmic Evolution, Ehime University, 2-5 Bunkyo-cho, Matsuyama 790-8577, Japan
          \and
             European Southern Observatory, Alonso de Cordova 3107, Casilla 19, Santiago 19001, Chile
             }

   \date{Received 19 June 2018 / Accepted 3 August 2018}

 
  \abstract
   {Ionized outflows, revealed by broad asymmetric wings of the [O\,\textsc{iii}] $\lambda5007$ line, are commonly observed in active galactic nuclei (AGN) but the low intrinsic spatial resolution of the observations has generally prevented a detailed characterization of their properties. The MAGNUM survey aims at overcoming these limitations by focusing on the nearest AGN, including 
    NGC 1365, a nearby Seyfert galaxy ($D$ $\sim$ 17 Mpc), hosting a low-luminosity active nucleus ($L_\textrm{bol}$ $\sim$ $2 \times 10^{43}$ erg s$^{-1}$).
}
   {We want to obtain a detailed picture of the ionized gas in the central $\sim$5 kpc of NGC 1365 in terms of physical properties, kinematics, and ionization mechanisms. We also aim to characterize the warm ionized outflow as a function of distance from the nucleus and its relation with the nuclear X-ray wind.
   }
   {We employed optical integral-field spectroscopic observations from VLT/MUSE to investigate the warm ionized gas and {\it Chandra} ACIS-S X-ray data for the hot highly-ionized phase.
   We obtained flux, kinematic, and diagnostic maps of the optical emission lines, which we used to disentangle outflows from gravitational motions in the disk and measure the gas properties down to a spatial resolution of $\sim$70 pc.
   We then performed  imaging spectroscopy on {\it Chandra} ACIS-S data guided by the matching with MUSE maps.
   }
   {The [O\,\textsc{iii}] emission mostly traces a kpc-scale biconical outflow ionized by the AGN having velocities up to $\sim$200 km s$^{-1}$. 
   H$\alpha$ emission traces instead star formation in a circumnuclear ring and along the bar, where we detect non-circular streaming gas motions.
   Soft X-rays are predominantly due to thermal emission from the star-forming regions, but we manage to isolate the AGN photoionized component which nicely matches the [O\,\textsc{iii}] emission.
   The mass outflow rate of the extended ionized outflow is similar to that of the nuclear X-ray wind and then decreases with radius, implying that the outflow either slows down or that the AGN activity has recently increased. 
   However, the hard X-ray emission from the circumnuclear ring suggests that star formation might in principle contribute to the outflow. 
   The integrated mass outflow rate, kinetic energy rate, and outflow velocity are broadly consistent with the typical relations observed in more luminous AGN.
   }
{}

   \keywords{Galaxies: individual: NGC 1365, Galaxies: Seyfert, ISM: jets and outflows, Techniques: imaging spectroscopy, X-rays: individuals: NGC 1365, X-rays: ISM
\vspace{-0.2cm}               }	
               
	\titlerunning{MAGNUM survey: A MUSE-Chandra resolved view on ionized outflows and photoionization in  NGC 1365}	
	\authorrunning{G. Venturi et al.}	
	
   \maketitle
%

\section{Introduction}
Quasar-driven outflows are thought to affect the evolution of the host galaxy, by removing gas and quenching star formation (e.g., \citealt{Fabian:2012aa}, \citealt{Cresci:2018aa}).
Most of the extended outflows have been observed in the ionized phase in luminous AGN ($L_\textrm{bol}$ $\gtrsim$ $10^{46}$ erg s$^{-1}$) at redshifts 1-3 around the peak of AGN activity, showing up as prominent broad wings in the strongest optical nebular lines (e.g., \citealt{Cano-Diaz:2012aa}, \citealt{Harrison:2012aa}, \citealt{Cresci:2015ab}, \citealt{Carniani:2016aa}, \citealt{Vayner:2017aa}). Mass-loss rates and velocities of these outflows are found to correlate with the AGN bolometric luminosities, indicating that such powerful winds are AGN-driven (\citealt{Carniani:2015aa}, \citealt{Fiore:2017aa}).
However, because of the limited spatial resolution of such observations, it is not possible to study high-z outflows on small spatial scales ($\lesssim$100 pc), since even with adaptive optics we achieve a spatial resolution of $\sim$1 kpc at $z$ = 2.4 (\citealt{Williams:2017aa}).

Although the nearby Universe mainly hosts low-luminosity AGN ($L_\textrm{bol}$ $\lesssim$ 10$^{44}$ erg s$^{-1}$), likely having slower and/or less massive outflows, these objects are ideal laboratories to perform a spatially resolved kinematic analysis down to 10-100 parsec scales (\citealt{Cresci:2015aa}).
Their vicinity allows us to probe in detail the spatial structure of the outflows and to inspect their properties (such as velocity and mass outflow rate) as a function of distance from the nucleus, as well as their interplay with the star-forming processes in the host galaxy.

\begin{figure*}[t]
	\centering
	\includegraphics[width=0.9\columnwidth,valign=b, trim={0 -2.6cm 0 0},clip,]{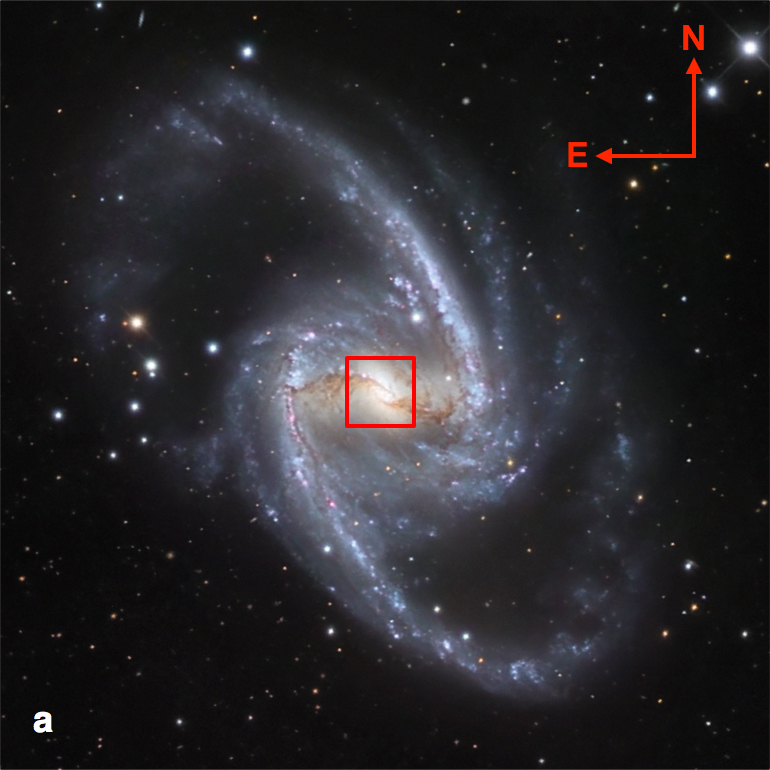}
	\hfill
	\includegraphics[width=1.05\columnwidth, trim={2.5cm 0.5cm 4cm 0.5cm},clip,valign=b]{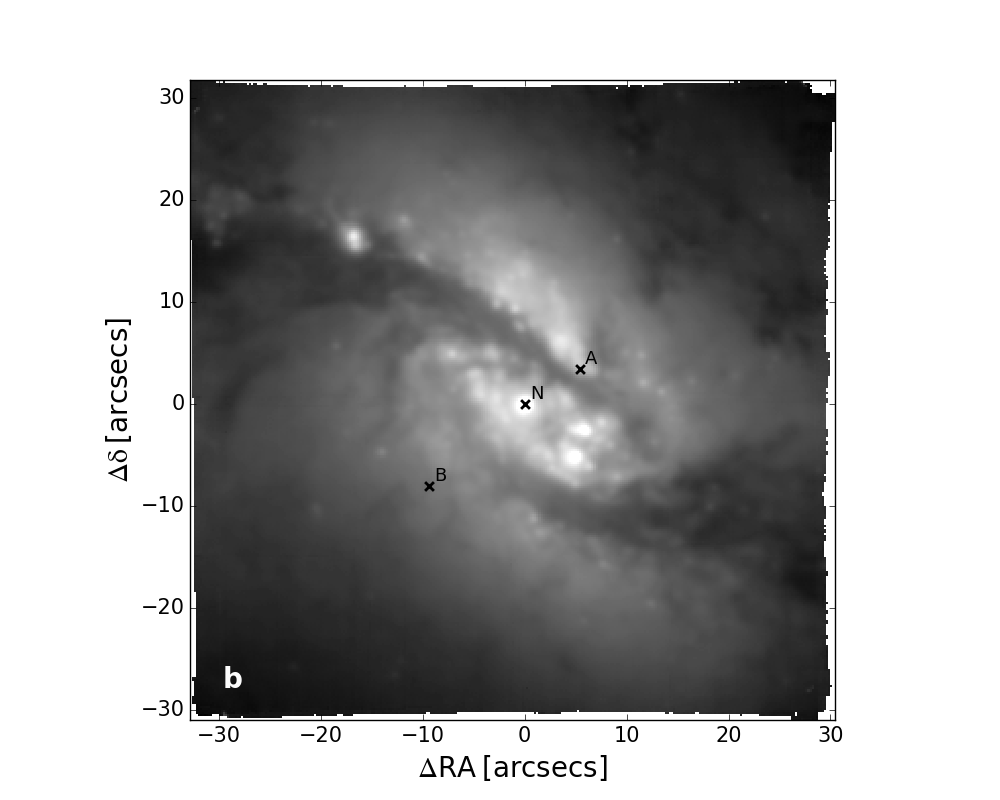}
	\caption{{\bf (a)} Three-color optical image of NGC 1365 combining observations performed through three different filters ($B, V, R$) with the 1.5-meter Danish telescope at the ESO La Silla Observatory in Chile. The red box indicates the region analyzed in this work, corresponding to the MUSE FOV ($\sim$1$'\times1'$). East is to the left. Credit: ESO/IDA/Danish 1.5 m/ R. Gendler, J-E. Ovaldsen, C. Thöne, and C. Feron.
	{\bf (b)} Continuum emission from MUSE having collapsed the data cube in the range $\sim$5100-6470 \AA, excluding emission lines and the sky residuals.  The crosses mark the extraction regions of the spectra shown in Fig. \ref{fig:spectra}.}	
	\label{fig:n1365_eso}
\end{figure*}

In this paper we present a study of the ionized gas in the central regions of NGC 1365, a local ($z$ = 0.005457, \citealt{Bureau:1996aa}) barred spiral (SB(s)b according to \citealt{de-Vaucouleurs:1991aa}) Seyfert 1.8 galaxy (\citealt{Veron-Cetty:2006aa}), belonging to the Fornax cluster (\citealt{Jones:1980aa}). 
A three-color optical image ($B, V, R$ bands) of NGC 1365 is displayed in Fig. \ref{fig:n1365_eso}a, showing the two main spiral arms, connected by the bar, along which two main dust filaments can be seen. The galaxy has an angular size in the sky of $11.2' \times 6.2'$, implying an average diameter of $\sim$44 kpc (having considered a distance of 17.3 Mpc\footnote{From \href{http://leda.univ-lyon1.fr/}{HyperLeda} best distance modulus, that is the weighted average between the redshift distance modulus corrected for infall of the Local Group toward Virgo and the weighted average of the published redshift-indipendent distance measurements.}), almost twice the size of the Milky Way.
NGC 1365 has an inclination of 40$^{\circ}$ with respect to the line of sight\footnote{L.o.s. hereafter.} (\citealt{Jorsater:1995aa}). 

NGC 1365 has been extensively studied over the years and a comprehensive review about the early works on this object is given in \cite{Lindblad:1999aa}. 
The galaxy presents both AGN activity and star formation in its central regions. Nuclear activity emerges both at X-ray and optical wavelengths, from the unresolved hard X-ray emission and strong Fe\,\textsc{i} K$\alpha$ feature (e.g., \citealt{Iyomoto:1997aa}, \citealt{Komossa:1998aa}), which are spatially coincident with the optical nucleus, characterized by broad Balmer lines (e.g., \citealt{Schulz:1999aa}).  
Extended ($\sim$1 kpc) conical [O\,\textsc{iii}] emission was observed to the SE of the nucleus (e.g., \citealt{Edmunds:1988aa}, \citealt{Storchi-Bergmann:1991aa}, \citealt{Kristen:1997aa}) and a complementary, fainter cone to the NW was later revealed (e.g., \citealt{Veilleux:2003aa}). \cite{Sharp:2010aa}, exploting integral field spectroscopic (IFS) observations, found enhanced [O\,\textsc{iii}]- and [N\,\textsc{ii}]-to-Balmer line ratios in correspondence with the two cones, suggesting that they are ionized by the AGN. This was confirmed with diagnostic diagrams only for the inner part of the SE cone, due to the limited signal-to-noise ratio available. The kinematics of [O\,\textsc{iii}] in the cones from long-slit spectroscopy revealed the presence of an outflow (e.g., \citealt{Phillips:1983aa}, \citealt{Hjelm:1996aa}), whose SE inner part has been recently mapped in [N\,\textsc{ii}] by \cite{Lena:2016aa} with 13$''\times$6$''$ IFS observations.

From the X-ray point of view, the unresolved active nucleus exhibits a large variability, with highly ionized blueshifted absorption lines indicating a fast wind ($v$ $\sim$ 3000 km s$^{-1}$; e.g., \citealt{Risaliti:2005aa}, \citealt{Braito:2014aa}). The extended soft X-ray emission is dominated by thermal emission from star formation-related processes, and its spatial distribution shows a roughly biconical morphology as well (\citealt{Wang:2009aa}). By using the high-resolution gratings onboard {\it Chandra} and {\it XMM-Newton}, \cite{Nardini:2015ab} and \cite{Whewell:2016aa} were able to isolate the AGN-photoionized emission lines in the soft X-ray spectrum.

Star formation mostly occurs in an elongated circumnuclear ring, which was reported for instance by \cite{Kristen:1997aa} in H$\alpha$+[N\,\textsc{ii}], \cite{Forbes:1998aa} and \cite{Stevens:1999aa} in radio, and has been recently studied in detail in different IR bands by \cite{Alonso-Herrero:2012aa}. The gas ionization in the corresponding region is dominated by star formation processes, as already suggested from emission-line diagnostics by \cite{Sharp:2010aa} with their IFS observations.
Non-circular motions associated with the bar were found by \cite{Teuben:1986aa} in H$\alpha$ and \cite{Jorsater:1995aa} in H\,\textsc{i}. Using H\,\textsc{i} and optical data \cite{Lindblad:1996aa} inferred gas motions parallel to the bar, as found from Fabry-Perot interferometric H$\alpha$ data by \cite{Zanmar-Sanchez:2008aa} and \cite{Speights:2016aa}. 

\cite{Sandqvist:1995aa} suggested that NGC 1365 hosts a radio jet, but \cite{Stevens:1999aa} showed that radio emission is dominated by the elongated star-forming ring in the direction NE-SW, and that the evidence of a jet emanating from the nucleus is at best marginal.

NGC 1365 was observed with MUSE (Multi Unit Spectroscopic Explorer; \citealt{Bacon:2010aa}), the optical and near-IR large-field, integral-field spectrograph located at the Very Large Telescope (VLT), as part of the MAGNUM survey (Measuring Active Galactic Nuclei Under MUSE Microscope; \citealt{Venturi:2018aa}). This program aims at observing nearby AGN to study the physical conditions of the narrow line region (NLR), the interplay between nuclear activity and star formation, and the effects and acceleration mechanisms of outflows. The work done on one of the targets, NGC 5643, has been already presented in \cite{Cresci:2015aa}, suggesting a case of positive AGN feedback in action. A brief description of the survey is given in \cite{Venturi:2017aa}, together with preliminary results on NGC 1365 itself and another MAGNUM galaxy, namely NGC 4945.

MUSE observations of NGC 1365 have been complemented with X-ray data acquired by the Chandra X-ray Observatory (CXO). The galaxy is known to host an X-ray nuclear wind (e.g., \citealt{Risaliti:2005aa}, \citealt{Braito:2014aa}) and to have a complex circumnuclear medium giving rise to heterogeneous soft X-ray line emission (e.g., \citealt{Guainazzi:2009aa}, \citealt{Nardini:2015ab}, \citealt{Whewell:2016aa}). \cite{Wang:2009aa} performed a multiband spatially resolved analysis of this galaxy employing the same {\it Chandra} data.

In our work we took advantage of the unprecedented capabilities of MUSE to carry out a highly detailed spatially and spectrally resolved study of the central regions of NGC 1365, both in the optical and in the X-ray band. The combination of MUSE and {\it Chandra} is extremely powerful, as they both provide a wide spatial coverage and a similar subarcsecond spatial resolution, as well as extensive spectral information, allowing us to study the circumnuclear gas in its different phases and compare nuclear and extended gas outflows.

Throughout the paper, wavelengths are given rest-frame. When performing our data analysis they were redshifted according to the receding velocity of the galaxy ($\sim$1630 km s$^{-1}$, in agreement with \citealt{Jorsater:1995aa}, \citealt{Zanmar-Sanchez:2008aa}).

\section{MUSE data analysis and maps}\label{sec:muse_maps}
\subsection{Data reduction}
NGC 1365 was observed with MUSE on October 12, 2014, under program 094.B-0321(A) (P.I.\ A.\ Marconi). The data consisted of two Observing Blocks (OBs) for a total of eight dithered and 90\degree -rotated 500s exposures of the central regions of the galaxy, together with an equal number of 100s sky exposures. Each sky exposure is employed in the data reduction to create a model of the sky lines and sky continuum to be subtracted from the closest science exposure in time. The average seeing during the observations was FWHM = 0.76$''$ $\pm$ 0.02$''$, as derived from the unresolved nuclear Balmer line emission from the broad line region, because of the absence of foreground stars in the FOV. 

The data reduction has been carried out using ESO reflex, which gives a graphical and automated way to execute with EsoRex the Common Pipeline Library (CPL) reduction recipes, within the Kepler workflow engine (\citealt{Freudling:2013aa}). MUSE pipeline v1.6 has been employed for our reduction. The final data cube consists of $320 \times 317$ spaxels, corresponding to $64'' \times 63.4''$, as MUSE spatial sampling is $0.2''$/spaxel. This FOV covers the central $\sim$5.3 kpc of NGC 1365. MUSE spectral binning is 1.25 \AA/channel and its resolving power in Wide Field Mode (WFM) (the only available so far) is equal to 1770 at 4800 \AA \ and to 3590 at 9300 \AA.

Fig. \ref{fig:n1365_eso}b shows an image of the continuum obtained by collapsing the MUSE data cube in the wavelength range $\sim$5100-6470 \AA, having masked the emission lines and the sky residuals included in the range. The continuum exhibits diffuse and clumpy emission as well as the two main dust filaments (compare with Fig. \ref{fig:n1365_eso}a).

\begin{figure*}[t]
\includegraphics[width=\textwidth]{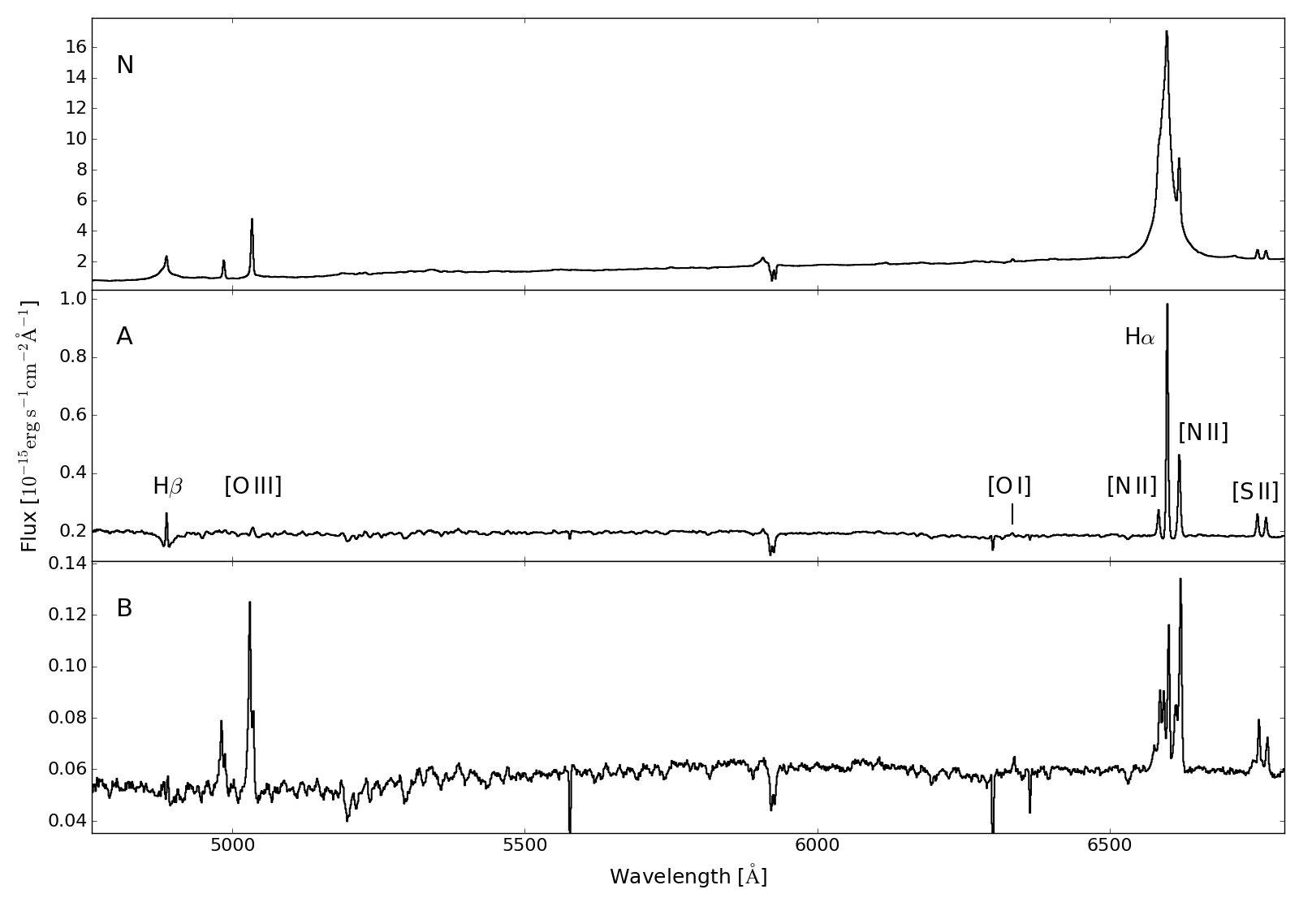}
\caption{Representative spectra extracted from circular regions with radius of 3 spaxels as defined in Fig. \ref{fig:n1365_eso}b. N indicates the nucleus, where broad Balmer emission is present. The spectrum of region A shows narrow emission lines on top of the stellar continuum. The spectrum of region B exhibits complicated line profiles indicative of outflowing gas. The deep narrow lines in absorption are sky residuals.}
\label{fig:spectra}
\end{figure*}

Three representative spectra, extracted from circular regions having radius of 3 spaxels whose position is indicated in Fig. \ref{fig:n1365_eso}b, are displayed in Fig. \ref{fig:spectra}. The nucleus (N) clearly exhibits broad Balmer emission typical of the broad line region (BLR). The spectrum from the region of intense stellar emission A is characterized by narrow emission lines, superimposed on the stellar continuum with its absorption features, among which also the H$\beta$ can be seen. The spectrum from region B shows instead complicated line profiles with multiple kinematic components, indicative of outflowing gas, together with clearly different emission-line ratios compared to the spectrum from A, suggesting different ionization mechanisms for the gas in the two regions.

\subsection{Data analysis}\label{ssec:data_anal}
For the data analysis we used our own suite of python scripts and followed the same scheme adopted for all the other galaxies belonging to the MAGNUM survey so far, which will be extensively described in a forthcoming paper (\citealt{Venturi:2018aa}). In brief, we fitted and subtracted the stellar continuum using a linear combination of \cite{Vazdekis:2010aa} synthetic spectral energy distributions (SEDs) for single-age, single-metallicity stellar populations (SSPs), after having performed a Voronoi adaptive binning (\citealt{Cappellari:2003aa}) to get an average signal-to-noise ratio on the stellar continuum of at least 50 per wavelength channel in each bin. The fit has been performed in the range 4770-6800 \AA\ using the \textsc{ppxf} code (Penalized PiXel-Fitting; \citealt{Cappellari:2004aa}), which convolves the linearly combined stellar templates with a Gaussian profile so as to reproduce the systemic velocity and the velocity dispersion of the stellar absorption lines\footnote{We set in fact to two the number of the Gauss-Hermite moments to fit with \textsc{ppxf}, corresponding to the velocity and velocity dispersion.}.
Since NGC 1365 hosts a Seyfert 1 nucleus, an additional template has been included when fitting the bins residing within a radius of 12 spaxels from the center, so as to reproduce the BLR features (broad Balmer lines + FeII forest) and the disk continuum. This template, previously obtained by fitting the collapsed spectrum of central spaxels, has been at this stage fixed in shape and velocity and allowed to vary in flux only. We then subtracted the stellar and the nuclear emission from the original unbinned data cube, by rescaling the fitted continuum -- constant within each Voronoi bin -- to the original unbinned observed continuum flux of each spaxel before subtracting it.

Before proceeding with the fit of the gas emission lines we generated two cubes from the star-subtracted one, namely a spatially-smoothed one and a Voronoi binned one. The former, used to get flux and kinematic maps, has been obtained by spatially smoothing the cube with a Gaussian kernel having $\sigma=1$ spaxel (i.e., $0.2''$). Such a smoothing does not degrade much the spatial resolution, which goes from FWHM$_{\textrm{seeing}}$ $\simeq$ 0.76$''$ to FWHM $\simeq$ 0.89$''$ (corresponding to $\sim$70 pc). The latter cube has been produced by performing a Voronoi binning around H$\beta$, in the range 4850-4870 \AA, and requesting an average signal-to-noise ratio per wavelength channel of at least 4. The reason for having created such a cube is to get reliable line ratios in each bin for spatially resolved BPT diagrams (\citealt{Baldwin:1981aa}, \citealt{Veilleux:1987aa}). We have chosen H$\beta$ for the Voronoi binning as it is the weakest line among the ones used in every BPT diagram, namely [O\,\textsc{iii}], H$\beta$, and H$\alpha$.

Gas emission lines have been fitted in each spaxel or bin of the two cubes described above in the same spectral range used for the stellar fitting (4770-6800 \AA) making use of \textsc{mpfit} (\citealt{Markwardt:2009aa}), adopting one, two, or three Gaussians to reproduce each line profile, forcing each Gaussian component separately to have the same velocity and velocity dispersion for all the emission lines. A $\chi^2$/d.o.f.-based algorithm has been employed to select whether to adopt one, two, or three components in each spaxel or bin, with the idea of keeping the number of fit parameters as low as possible and using multiple components only in case of complex non-Gaussian line profiles (for further details see \citealt{Venturi:2018aa}).

\begin{figure*}
	\centering
	\begin{subfigure}[t]{0.01\textwidth}
	\textbf{a}
	\end{subfigure}
	\begin{subfigure}[t]{0.47\textwidth}
	\includegraphics[trim={1cm 0.5cm 2cm 0.5cm},clip,width=\textwidth,valign=b]{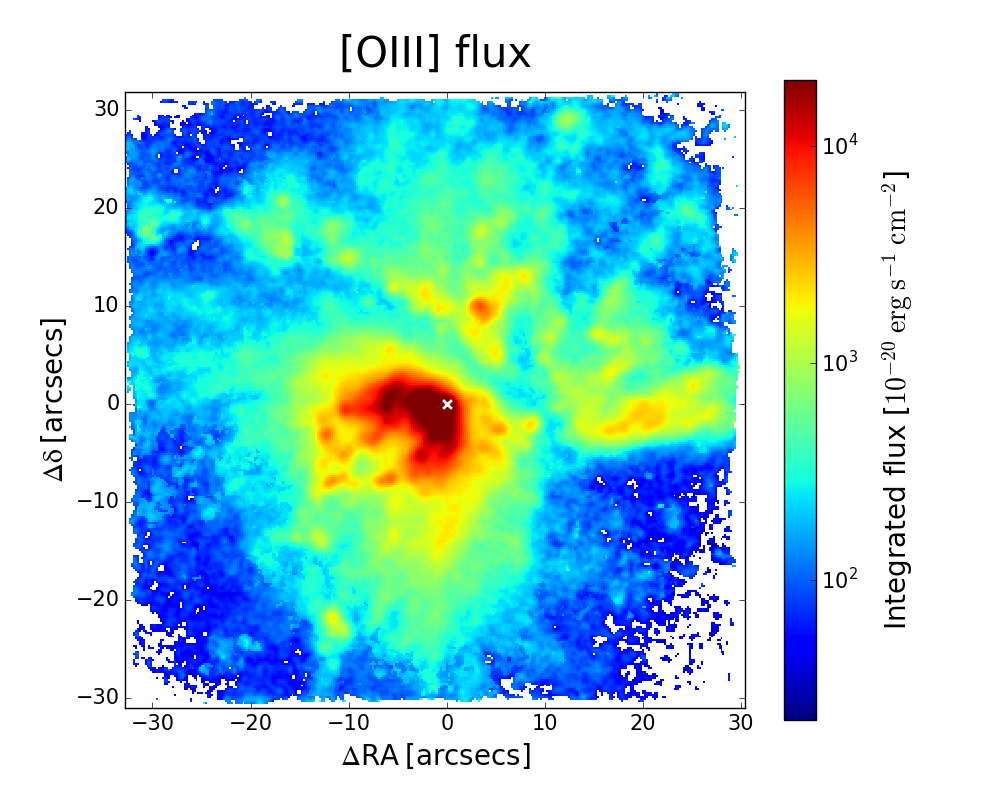}
	\end{subfigure}
	\hfill
	\begin{subfigure}[t]{0.01\textwidth}
	\textbf{b}
	\end{subfigure}
	\begin{subfigure}[t]{0.47\textwidth}
	\includegraphics[trim={1cm 0.5cm 2cm 0.5cm},clip,width=\textwidth,valign=b]{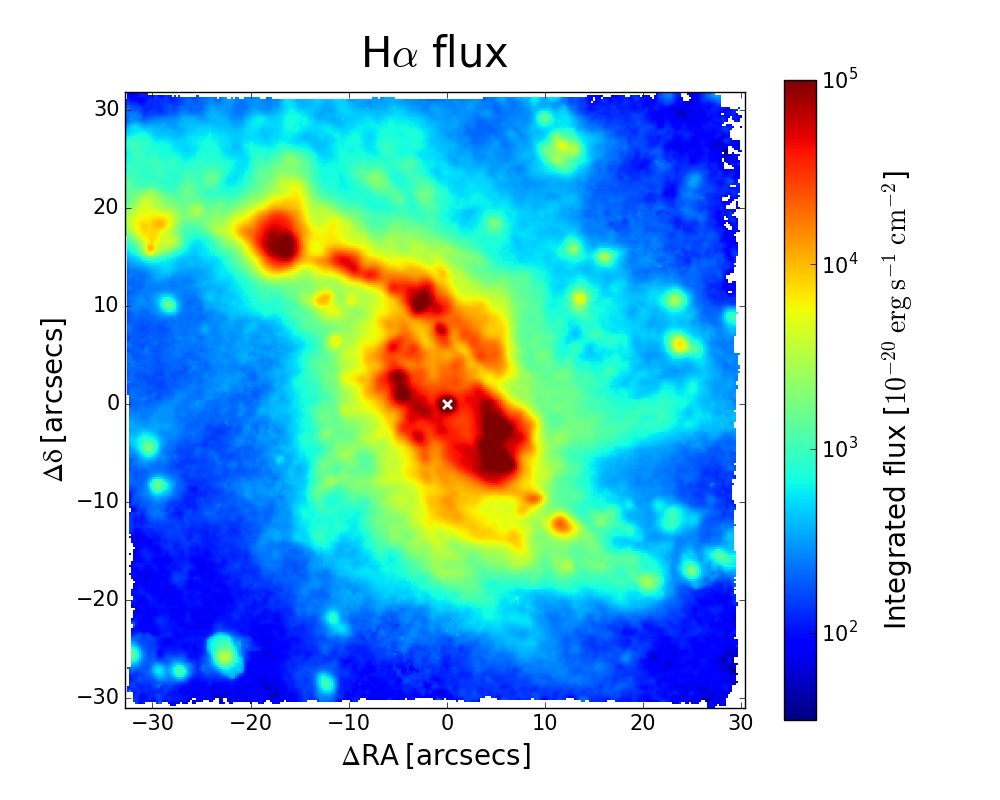}
	\end{subfigure}
	\caption{{\bf(a)} [O\,\textsc{iii}] and {\bf(b)} H$\alpha$ emission of the total fitted line profile. Both maps have been obtained from the fit of the star-subtracted smoothed (with 1-spaxel $\sigma$ gaussian kernel) data cube; a signal-to-noise cut of 3 has been applied. The FOV covers $\sim$ $5.3\times5.3$ kpc$^2$. The reported flux is per spaxel.}
	\label{fig:maps_flux}
\end{figure*}

\subsection{Flux maps and BPT diagrams}\label{ssec:flux+bpt}
In this section we present the emission line maps of the fitted gas lines. For a better visual output, all the maps are spatially smoothed using a Gaussian kernel with $\sigma=1\,$px (i.e., $0.2''$), except for the flux maps (due to their high signal-to-noise ratio), for the maps derived from the fit of the star-subtracted Voronoi-binned cube, and for the stellar ones obtained from the Voronoi-binned cube.

\begin{figure*}[th]
	\centering
	\begin{subfigure}[t]{0.01\textwidth}
	\textbf{a}
	\end{subfigure}
	\begin{subfigure}[t]{0.47\textwidth}
	\includegraphics[trim={0 0.5cm 0 0},clip,width=0.895\textwidth,valign=b]{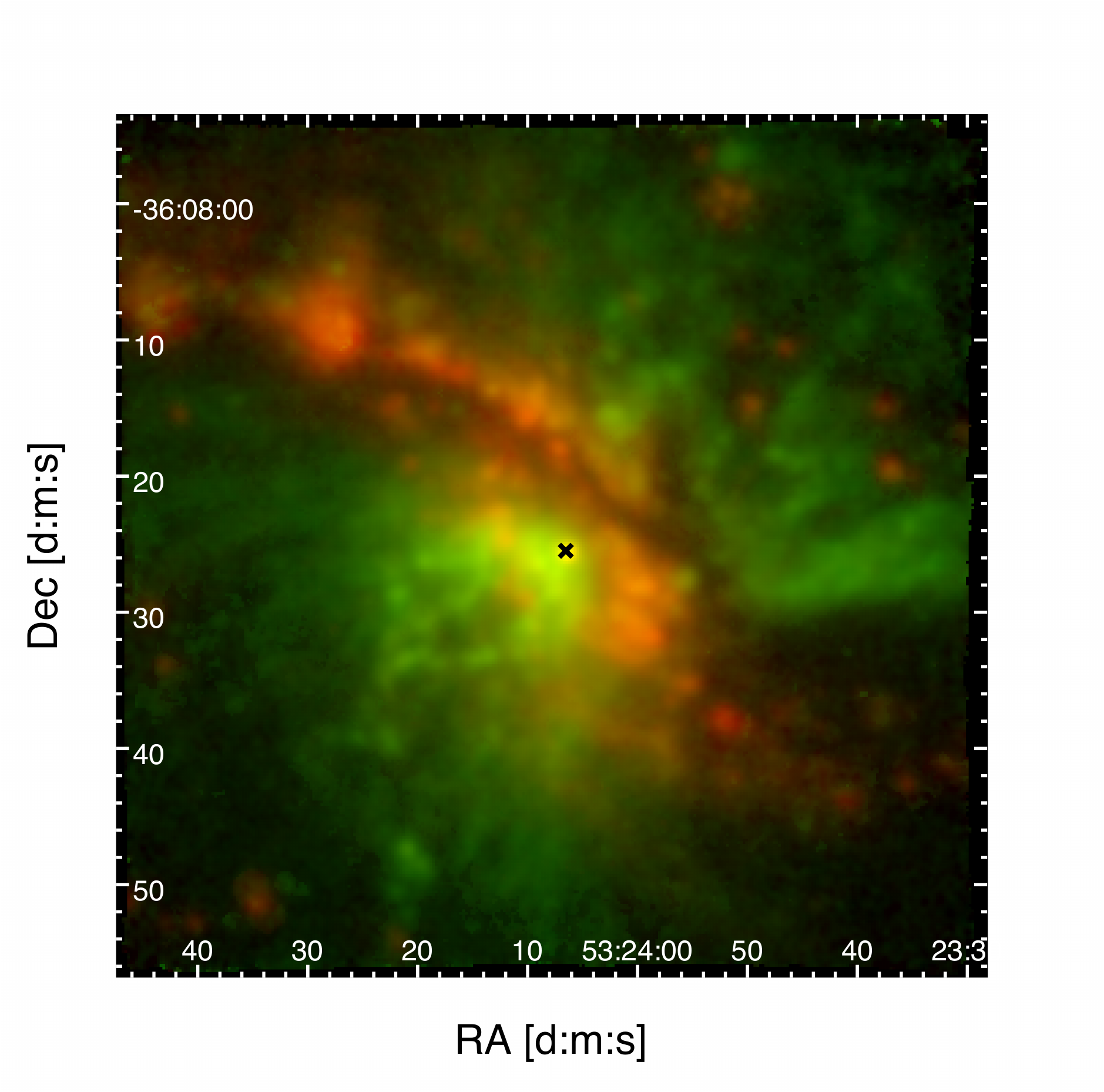}
	\end{subfigure}
	\hfill
	\begin{subfigure}[t]{0.01\textwidth}
	\textbf{b}
	\end{subfigure}
	\begin{subfigure}[t]{0.47\textwidth}
	\includegraphics[trim={0 0.5cm 0 0},clip,width=0.895\textwidth,valign=b]{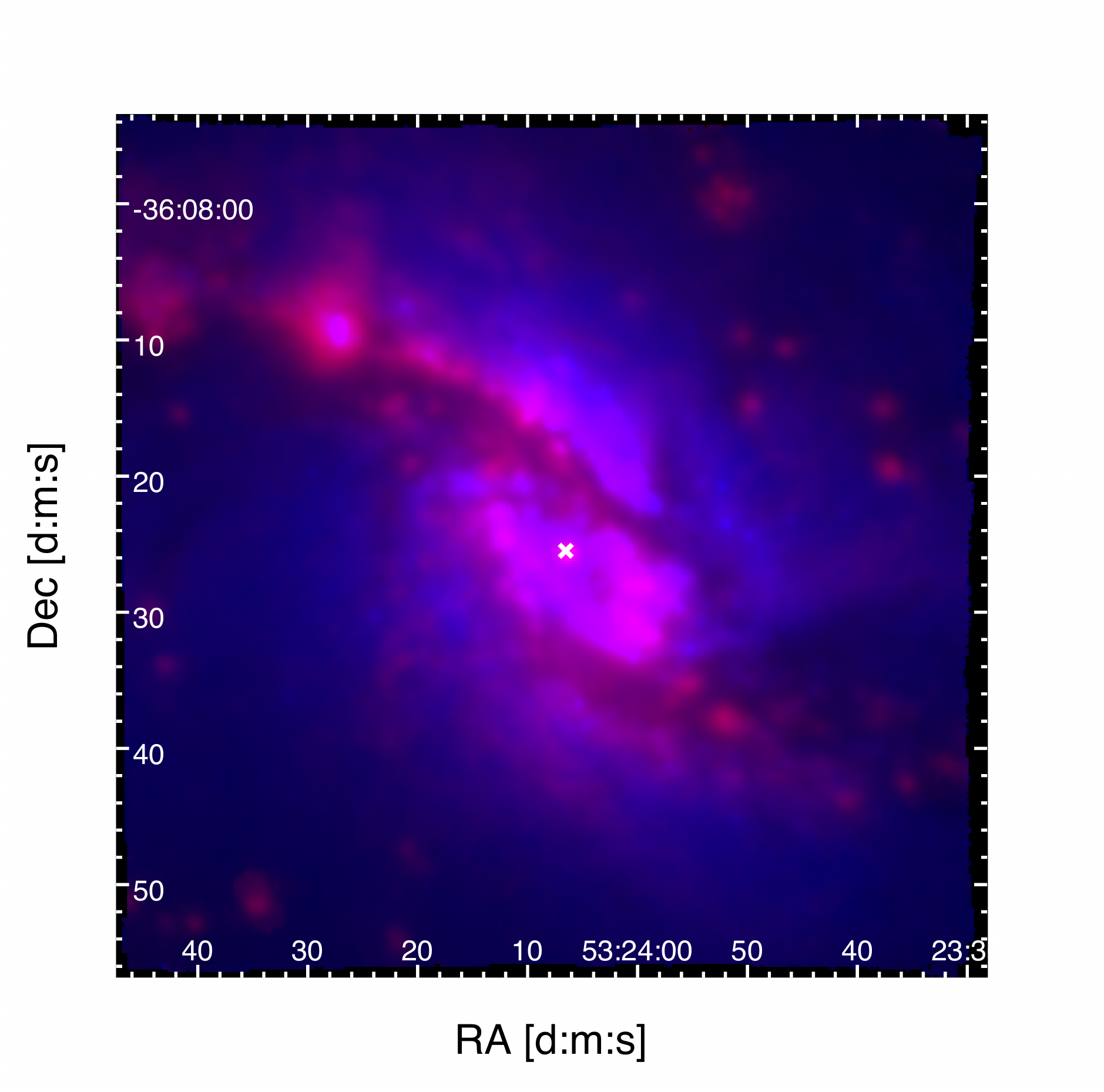}
	\end{subfigure}\\
	\vspace{0.3cm}
	\begin{subfigure}[t]{0.01\textwidth}
	\textbf{c}
	\end{subfigure}
	\begin{subfigure}[t]{0.47\textwidth}
	\includegraphics[trim={1cm 0.5cm 2cm 0.5cm},clip,width=\textwidth,valign=b]{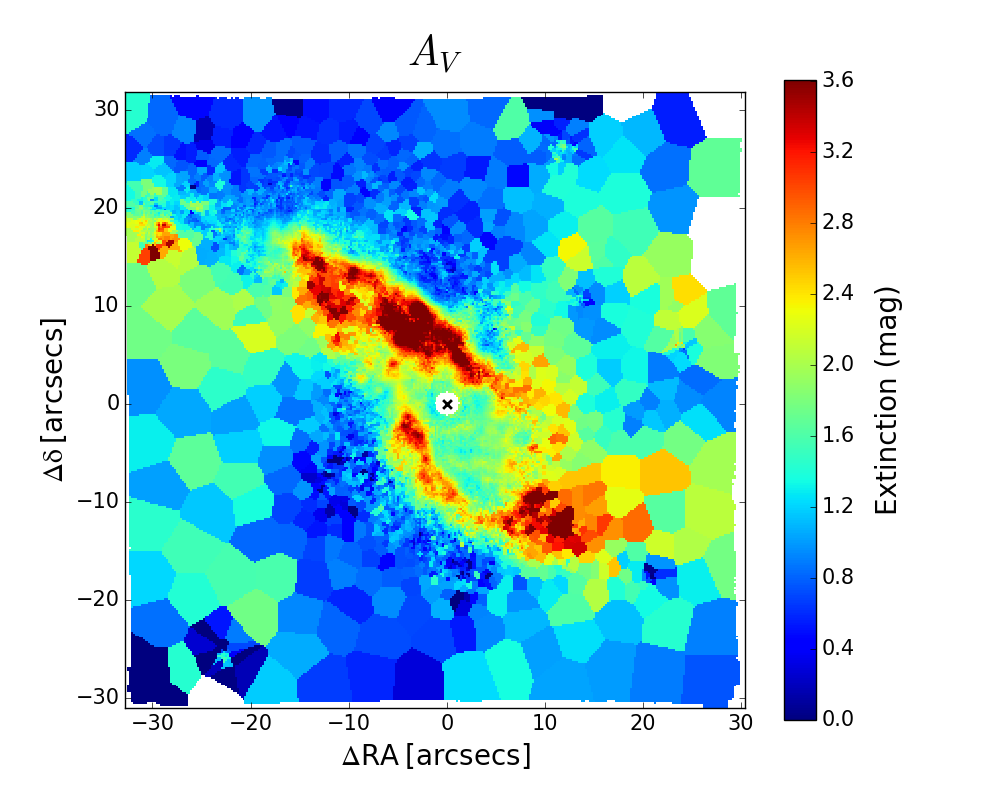}
	\end{subfigure}
	\hfill
	\begin{subfigure}[t]{0.01\textwidth}
	\textbf{d}
	\end{subfigure}
	\begin{subfigure}[t]{0.47\textwidth}
	\includegraphics[trim={1cm 0.5cm 2cm 0.5cm},clip,width=\textwidth,valign=b]{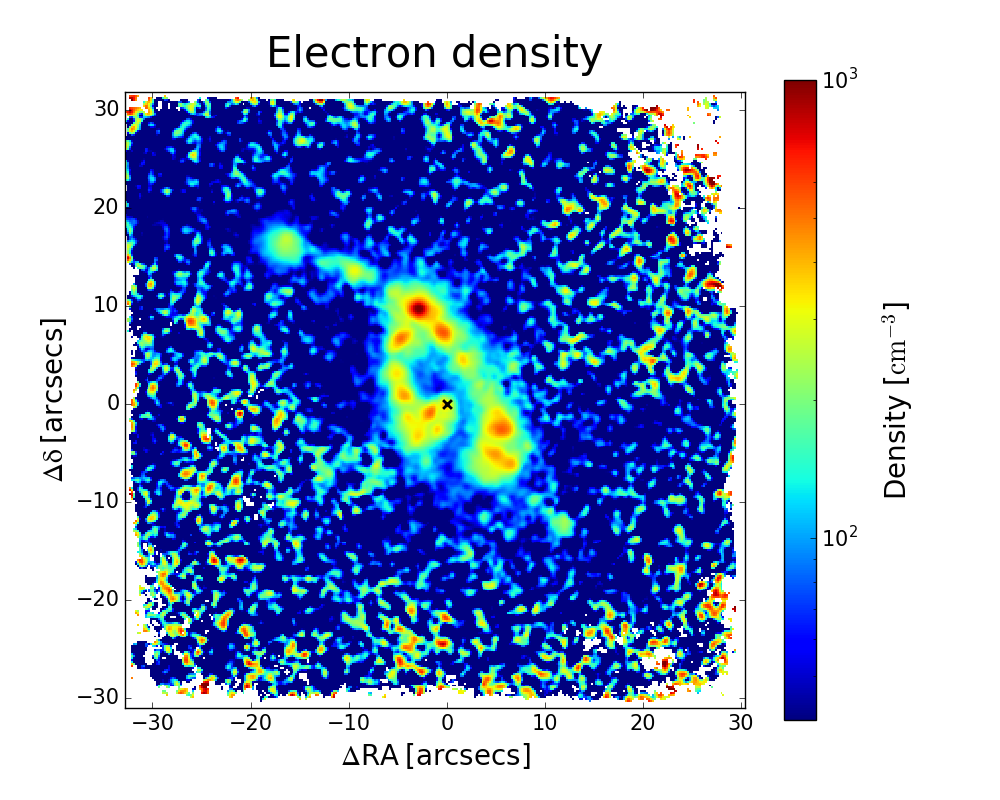}
	\end{subfigure}
	\caption{{\bf(a)} Two-color image of [O\,\textsc{iii}] (green) and H$\alpha$ (red), from the fit of the star-subtracted 1px-$\sigma$ smoothed data cube. {\bf(b)} Two-color image of H$\alpha$ emission (red) and stellar continuum between $\sim$5100-5800 \AA\ (blue). {\bf(c)} Map of the total dust extinction in $V$ band $A_V$ from the measured Balmer decrement (H$\alpha$/H$\beta$ flux ratio), from the fit of the star-subtracted Voronoi-binned data cube (produced to have an average signal-to-noise ratio per wavelength channel around H$\beta$ of at least 4 in each bin). {\bf(d)} Electron density from [S\,\textsc{ii}] $\lambda6716/\lambda6731$ diagnostic line ratio (a typical value of $T_e$ = 10$^4$ K is assumed for the temperature of the gas).}
	\label{fig:maps_redd_ratio}
\end{figure*}

\begin{figure*}[p]
	\centering
	\begin{subfigure}[t]{0.46\textwidth}
	\includegraphics[width=\textwidth]{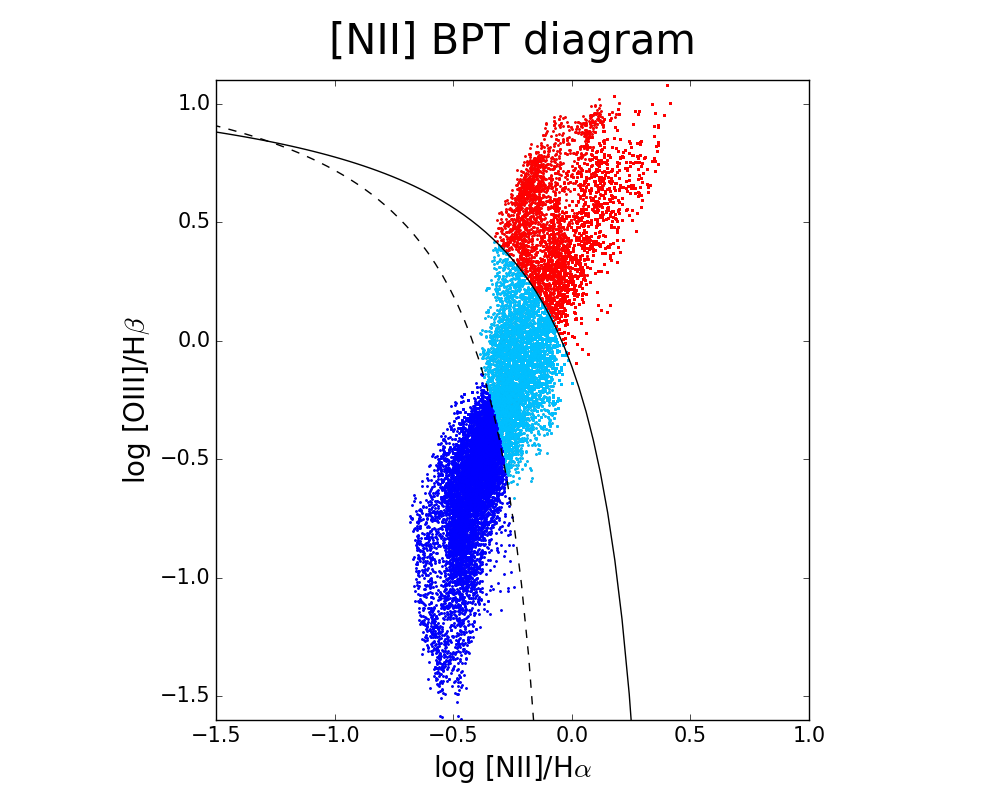}
	\end{subfigure}
	\begin{subfigure}[t]{0.46\textwidth}
	\includegraphics[width=\textwidth]{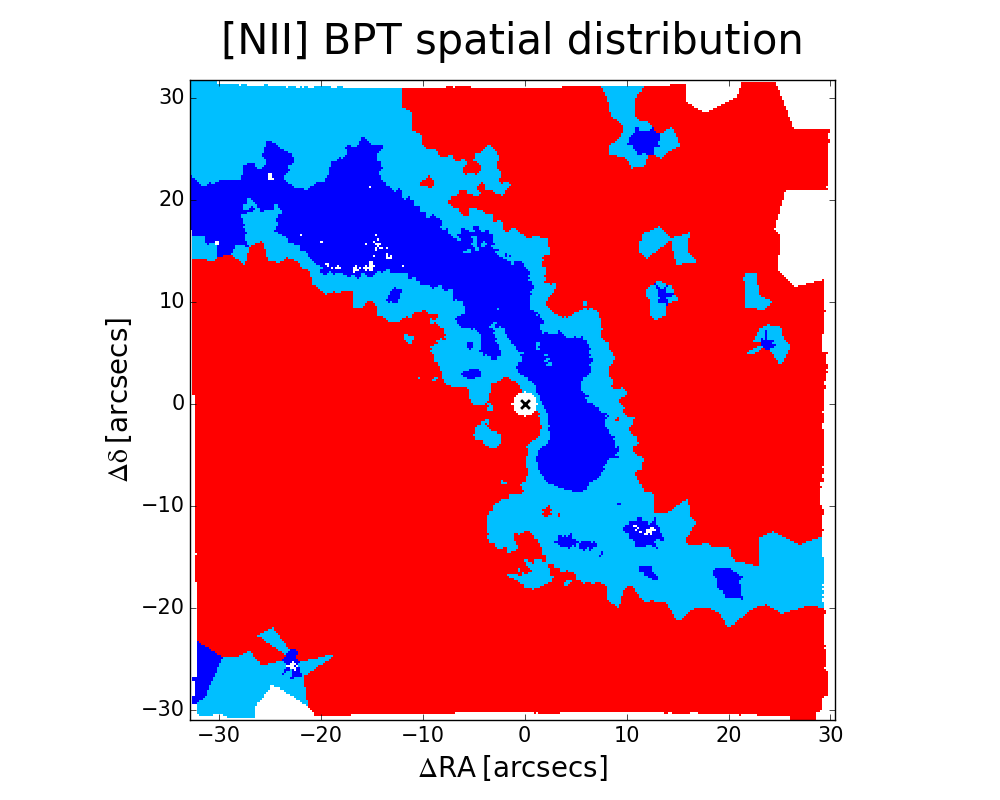}
	\end{subfigure}
	\begin{subfigure}[t]{0.46\textwidth}
	\includegraphics[width=\textwidth]{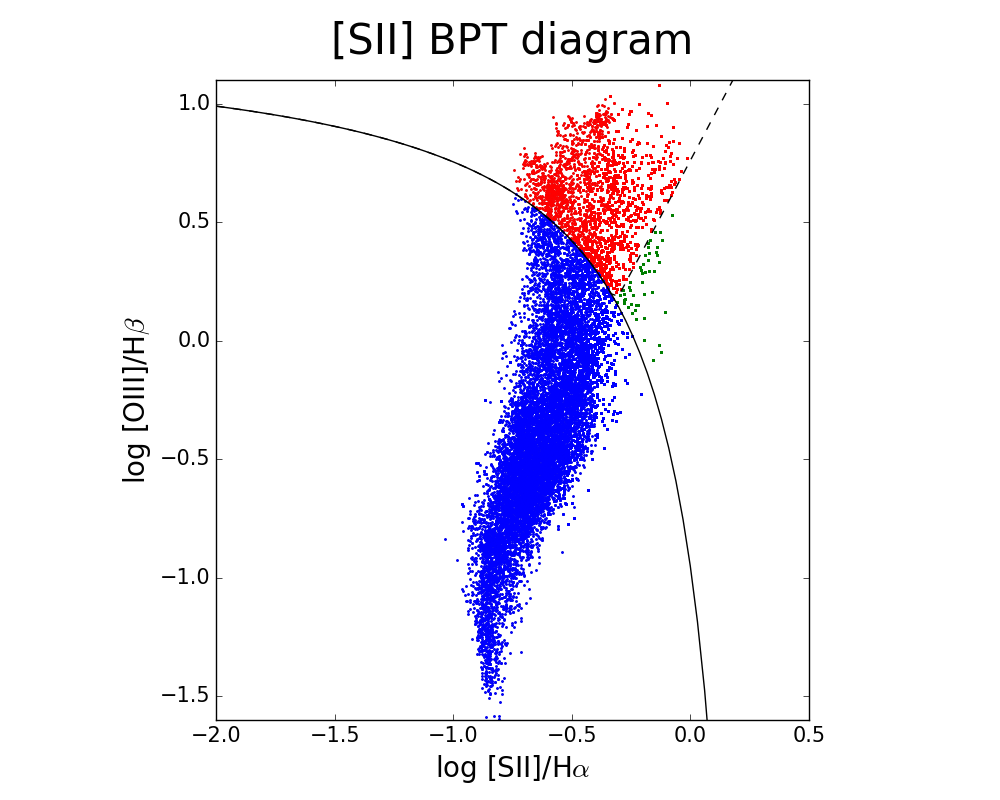}
	\end{subfigure}
	\begin{subfigure}[t]{0.46\textwidth}
	\includegraphics[width=\textwidth]{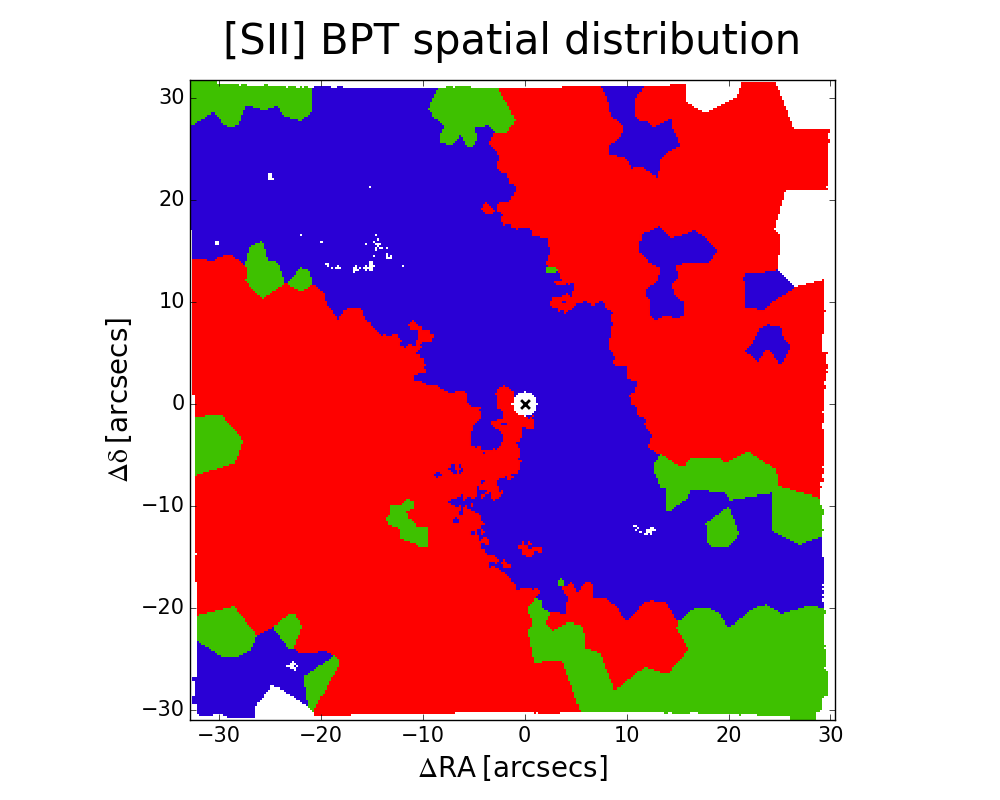}
	\end{subfigure}
	\begin{subfigure}[t]{0.46\textwidth}
	\includegraphics[width=\textwidth]{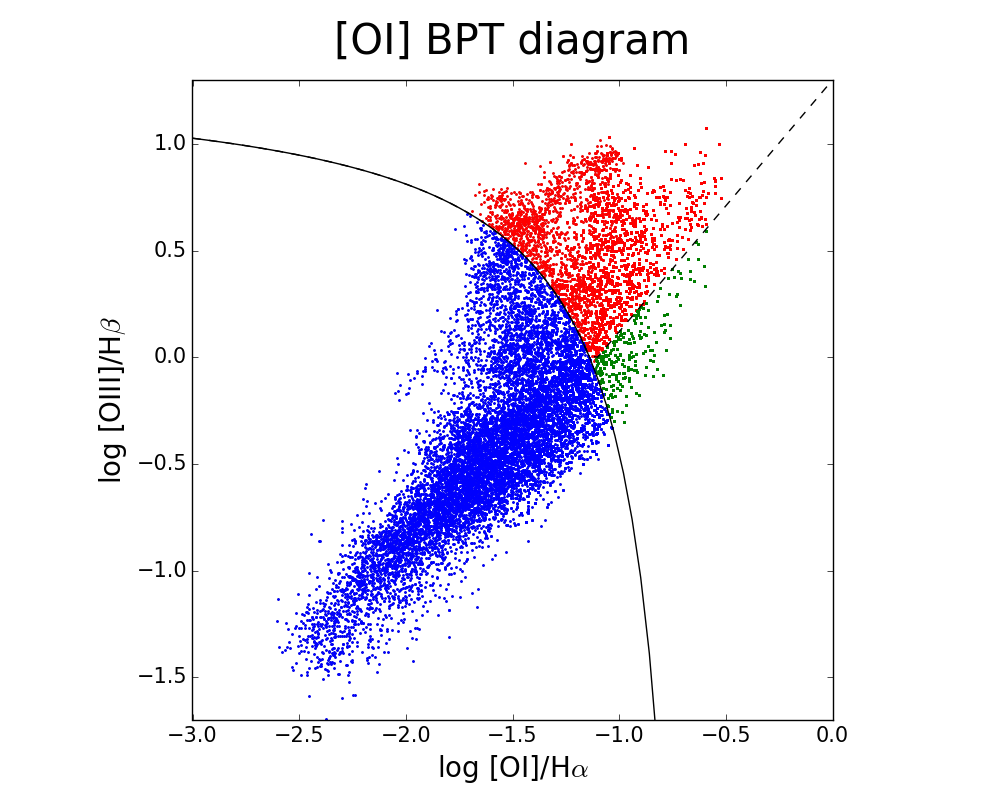}
	\end{subfigure}
	\begin{subfigure}[t]{0.46\textwidth}
	\includegraphics[width=\textwidth]{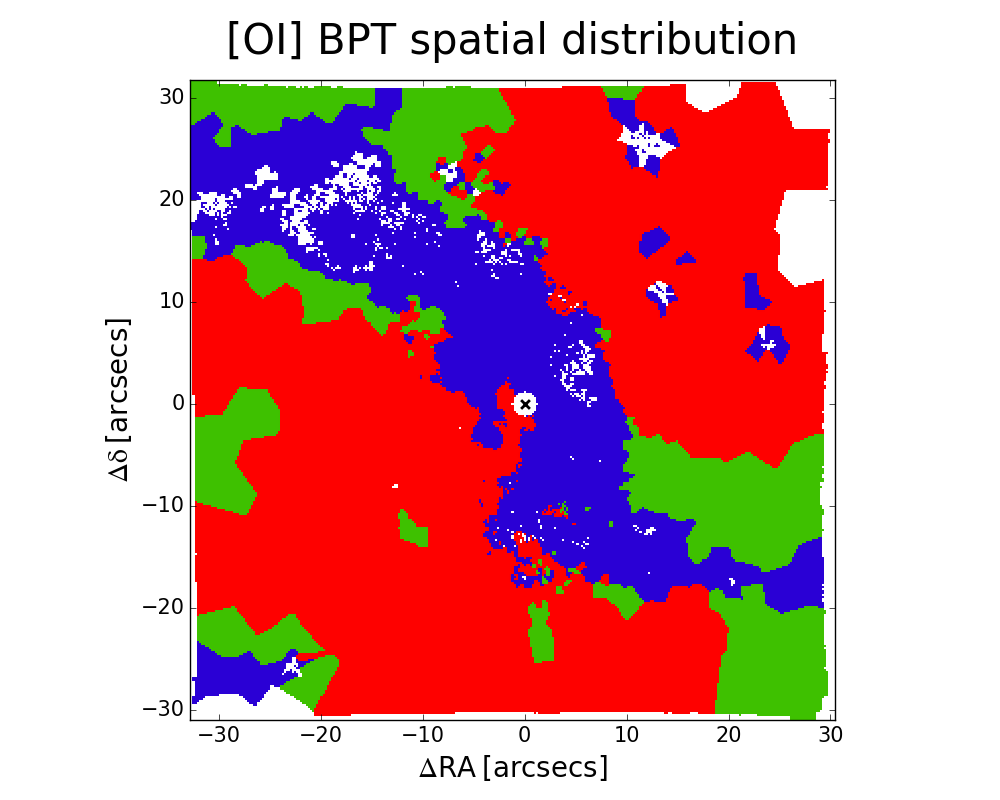}
	\end{subfigure}
	\caption{Spatially resolved BPT diagrams (left) of NGC 1365 and corresponding spatial distribution (right), obtained from the fit of the star-subtracted Voronoi-binned cube (so as to have an average signal-to-noise ratio of at least 4 per bin in each wavelength channel around H$\beta$). The N-BPT ([N\,\textsc{ii}]$\lambda$6584/H$\alpha$ vs. [O\,\textsc{iii}]$\lambda$5007/H$\beta$, upper panels), S-BPT ([S\,\textsc{ii}]$\lambda\lambda$6716,6731/H$\alpha$ vs. [O\,\textsc{iii}]$\lambda$5007/H$\beta$, central panels), and O-BPT ([O\,\textsc{i}]$\lambda$6300/H$\alpha$ vs. [O\,\textsc{iii}]$\lambda$5007/H$\beta$, lower panels) diagrams for each bin with S/N $>$ 3 in each line are shown on the left, while the corresponding maps are reported on the right (each bin in the maps corresponds to a single dot in the diagrams). The solid curves in the diagrams define the theoretical upper bound for pure star formation (\citealt{Kewley:2001ab}), while the dashed one in the N-BPT is the \cite{Kauffmann:2003aa} empirical classification, and all bins under this curve are star-formation dominated. Finally, the dashed line in S-BPT and O-BPT diagrams separates Seyfert galaxies from LI(N)ERs (\citealt{Kewley:2006aa}). SF-dominated regions are then marked in blue, Seyfert-type ionization is displayed in red, green marks LI(N)ER regions in the S-BPT and O-BPT, while light blue denotes composite regions in the N-BPT lying between \cite{Kewley:2001ab} and \cite{Kauffmann:2003aa} curves. A 6px-radius circle around the center has been masked, because of BLR contamination (see text).}
	\label{fig:bpt}
\end{figure*}

Hereafter, the cross in the maps marks the position of the nucleus, that is, of the peak of the broad H$\alpha$ emission and of the continuum between $\sim$6800-7000 \AA. In all the line maps reported from now on we exclude spaxels or bins having a signal-to-noise ratio lower than 3 on any of the lines involved. We define the signal-to-noise ratio of a line as the ratio between the peak value of the fitted line profile and the standard deviation of the data--model residuals of the fit around that line (within a range about 60 to 110 \AA\ wide, depending on the line).

[O\,\textsc{iii}] $\lambda$5007 ([O\,\textsc{iii}] hereafter) and H$\alpha$ flux maps are reported in Fig. \ref{fig:maps_flux} and in the two-color image in Fig. \ref{fig:maps_redd_ratio}a, where they are superimposed ([O\,\textsc{iii}] in green and H$\alpha$ in red) to better inspect their different spatial distributions, tracing the two distinct ionizing sources contributing to the line fluxes, namely the active nucleus and young stars in star-forming regions (this is quantitatively studied later on in this section). H$\alpha$ dominates in an elongated region spanning from NE to SW parallel to the bar direction, while [O\,\textsc{iii}] is higher in the perpendicular direction.

The [O\,\textsc{iii}] emission has a clear conical shape to the SE of the center and, even though a little bit less defined in shape, to the NW, too. 
The SE cone is in fact the strongest one of the two by far (see also \citealt{Edmunds:1988aa}, \citealt{Storchi-Bergmann:1991aa}, \citealt{Veilleux:2003aa}, \citealt{Sharp:2010aa}), being, according to \cite{Hjelm:1996aa}, in the near side, above the disk.

The H$\alpha$ spatial distribution is completely different from [O\,\textsc{iii}], being dominant in a $\sim$20$''\times$10$''$ (1.6 kpc $\times$ 0.8 kpc) elongated circumnuclear ring hosting hotspots, consistent with the star-forming ring reported for example by \cite{Kristen:1997aa} in narrowband {\it HST} images and by \cite{Forbes:1998aa} and \cite{Stevens:1999aa} in radio. 
As pointed out in \cite{Lindblad:1999aa}, circumnuclear rings of intense star formation are not infrequently observed in barred spirals and are ascribed to the presence of the inner Lindblad resonance (ILR; \citealt{Binney:2008aa}), with inflowing gas slowing down and accumulating between the outer and the inner ILR (having radius of $\sim$30$''$ and $\sim$3$''$ from the nucleus in NGC 1365, respectively; \citealt{Lindblad:1996ab}, \citealt{Lindblad:1999aa}), forming a massive ring with enhanced star formation (e.g.,\ \citealt{Shlosman:1996aa}).
\cite{Alonso-Herrero:2012aa} also identify the starburst ring in their IR study of NGC 1365, inferring that $\sim$85$\%$ of the star formation within the ILR ($r\sim30''$, fully covered by our $\sim$60$''\times$60$''$ MUSE data) is taking place in dust-obscured regions inside the star-forming ring. 
The effect of dust obscuration can be appreciated in Fig. \ref{fig:maps_redd_ratio}b, where we report the H$\alpha$ emission in red and the stellar continuum (collapsed between $\sim$5100-5800 \AA) in blue.
Here the dust lanes can be clearly seen (see also Fig. \ref{fig:n1365_eso}b), as well as some stronger H$\alpha$ blobs in correspondence with the lanes, indicating ongoing star formation inside them. Besides the circumnuclear ring, H$\alpha$ is very strong also at the leading edge of the northern dust lane, while it is not in correspondence with the southern dust lane, which is probably obscuring the H$\alpha$ emission at its leading edge on the other side with respect to the observer. There are also some isolated H$\alpha$-emitting blobs far from the central diagonal emission.

For a quantitative inspection of the dust reddening, we report in  Fig. \ref{fig:maps_redd_ratio}c the total extinction in $V$ band $A_V$, obtained from the Balmer decrement (H$\alpha$/H$\beta$ flux ratio) from the fit of the stellar continuum-subtracted Voronoi-binned cube. This cube provides information on regions which otherwise would have had low signal-to-noise ratio in H$\beta$. We employed the \cite{Calzetti:2000aa} attenuation law for galactic diffuse ISM ($R_V$ = 3.12) and an intrinsic ratio (H$\alpha$/H$\beta$)$_0$ = 2.86 (for an electron temperature of $T_e$ = 10$^4$ K; \citealt{Osterbrock:2006aa}). 
The nucleus has been masked since the bin-by-bin fit of the broad Balmer lines in the central bins (see Sect. \ref{ssec:data_anal}) was not optimal,
preventing a reliable estimate of the H$\alpha$/H$\beta$ ratio for the narrow line components.

Fig. \ref{fig:maps_redd_ratio}d shows the electron density map, obtained from the [S\,\textsc{ii}] $\lambda6716/\lambda6731$ diagnostic line ratio (\citealt{Osterbrock:2006aa}), assuming a typical value for the temperature of [S\,\textsc{ii}]-emitting ionized gas of $T_e$ = 10$^4$ K. 
As the [S\,\textsc{ii}] doublet line ratio is sensitive to the electron density only for intermediate values of the density, flattening at low and high values (see Fig. 5.8 in \citealt{Osterbrock:2006aa}), we set to 40 cm$^{-3}$ all the measured densities falling below this value (corresponding to $\sim$5$\%$ below the peak of the total range spanned by the ratios among the asymptotic extremes).
The map shows that the circumnuclear star-forming ring has the highest densities in the FOV, reaching a peak of $\sim$10$^3$ cm$^{-3}$ in a region to the north of the center.

Spatially resolved BPT diagnostic diagrams (Fig. \ref{fig:bpt}) allow us to distinguish the dominant contribution to ionization in each spaxel. As suggested by Fig. \ref{fig:maps_flux} and \ref{fig:maps_redd_ratio}, they show that star formation dominates gas ionization in the elongated region in the direction NE-SW,
 where [O\,\textsc{iii}]/H$\beta$ and [N\,\textsc{ii}]/H$\alpha$ are lower, while AGN dominates in a perpendicular double-conical region. 
This clear distinction between AGN- and SF-ionized regions is not reported in previous studies (e.g., \citealt{Sharp:2010aa}) due to the  lower sensitivity of their observations. 
We can also identify some star-forming (or composite) bins outside the main diagonal SF-dominated region, corresponding to strong H$\alpha$-emitting blobs showing up in the [O\,\textsc{iii}]-dominated double cone (see Fig. \ref{fig:maps_redd_ratio}a). 
Few bins are marked as LI(N)ER-like regions, whose line ratios can be either due to shock excitation (e.g., \citealt{Dopita:1995aa}) or to diffuse gas possibly ionized by hot evolved (post-asymptotic giant branch) stars (e.g., \citealt{Singh:2013aa}, \citealt{Belfiore:2016aa}). 
The circumnuclear star-forming ring seen in H$\alpha$ is not completely SF-dominated in the BPT maps, but some of its parts, to the SE of the center, result to be AGN-dominated. This likely happens because the AGN-ionized cone and the SF ring belonging to the disk reside in different parts of the 3D physical space but overlap in the l.o.s. Depending on which of the two components dominates the total line profiles, each bin then occupies the SF or the AGN part of the BPT diagram.

\begin{figure*}[p]
	\centering
	\null\hfill
	\begin{subfigure}[t]{0.01\textwidth}
	\textbf{a}
	\end{subfigure}
	\begin{subfigure}[t]{0.43\textwidth}
	\includegraphics[trim={2cm 0.46cm 0.5cm 0},clip,width=\textwidth,valign=b]{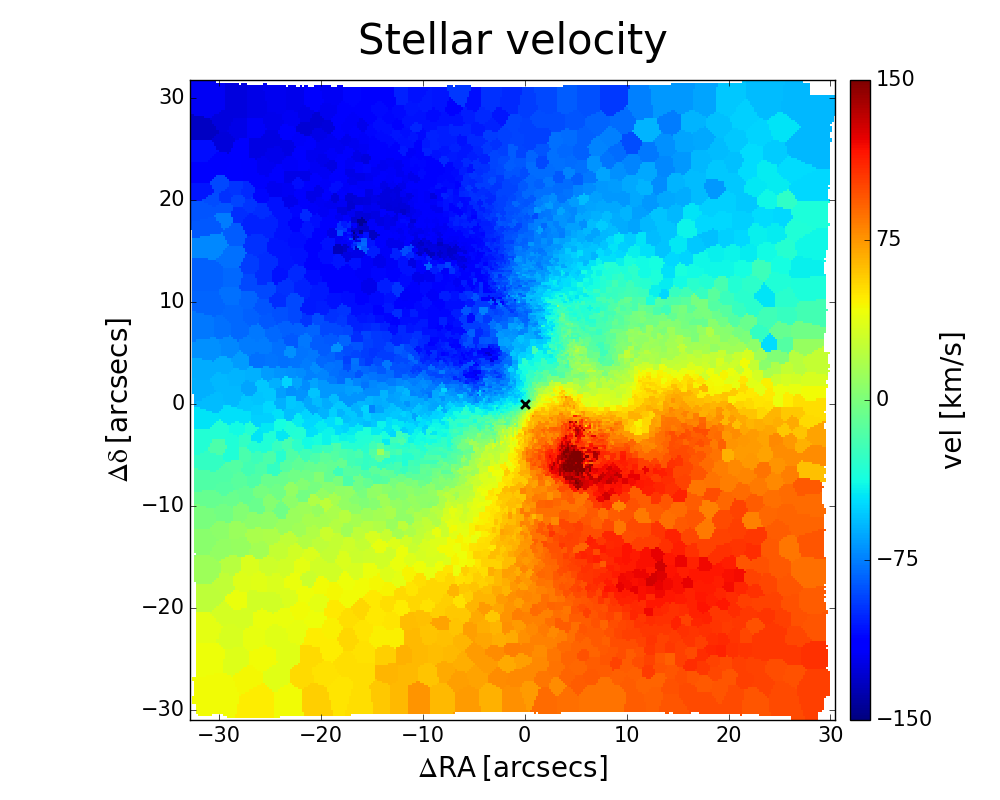}
	\end{subfigure}
	\hfill
	\begin{subfigure}[t]{0.01\textwidth}
	\textbf{b}
	\end{subfigure}
	\begin{subfigure}[t]{0.43\textwidth}
	\includegraphics[trim={2cm 0.46cm 0.5cm 0},clip,width=\textwidth,valign=b]{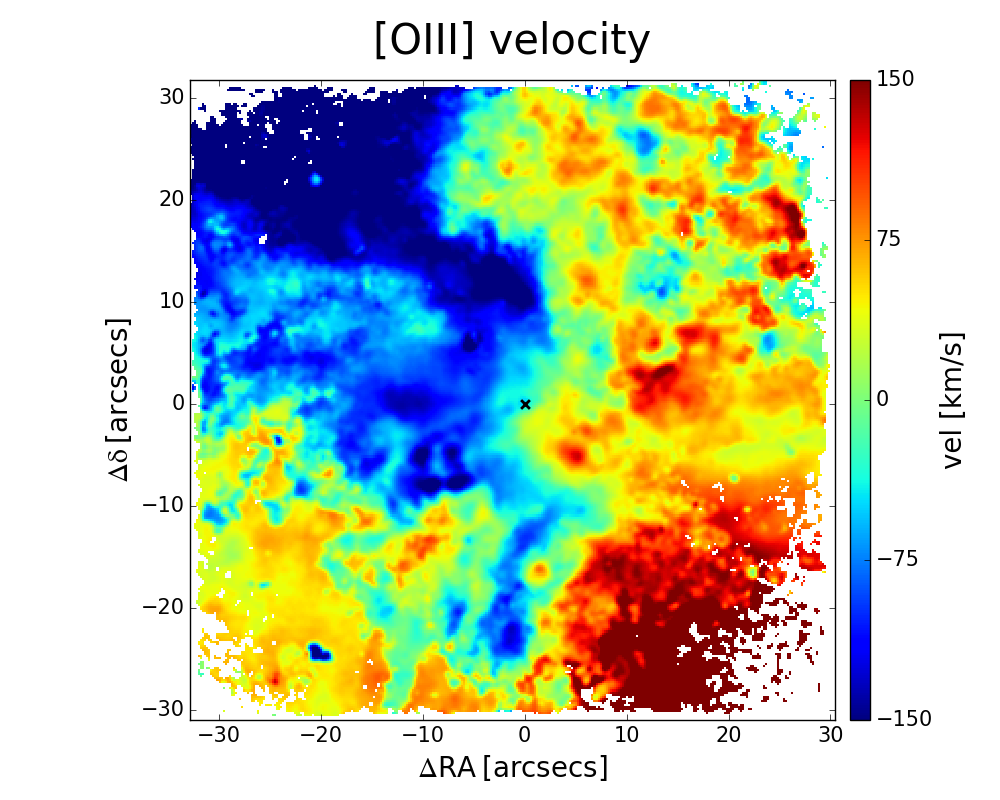}
	\end{subfigure}
	\hfill\null\\
	\null\hfill
	\begin{subfigure}[t]{0.01\textwidth}
	\textbf{c}
	\end{subfigure}
	\begin{subfigure}[t]{0.43\textwidth}
	\includegraphics[trim={2cm 0.46cm 0.5cm 0},clip,width=\textwidth,valign=b]{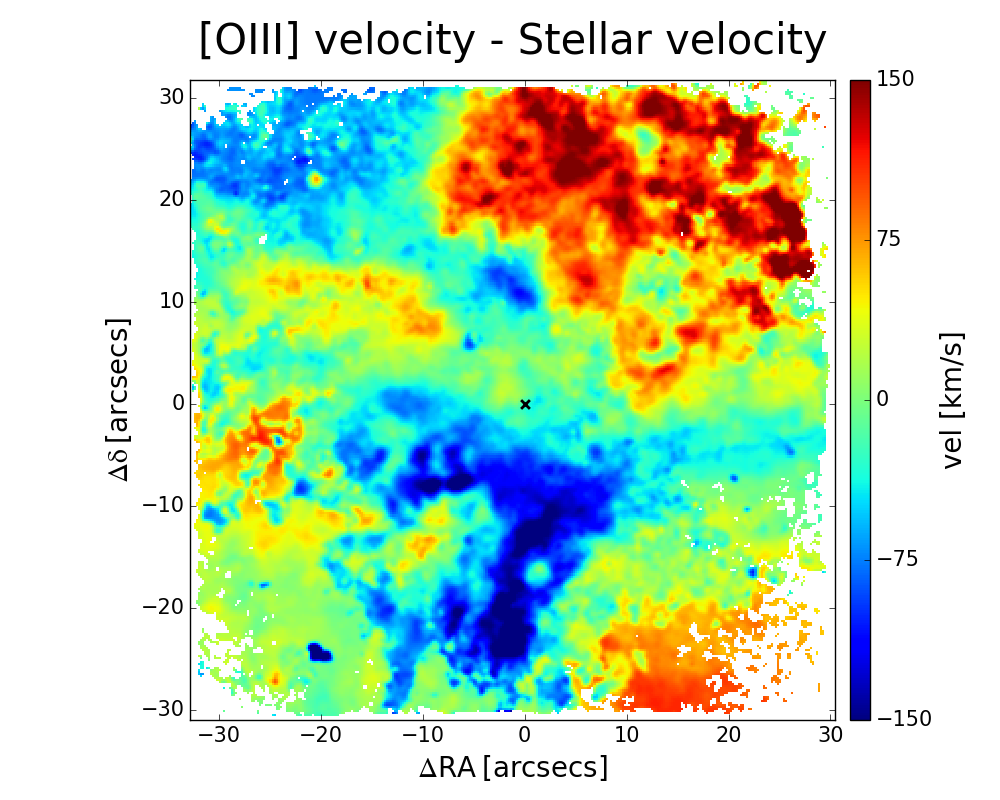}
	\end{subfigure}
	\hfill
	\begin{subfigure}[t]{0.01\textwidth}
	\textbf{d}
	\end{subfigure}
	\begin{subfigure}[t]{0.43\textwidth}
	\includegraphics[trim={2cm 0.46cm 0.5cm 0},clip,width=\textwidth,valign=b]{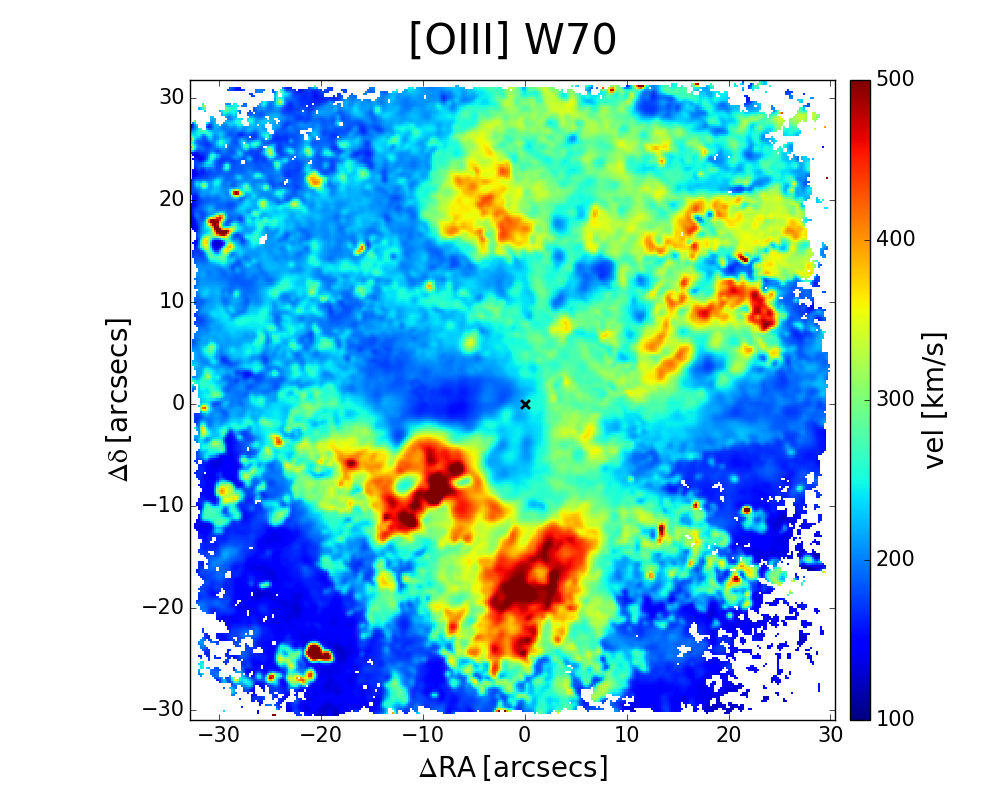}
	\end{subfigure}
	\hfill\null\\
	\null\hfill
	\begin{subfigure}[t]{0.01\textwidth}
	\textbf{e}
	\end{subfigure}
	\begin{subfigure}[t]{0.43\textwidth}
	\includegraphics[trim={2cm 0.46cm 0.5cm 0},clip,width=\textwidth,valign=b]{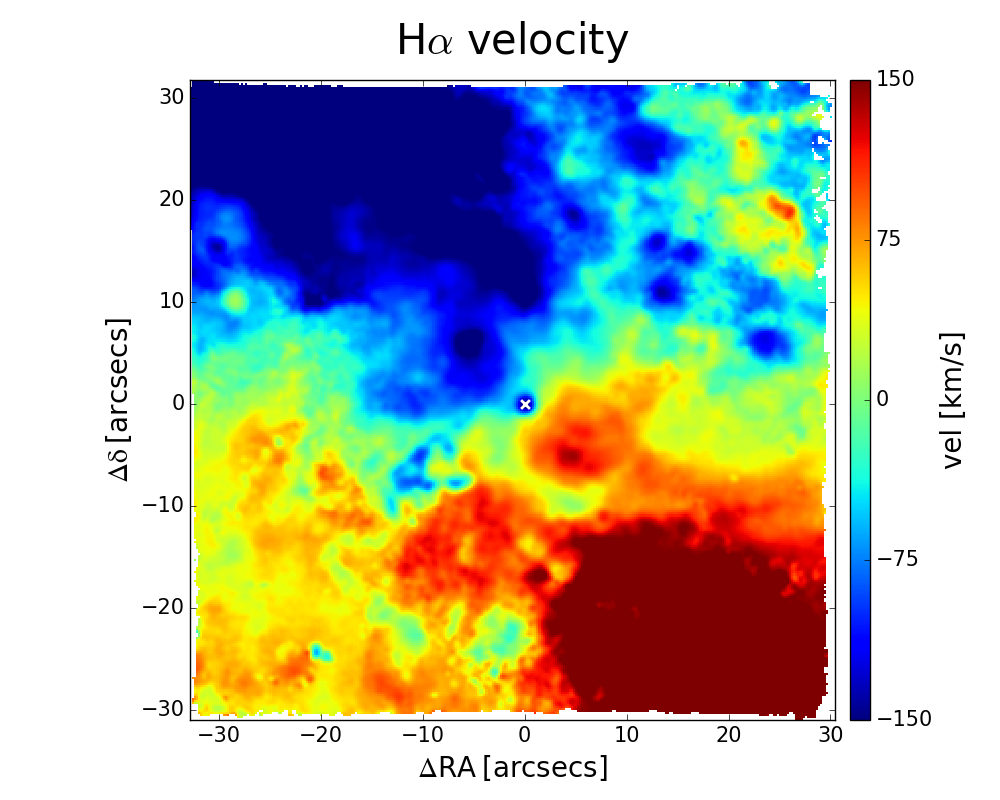}
	\end{subfigure}
	\hfill
	\begin{subfigure}[t]{0.01\textwidth}
	\textbf{f}
	\end{subfigure}
	\begin{subfigure}[t]{0.43\textwidth}
	\includegraphics[trim={2cm 0.46cm 0.5cm 0},clip,width=\textwidth,valign=b]{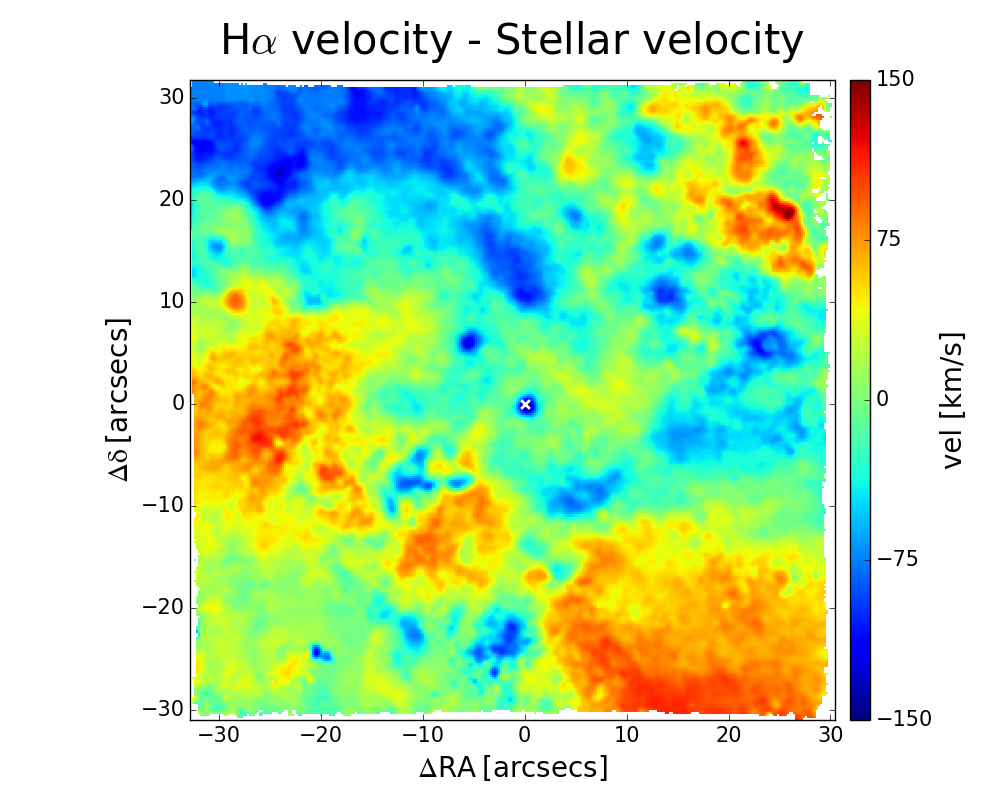}
	\end{subfigure}
	\hfill\null
	\caption{{\bf(a)} Stellar velocity map of NGC 1365, with respect to its systemic velocity, for which a value of $1630$ km s$^{-1}$ with respect to Earth has been considered. The map has been obtained from the fit of stellar continuum and absorption lines carried out on the Voronoi-binned data cube. {\bf(b)} [O\,\textsc{iii}] velocity map with respect to the systemic velocity. The velocities are the 1st-order moments of the total line profile resulting from the fit of the star-subtracted smoothed (with 1-spaxel $\sigma$ gaussian kernel) data cube. The map has been re-smoothed with a Gaussian kernel having $\sigma$ = 1 spaxel, for a better visual output, and a signal-to-noise cut of 3 has been applied (this holds for the other maps reported in panels c, d, and e of this figure, too). {\bf(c)} [O\,\textsc{iii}] velocity subtracted spaxel-by-spaxel by the stellar velocity. {\bf(d)} [O\,\textsc{iii}] W70 map, i.e\ difference between the 85th-percentile and 15th-percentile velocities of the fitted line profile. {\bf(e)} H$\alpha$ velocity map with respect to the systemic velocity. {\bf(f)} H$\alpha$ velocity map spaxel-by-spaxel subtracted by the stellar velocity.}
	\label{fig:maps_kin}
\end{figure*}

\subsection{Kinematic maps}\label{ssec:kinmaps}
The MUSE data allow us to trace the kinematics of both the warm ionized gas and the stars in NGC 1365 (Fig. \ref{fig:maps_kin}), from the Doppler shift of emission and absorption lines, respectively. 
Stellar and [O\,\textsc{iii}] velocity maps are shown in Fig. \ref{fig:maps_kin}a and \ref{fig:maps_kin}b, respectively, and, in both cases, the velocity is reported in the galaxy rest frame. 
From the stellar velocity map one can clearly see the rotation of the stars in the galactic disk, approaching the observer to the NE, receding to the SW. The boundary where the velocity changes sign shows a twisted shape, likely associated to the presence of the bar. 
The [O\,\textsc{iii}] velocity map (Fig. \ref{fig:maps_kin}b) shows other complex motions in addition to the rotational field. In order to isolate and highlight these further motions of the gas with respect to stellar rotation, we report in Fig. \ref{fig:maps_kin}c the map of [O\,\textsc{iii}] velocity subtracted spaxel-by-spaxel by the stellar velocity. This reveals a clumpy double-conical outflow approximately in the direction SE to NW, the SE cone having approaching velocities, the NW having receding ones. The outflow shape broadly corresponds to the double cone observed in [O\,\textsc{iii}] emission (Fig. \ref{fig:maps_flux}a and \ref{fig:maps_redd_ratio}a), which is AGN-photoionized according to the BPT diagrams (Fig. \ref{fig:bpt}).

The shape and structure of the outflow can be well appreciated also from the map of the [O\,\textsc{iii}] W70 velocity\footnote{That is, the difference between the velocities including 85$\%$ and 15$\%$ of the flux of the total fitted line profile (without any deconvolution for the instrumental spectral resolution).}, where values exceeding 500 km s$^{-1}$ are present along the outflow. The high values of the W70 are due to the fact that it is calculated from the total fitted profile of [O\,\textsc{iii}], which, in the regions of the outflow, is composed by the kinematic component belonging to disk and bar and by the outflowing one. Double-peaked profiles are in fact ubiquitous in our data in correspondence with the regions having higher W70 values (see e.g., the spectra in Fig. \ref{fig:moutrate_spectra}).
\cite{Phillips:1983aa} observed line splitting of [O\,\textsc{iii}] for the first time to the SE of the center, proposing a model of a hollow conical outflow perpendicular to the disk to explain them. \cite{Hjelm:1996aa} suggested that line splitting could be due to two spectral components, one from the rotating disk, one from an outflow. They modeled the outflow as a wide-angle cone ($50^{\circ}$ half-opening angle), inclined of $5^{\circ}$ with respect to the rotation axis and of $35^{\circ}$ with respect to the l.o.s. (being the disk inclined of 40$^{\circ}$ with respect to the l.o.s.; \citealt{Jorsater:1995aa}). 
This picture seems to be compatible with the [O\,\textsc{iii}] maps presented in this work, with the approaching SE cone residing between the observer and the disk of the galaxy and the receding NW one being behind the disk, thus having a lower observed [O\,\textsc{iii}] flux compared to the SE one. Recently, \cite{Lena:2016aa} identified the inner part of the SE blueshifted outflow from [N\,\textsc{ii}] kinematics maps using 13$''\times$6$''$ IFS data, observing line splitting also in other low-ionization lines (H$\alpha$, [S\,\textsc{ii}]) in a few nearby spaxels. 

Ionized gas kinematics in the disk can be better inspected from the velocity of H$\alpha$ rather than from [O\,\textsc{iii}], as the total profile of Balmer lines is less affected by the outflow compared to the [O\,\textsc{iii}]. 
This happens because the [O\,\textsc{iii}]-to-Balmer lines flux ratio is higher in AGN-dominated regions than in star-forming regions and in NGC 1365 the outflowing component is much more dominated by the AGN ionization rather than the disk one (see e.g., the spectra in Fig. \ref{fig:moutrate_spectra} from following Sect.).
The H$\alpha$ velocity map is reported in Fig. \ref{fig:maps_kin}e and Fig. \ref{fig:maps_kin}f, in the latter subtracted spaxel-by-spaxel by the stellar rotational motion. 
The maps show that the double-conical outflow, which is very prominent in [O\,\textsc{iii}], 
has a much lower impact on the line profile of H$\alpha$\footnote{We stress that the gas velocity reported for a given line is the 1st-order moment of the fitted line profile and can then change from line to line depending on the shape of the profile.}. 
The H$\alpha$ kinematics is instead dominated by two thick lanes aligned with the bar (in the direction NE-SW; see Fig. \ref{fig:n1365_eso}a) having opposite velocities in excess to stellar rotation, the one in the lower-left part of the image receding, the one in the upper-right part approaching. Non-circular motions associated with the bar of NGC 1365 were already identified by \cite{Teuben:1986aa} in H$\alpha$ and then by \cite{Jorsater:1995aa} in H\,\textsc{i} velocities, which \cite{Lindblad:1996aa} combined with optical ones inferring gas motions parallel to the bar, with strong velocity gradients across its leading side. 
\cite{Lena:2016aa} found no obvious evidence for an inflow from their [N\,\textsc{ii}] kinematic map in the central 13$''\times$6$''$, but they do not exclude the possibility that gas is actually slowly migrating toward the nucleus. By using large-scale Fabry-Perot interferometric H$\alpha$ data,
\cite{Speights:2016aa} found clear non-circular motions consistent with elliptical streaming in correspondence with the bar, compatible with what we see in our H$\alpha$ kinematic map. This non-circular motion likely indicates an inflow of material along the bar of NGC 1365.

\subsection{Mass outflow rate mapping}\label{ssec:moutrate}
\begin{figure}
\includegraphics[width=0.5\textwidth]{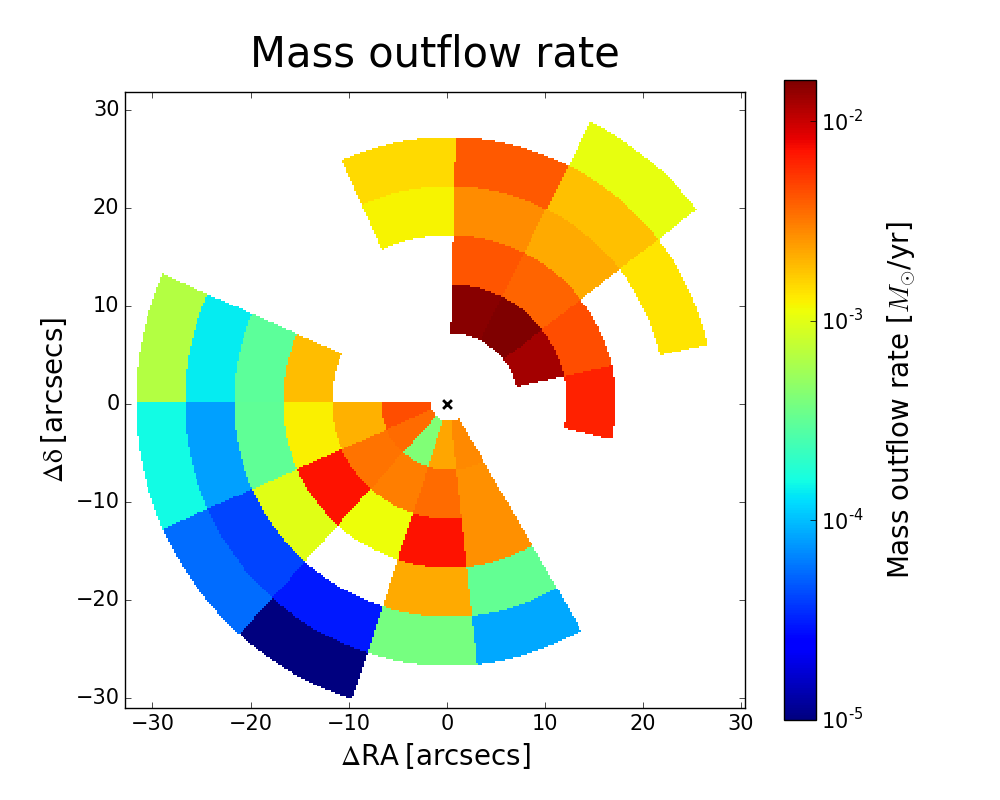}
\caption{Mass outflow rate map of the ionized gas, from the fitted H$\alpha$ flux of the outflow kinematic component (see example spectra in Fig. \ref{fig:moutrate_spectra}).}
\label{fig:moutrate}
\end{figure}
The extension of the outflow across the FOV of MUSE allowed us to carry out a radial and angular analysis of the mass outflow rate.
We grouped the spaxels in the two outflowing cones in a grid made up of radial and angular slices, excluding a circular area of radius 10 spaxels centered on the active nucleus, where the BLR contributes to Balmer line emission. The grid is made up of six radial slices per cone, 25 spaxels in size each (i.e., 5 arcsec or $\simeq$ 420 pc) and six angular slices spanning 24$''$ each for the SE cone and five spanning 25.4$''$ each for the NW one (see Fig. \ref{fig:moutrate}).

\begin{figure*}[t]
\centering
\begin{subfigure}[t]{0.01\textwidth}
\textbf{a}
\end{subfigure}
\begin{subfigure}[t]{0.8\textwidth}
\includegraphics[trim={0 2cm 0 2cm},clip,width=\textwidth,valign=b]{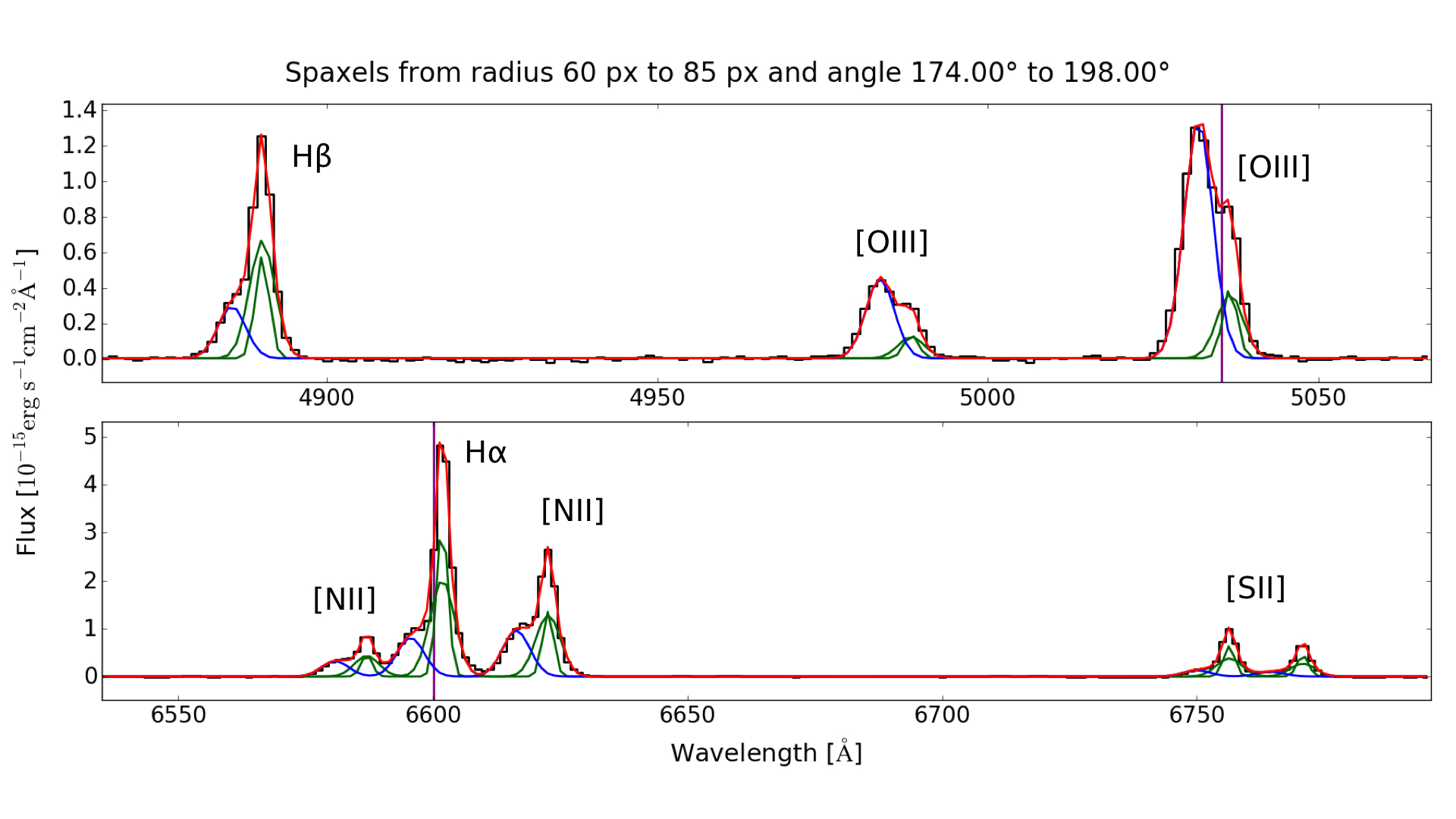}
\end{subfigure}\\
\vspace{0.5cm}
\begin{subfigure}[t]{0.01\textwidth}
\textbf{b}
\end{subfigure}
\begin{subfigure}[t]{0.8\textwidth}
\includegraphics[trim={0 2cm 0 2cm},clip,width=\textwidth,valign=b]{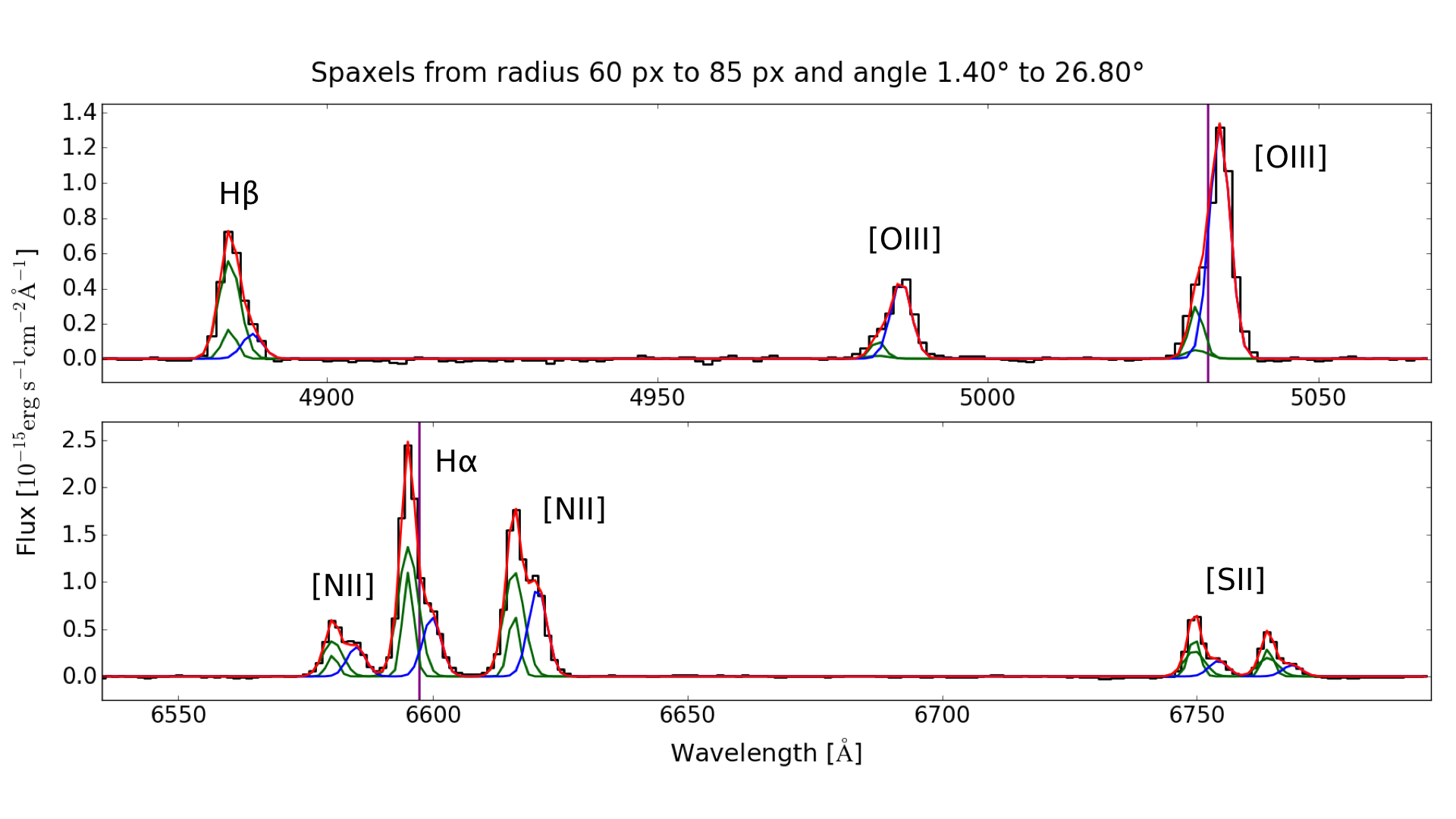}
\end{subfigure}
\caption{Fit of [O\,\textsc{iii}], H$\beta$, [N\,\textsc{ii}], H$\alpha$, and [S\,\textsc{ii}] lines for two elements of the grid through which we divided the outfowing cones (see Fig. \ref{fig:moutrate}). The spectrum in \textbf{(a)} belongs to the SE blueshifted cone, the spectrum in \textbf{(b)} to the NW redshifted one. The blue Gaussian corresponds to the outflow kinematic component, the two green Gaussians (having the same velocity) to the disk; the velocity and velocity dispersion of all the Gaussians are tied between the lines. The purple vertical lines at [O\,\textsc{iii}] and H$\alpha$ wavelengths mark the stellar velocity for the element of the grid considered, obtained from the stellar kinematics presented in Fig. \ref{fig:maps_kin}a. The reported wavelengths are in the Earth rest frame.}
\label{fig:moutrate_spectra}
\end{figure*}

We fitted [O\,\textsc{iii}], H$\beta$, [N\,\textsc{ii}], H$\alpha$, and [S\,\textsc{ii}] emission lines all together using three Gaussians per line, forcing each Gaussian separately to have the same velocity and velocity dispersion among all the emission lines.
One component has been dedicated to fit the outflow (blueshifted in the SE cone, redshifted in the NW one), the other two - tied to the same velocity - to fit the disk component. 
In fact, the disk component, despite being symmetric, was not reproduced with a simple Gaussian profile, probably because of the presence of the bar.

Due to the limited number of grid elements we could inspect the fits one-by-one and keep them under control.
We immediately excluded those fits where the outflow component had null flux in H$\alpha$ (as we use H$\alpha$ to calculate the outflow mass) or where the line profiles showed no significant asymmetry, meaning that either no outflow is present or that it has the same l.o.s. velocity of the disk and it is thus not possible to disentangle the two.

As an example, Fig. \ref{fig:moutrate_spectra} displays a fit for the SE blueshifted cone and a fit for the NW redshifted cone. The two spectra show that the [O\,\textsc{iii}] is largely dominated by the outflowing component (blue Gaussian), whereas in the other lines the dominant component is the disk one (the two green Gaussians). [O\,\textsc{iii}]/H$\beta$ is thus much higher in the outflow component compared to the disk one. [N\,\textsc{ii}]/H$\alpha$ is also higher in the outflow component. This holds true in general also for all the other grid elements, whose [O\,\textsc{iii}]/H$\beta$ and [N\,\textsc{ii}]/H$\alpha$ line ratios for the outflow component reside in the AGN-ionized part of the BPT diagrams. 
The disk component does not have the same velocity of the stars, indicating that, besides the outflowing gas, the gas in the disk does not follow the stellar rotation either, having redder velocities in the SE part (panel a) and bluer in the NW part (panel b). This effect is found also in the other grid elements and is exactly what emerged from the H$\alpha$ velocity map subtracted by stellar velocity in Fig. \ref{fig:maps_kin}f, where non-circular motions in excess to stellar rotation came out in the direction of the bar of the galaxy, red in the SE, blue in the NW.

We calculated the outflow mass of ionized gas in each element of the grid from the flux of the outflowing component of H$\alpha$, using the following relation from \cite{Cresci:2017aa}, which assumes ``Case B'' recombination in fully ionized gas with electron temperature $T_e$ = 10$^4$ K: 
\begin{equation}
M_{\textrm{out}} = 3.2 \times 10^5 M_{\odot} \left( \dfrac{L_{\textrm{H}\alpha\textrm{,out}}}{10^{40} \textrm{ erg s}^{-1}} \right) \left( \dfrac{n_e}{100 \textrm{ cm}^{-3}} \right) ^{-1}.
\label{eq:mout}
\end{equation}
In order to get the intrinsic H$\alpha$ flux to be converted to luminosity, the measured H$\alpha$ outflow component has been dereddened using the Balmer decrement H$\alpha$/H$\beta$ (of the outflow component itself). 
However, we adopted the Balmer decrement from the entire circular shells at the same distance (separately for the two cones), rather than from the single grid elements, to increase the accuracy of the reddening correction. Indeed, the fluctuations of the Balmer decrement within each shell were consistent with being roughly constant. 
As expected, the outflow component results to be much more reddened in the NW cone than in the SE one, the former lying behind the galactic disk, the latter residing above the disk, with values of H$\alpha$/H$\beta$ going from $\simeq$4.7 to $\simeq$7.2 ($A_V \simeq 1.3$ to $\simeq 2.5$) in the NW cone, while the SE cone has a typical ratio of 3.7 ($A_V \simeq$ 0.7), with only a maximum of $\simeq$4.2 ($A_V \simeq$ 1.1).
We stress that the SE outflowing cone is extincted even though it resides above the disk\footnote{The SE outflowing cone has in fact approaching velocities and, in addition, the SE part of the FOV corresponds to the far part of the galactic disk, residing behind the plane of the sky (this follows from the comparison of the large-scale morphology of the galaxy in Fig. \ref{fig:n1365_eso}a with the stellar rotational field in Fig. \ref{fig:maps_kin}a, together with the fact that the spiral arms are trailing in the rotation of the galaxy, as reported by \citealt{Lindblad:1999aa}).}, indicating the presence of dust inside it (signatures of a dusty phase in outflows are found e.g., in \citealt{Rupke:2015aa}, \citealt{Perna:2015aa}; polar MIR emission from dust associated with the NLR has been observed on scales of pc up to tens of pc in some AGN, see \citealt{Asmus:2016aa}, \citealt{Lopez-Gonzaga:2016aa}, \citealt{Stalevski:2017aa}). This is an interesting finding as the dust might be coupled with molecular gas, which would provide the material to potentially form stars within the outflow (as observed for the first time by \citealt{Maiolino:2017aa}).

The electron density $n_e$ has been obtained from the [S\,\textsc{ii}] $\lambda$6716/$\lambda$6731 line doublet ratio of the outflow component from the fit of the whole shells (still separately for the two cones), rather than from the single elements of the grid, for the same reason of the Balmer decrement.

Both for H$\alpha$/H$\beta$ and for $n_e$ a few values resulting from bad fits have been replaced with the average from the two nearest shells or with the value of the nearest shell in case the shell of interest was the most internal or the most external one. Moreover, as done in Sect. \ref{ssec:flux+bpt}, we set to 40 cm$^{-3}$ all the calculated densities falling below this value, being the [S\,\textsc{ii}] ratio insensitive to densities below this threshold.

We thus obtain a total ionized outflow mass from both cones of $\sim$ 9.2 $\times$ 10$^5$ $M_{\odot}$ and a total kinetic energy of $\sim$ 4.5 $\times$ 10$^{52}$ erg, by summing the values from the single grid elements.

The mass outflow rate through a spherical surface of radius $r$ subtended by a solid angle $\Omega$ is $\dot{M}(r) = \Omega r^2 \rho(r) v(r)$. Assuming that, within each grid element, density and outflow velocity are constant, the radial average of the outflow rate within a grid element is simply 
\begin{equation}
\dot{M}_{\textrm{out}} = \dfrac{M_{\textrm{out}} v_\textrm{out}}{\Delta R},
\label{eq:moutrate}
\end{equation}
where $v_\textrm{out}$ is the velocity of the outflow component in the galaxy rest frame and $\Delta R$ the radial width of each grid element (i.e., $\simeq$ 420 pc). This outflow rate can then be interpreted as the mass of gas $M_{\textrm{out}}$ flowing through each radial element of width $\Delta R$ in a time $\Delta R / v_\textrm{out}$.
When using the observed outflow velocities we do not take into account geometrical projection effects and therefore our estimates are more likely lower limits. 
To obtain de-projected velocities a detailed 3D kinematic model is required but this is beyond the scope of this paper.

The gridded map of the mass outflow rate can be inspected in Fig. \ref{fig:moutrate}. 
The mass outflow rate exhibits both radial and angular variations, as a consequence of the clumpiness and inhomogeneity of the outflowing gas (see kinematic maps in Fig. \ref{fig:maps_kin}).
We use this map to build radial profiles of the outflow physical quantities, which are presented later on in Sect. \ref{ssec:outflow} where they are extensively discussed in relation with the properties of the nuclear X-ray outflow.

\begin{figure}[t]
\centering
\includegraphics[width=\columnwidth]{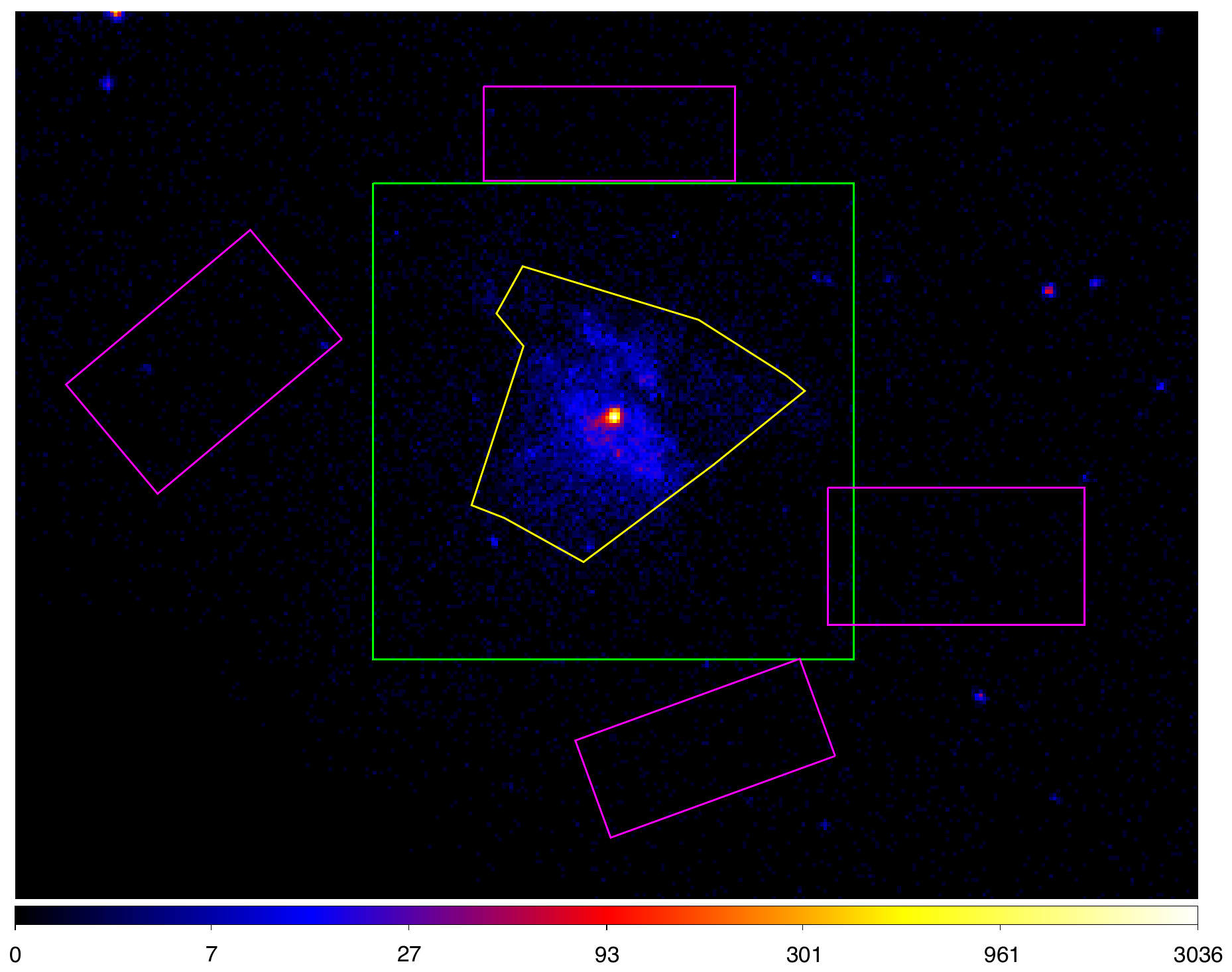}
\caption{Full-band (0.3--8 keV) image obtained by merging the six {\it Chandra} ACIS-S observations of NGC 1365 of April 2006. The green box indicates the MUSE FOV. A spectrum for each of the six observations has been extracted from the yellow polygonal region, while the background has been evaluated over the magenta regions.}
\label{fig:x_merged}
\end{figure}

\section{Chandra X-ray data analysis}\label{sec:x_rays}
NGC 1365 shows diffuse emission with roughly biconical morphology also in the soft X-rays (\citealt{Wang:2009aa}), so it is worth comparing the properties of this hot gas component with those of the [O\,\textsc{iii}]-emitting gas.

In this section we illustrate our analysis of six {\it Chandra} ACIS-S observations of NGC 1365 (ObsIDs 6868--6873). We note that {\it Chandra} is the only X-ray observatory affording comparable spatial resolution to our MUSE data.

\begin{figure*}[t]
\centering
\begin{subfigure}[t]{0.01\textwidth}
\textbf{a}
\end{subfigure}
\begin{subfigure}[t]{0.45\textwidth}
\includegraphics[width=\textwidth,valign=b]{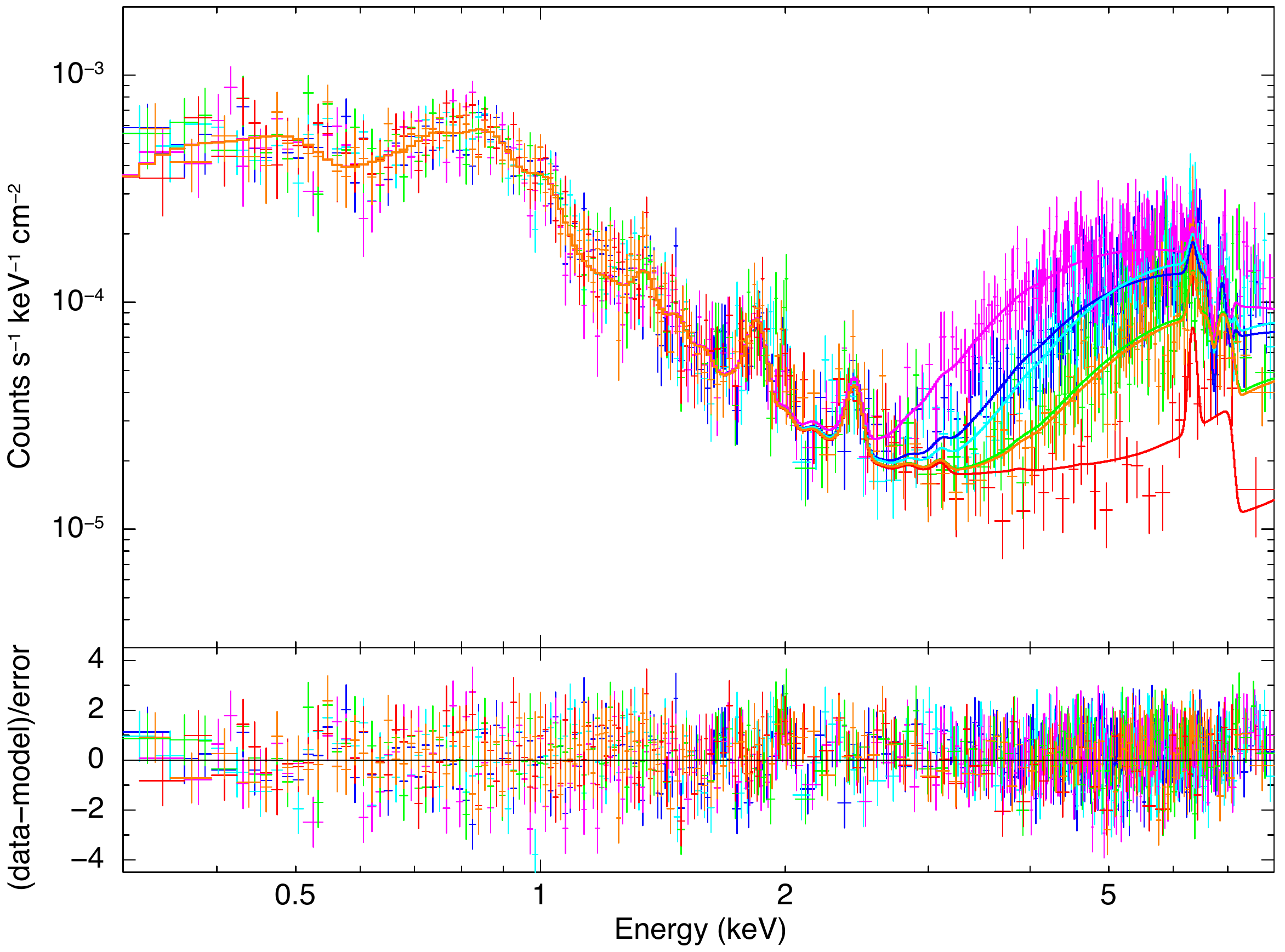}
\end{subfigure}
\qquad
\begin{subfigure}[t]{0.01\textwidth}
\textbf{b}
\end{subfigure}
\begin{subfigure}[t]{0.455\textwidth}
\includegraphics[width=\textwidth,valign=b]{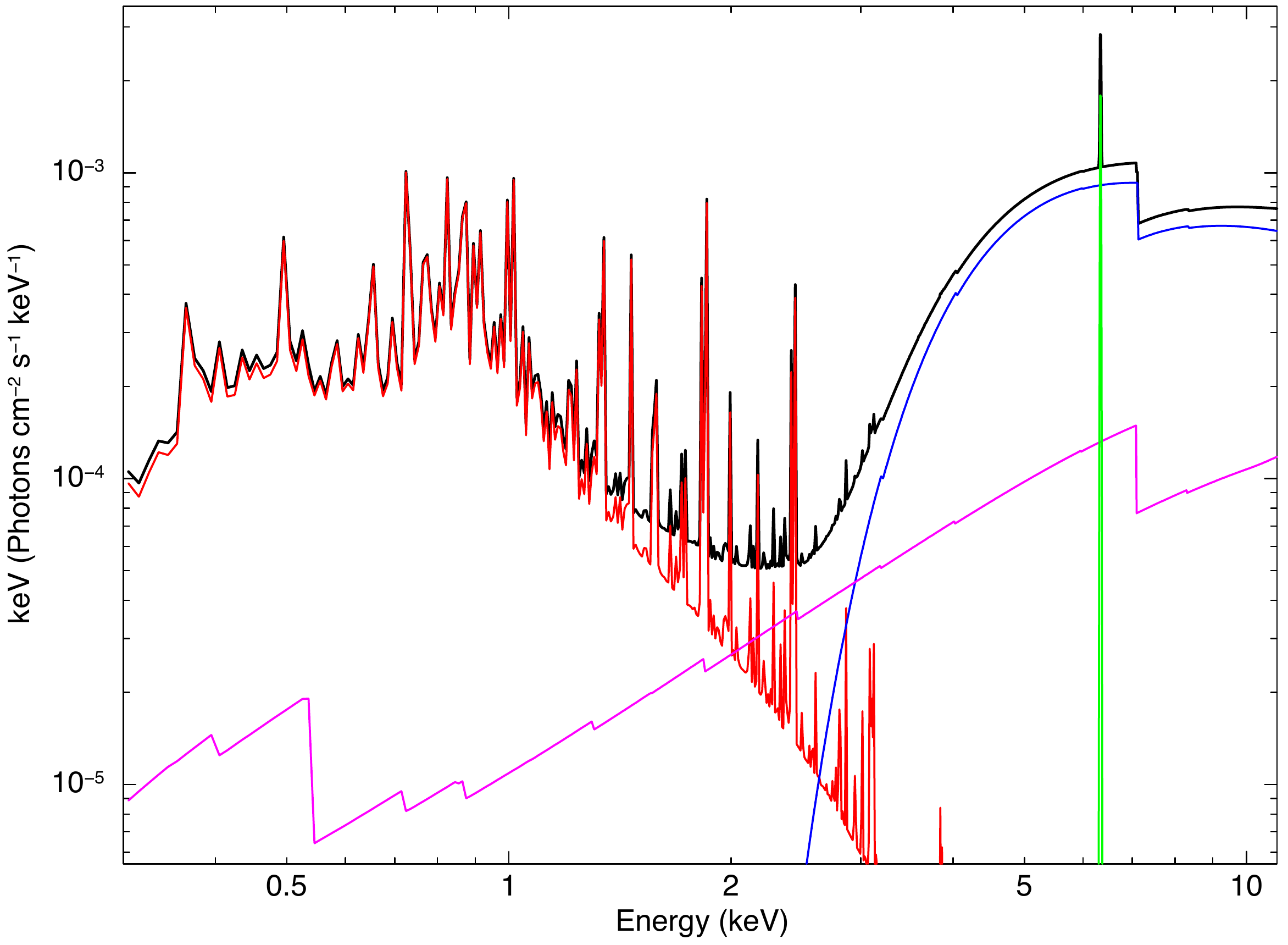}
\end{subfigure}
\caption{{\bf(a)} Spectra of each of the six {\it Chandra} observations (distinguished with different colors: blue, cyan, magenta, green, red, orange in chronological order), together with the best-fitting models (solid lines of the corresponding colors). The residuals are reported in the bottom panel. {\bf(b)} Physical model corresponding to the strongest spectrum in panel a (ObsID 6870). The single components are shown with the following color code: red for the absorbed \texttt{vapec}, thermal emission; blue for the absorbed \texttt{zpowerlaw}, primary X-ray emission; magenta for the \texttt{pexrav}, reflected continuum; green for the Gaussian emission line, fluorescent Fe\,\textsc{i} K$\alpha$ at 6.4 keV. Ionized absorption barely affects the continuum and is omitted for clarity.}
\label{fig:fit_x}
\end{figure*}

\subsection{Data description}
The data have been taken with the Advanced CCD Imaging Spectrometer (ACIS; \citealt{Garmire:2003aa}), positioning the galaxy inside the back-illuminated ACIS-S S3 chip, which among the ACIS-S chips has the highest energy resolution ($\simeq$74 eV at 1.49 keV and $\simeq$147 eV at 5.9 keV). The six snapshots, with exposure times ranging from $\sim$13 ks to $\sim$15 ks, for a total of $\sim$87 ks, were acquired between April 10 and April 23, 2006.
The data have been processed with the Chandra Interactive Analysis of Observations v4.6 (CIAO\footnote{\url{http://cxc.harvard.edu/ciao/}.}; \citealt{Fruscione:2006aa}), using the Chandra Calibration Database (CALDB\footnote{\url{http://cxc.harvard.edu/caldb/}.}) v4.6.3 calibration files and following the standard reduction thread. 

\subsection{Full-band spectral fitting}\label{ssec:fit_Xfull}
In Fig. \ref{fig:x_merged} we report the full-band (0.3--8 keV) X-ray image resulting from the combination of the six {\it Chandra} ACIS-S exposures of NGC 1365.
We extracted one spectrum per observation from the yellow polygonal area inside the FOV of MUSE (green box in Fig. \ref{fig:x_merged}), while the spectrum for the background-subtraction was taken from four rectangular regions free of strong point-like sources. 
The six spectra, which were rebinned so as to have a signal-to-noise ratio of at least 3 per energy channel, can be inspected in Fig. \ref{fig:fit_x}. 
The count rates in the plot are corrected for the energy-dependent effective area (this holds also for the other X-ray spectra presented in the following). 
We are mostly interested in the soft X-rays, as our goal is to match the extended X-ray emission with the optical one from MUSE, but we also fitted the hard X-ray part of the spectrum to isolate the unresolved emission from the active nucleus and its contribution at soft energies ($\lesssim$ 2 keV). The spectra show that the hard X-ray flux varies among the observations, as the nucleus of NGC 1365 is known to alternate between obscured and unobscured states (\citealt{Risaliti:2007aa}).

The spectra have been fitted using the X-ray spectral fitting package \textsc{xspec} v12.9.1 (\citealt{Arnaud:1996aa}), which is part of the NASA's 
HEASoft\footnote{\url{https://heasarc.gsfc.nasa.gov/lheasoft/}.} release v6.20. 
As no variation can arise from kpc-scale extended emission on short timescales, a constant model (red solid line in Fig. \ref{fig:fit_x}b) has been employed to reproduce the soft part of the spectrum of all the six observations, while the components adopted to fit the nuclear emission have been left free to vary among the observations. The Galactic column density, fixed to $N_\textrm{H} = 1.34 \times 10^{20}$ cm$^{-2}$ (\citealt{Kalberla:2005aa}), is included in each fit from now on.
The total model we employed is made up of the following \textsc{xspec} components:
\begin{itemize}
\item \texttt{vapec:} emission spectrum from collisionally-ionized optically-thin gas calculated from the AtomDB atomic database\footnote{\url{http://atomdb.org}.}, attenuated by foreground absorption (modeled with \texttt{phabs}). Following the approach of \cite{Wang:2009aa}, the abundances of O, Ne, Mg, Si, and Fe have been left free to vary, while Solar abundance has been assumed for all the other elements. This component has been introduced to fit the soft diffuse emission. 
\item \texttt{zpowerlw:} simple power law for the primary X-ray continuum from the AGN, in the rest frame of the galaxy. This is multiplied by \texttt{phabs} to account for the absorption due to neutral material along the line of sight. A common photon index has been used for all observations, while the normalization of the continuum and the column density of the absorber have been left free to vary to reproduce the different flux states.
\item \texttt{pexrav:} power-law spectrum reprocessed by neutral material (\citealt{Magdziarz:1995aa}), accounting for the fraction of X-ray continuum reflected by distant material into the line of sight. 
\item One \texttt{zgauss:} one Gaussian profile, 
employed to fit the Fe\,\textsc{i} K$\alpha$ fluorescent line at 6.4 keV (rest frame) produced by the distant material irradiated by the nuclear emission. The width of the line is expected to be much smaller than the energy resolution of the instrument ($\sim$150 eV), so it has been fixed to 10 eV. 
\item Two \texttt{zgauss:} two Gaussian profiles to fit the ionized Fe\,\textsc{xxv} and Fe\,\textsc{xxvi} K$\alpha$ absorption lines (at 6.7 keV and 6.97 keV rest-frame energy, respectively) due to foreground highly-ionized gas absorbing the primary nuclear continuum, which are usually blueshifted due to their origin within an accretion-disk wind. The observed energy is forced to be the same for all the observations, while the normalizations were left free to vary.
\end{itemize}

The best-fitting models, matching in color with the respective spectra, are superimposed to the data in Fig. \ref{fig:fit_x}a, while the various components (corresponding to the highest nuclear state, in the third observation) are disentangled in Fig. \ref{fig:fit_x}b. 
The best fit parameters together with the associated errors (90\% confidence intervals) are presented in Table \ref{table:x}, and are in good agreement with previous studies.
The primary X-ray continuum from the AGN is fitted with a power-law of spectral index $\Gamma$ = $2.10 \pm 0.25$ and variable foreground column density, going from the lowest value of $N_\textrm{H} \simeq 27 \times 10^{22}$ cm$^{-2}$ (ObsID 6870) to the highest one of $N_\textrm{H} \simeq 189 \times 10^{22}$ cm$^{-2}$ (ObsID 6872), when the nucleus experienced a Compton-thick state. The  Fe\,\textsc{xxv} and Fe\,\textsc{xxvi} features are blueshifted, implying an outflow velocity of $v \simeq -4540$ km s$^{-1}$. 
The best fit for the thermal model used for the soft part of the spectrum gives a temperature for the collisionally-ionized optically-thin gas of $kT$ $\simeq$ 0.66 keV (corresponding to $T$ $\sim$ $8 \times 10^6$ K). Overall, the fit is very good, with $\chi^2/\nu = 1305/1281$.

\begin{table*}[tp]
\centering
\caption{Best-fit parameters for the different \textsc{xspec} model components used to fit the spectra of Fig. \ref{fig:fit_x}: primary (1) and reflected (2) AGN emission, Fe\,\textsc{i} K$\alpha$ ($E$ = 6.4 keV) emission line (3), Fe\,\textsc{xxv} (4; $E$ = 6.7 keV) and Fe\,\textsc{xxvi} (5; $E$ = 6.97 keV) absorption lines, thermal emission (6). $N_\textrm{H}$: column density; $\Gamma$: power-law photon index; $A$: power-law normalization at 1 keV; $F$: integrated flux of the spectral line; $kT$: plasma temperature; $Z$: elemental abundance; EM: emission measure. (t) indicates tied parameters; (f) marks fixed parameters.}
	\setlength{\extrarowheight}{4pt}
	\begin{tabular*}{\textwidth}{l c c c c c c }
		\hline\hline
		Model			& ObsID 6868 & ObsID 6869 & ObsID 6870 & ObsID 6871 & ObsID 6872 & ObsID 6873 \\
		\hline
		(1) zpowerlw	&	& & & & &	\\
		\hspace{7mm} $N_\textrm{H}$ ($10^{22}$ cm$^{-2}$)	& 40$_{-4}^{+5}$ & 48$_{-4}^{+5}$ & 27$_{-2}^{+3}$ & 61$_{-9}^{+10}$ & 189$_{-35}^{+76}$ & 64$_{-9}^{+10}$ \\
		\hspace{7mm} $\Gamma$ & 2.10$_{-0.25}^{+0.25}$ & 2.10(t) & 2.10(t) & 2.10(t) & 2.10(t) & 2.10(t) \\
		\hspace{7mm} $A$ (10$^{-2}$ cm$^{-2}$ s$^{-1}$ keV$^{-1}$)	& 1.16$_{-0.48}^{+0.81}$ & 1.54$_{-0.64}^{+1.08}$ & 1.16$_{-0.45}^{+0.74}$ & 1.01$_{-0.47}^{+0.83}$ & 1.07(t) & 1.03$_{-0.48}^{+0.85}$ \\
		(2) pexrav	& 	& & & & &	\\
		\hspace{7mm} $A$ (10$^{-2}$ cm$^{-2}$ s$^{-1}$ keV$^{-1}$) & 1.07$_{-0.32}^{+0.42}$ 	& 1.07(t) & 1.07(t) & 1.07(t) & 1.07(t) & 1.07(t) \\
		(3) zgauss$_\textrm{(Fe\,\textsc{i} K$\alpha$ em.)}$ & & & & & & \\
		\hspace{7mm} $E$ (keV) & 6.38$_{-0.03}^{+0.02}$ & 6.38(t) & 6.38(t) & 6.38(t) & 6.38(t) & 6.38(t) \\
		\hspace{7mm} $F$ (10$^{-6}$ cm$^{-2}$ s$^{-1}$) & 8.2$_{-2.6}^{+2.6}$ & 8.2(t) & 8.2(t) & 8.2(t) & 8.2(t) & 8.2(t) \\
		(4) zgauss$_\textrm{(Fe\,\textsc{xxv} abs.)}$ & & & & & & \\
		\hspace{7mm} $v$ (km s$^{-1}$) & $-4540_{-885}^{+875}$ & $-4540$(t) & $-4540$(t) & $-4540$(t) & $-4540$(t) & $-4540$(t) \\
		\hspace{7mm} $F$ (10$^{-6}$ cm$^{-2}$ s$^{-1}$) & $-15.4_{-3.3}^{+3.3}$ & $-15.4$(t) & $-15.4$(t) & $>-9.06$ & 0 (f) & $>-9.06$(t) \\
		(5) zgauss$_\textrm{(Fe\,\textsc{xxvi} abs.)}$ & & & & & & \\
		\hspace{7mm} $v$ (km s$^{-1}$) & $-$4540(t) & $-4540$(t) & $-4540$(t) & $-4540$(t) & $-4540$(t) & $-4540$(t) \\
		\hspace{7mm} $F$ (10$^{-6}$ cm$^{-2}$ s$^{-1}$) & $-9.8_{-4.2}^{+4.2}$ & $-9.8$(t) & $-9.8$(t) & $>-6.35$ & 0 (f) & $>-6.35$(t) \\
		(6) vapec & & & & & & \\
		\hspace{7mm} $N_\textrm{H}$ ($10^{22}$ cm$^{-2}$)	& 0.059$_{-0.020}^{+0.016}$ & 0.059(t) & 0.059(t) & 0.059(t) & 0.059(t) & 0.059(t) \\
		\hspace{7mm} $kT$ (keV) & 0.66$_{-0.03}^{+0.06}$ & 0.66(t) & 0.66(t) & 0.66(t) & 0.66(t) & 0.66(t) \\
		\hspace{7mm} $Z_\textrm{O}$ ($Z_{\odot}$) & 0.08$_{-0.05}^{+0.08}$ & 0.08(t) & 0.08(t) & 0.08(t) & 0.08(t) & 0.08(t) \\
		\hspace{7mm} $Z_\textrm{Ne}$ ($Z_{\odot}$) & 0.21$_{-0.10}^{+0.12}$ & 0.21(t) & 0.21(t) & 0.21(t) & 0.21(t) & 0.21(t) \\
		\hspace{7mm} $Z_\textrm{Mg}$ ($Z_{\odot}$) & 0.36$_{-0.08}^{+0.09}$ & 0.36(t) & 0.36(t) & 0.36(t) & 0.36(t) & 0.36(t) \\
		\hspace{7mm} $Z_\textrm{Si}$ ($Z_{\odot}$) & 0.49$_{-0.08}^{+0.09}$ & 0.49(t) & 0.49(t) & 0.49(t) & 0.49(t) & 0.49(t) \\
		\hspace{7mm} $Z_\textrm{Fe}$ ($Z_{\odot}$) & 0.06$_{-0.01}^{+0.01}$ & 0.06(t) & 0.06(t) & 0.06(t) & 0.06(t) & 0.06(t) \\
		\hspace{7mm} EM ($10^{63}$ cm$^{-3}$) & 3.75$_{-0.50}^{+0.54}$ & 3.75(t) & 3.75(t) & 3.75(t) & 3.75(t) & 3.75(t) 
	\rule[-1.2ex]{0pt}{0pt} \\
	\hline
	\end{tabular*}
\label{table:x}
\end{table*}

In order to further inspect the nuclear outflow arising from the blueshifted Fe\,\textsc{xxv} and Fe\,\textsc{xxvi} absorption lines, we fitted them with a detailed absorption model instead of with two simple Gaussians, considering only the observations where such lines are clearly visible as the continuum is not heavily suppressed, namely ObsIDs 6868--6870. We used an \textsc{xstar} grid (\citealt{Kallman:2001aa}) to reproduce the high-ionization absorber responsible for the two lines, multiplied to the \texttt{zpowerlaw} for the primary nuclear continuum, tying the parameters of the absorber to be the same over the three observations. We left the continuum normalization and the column of the neutral absorber free to vary, guiding the intensity of the iron lines.
We obtained a velocity of the outflowing gas of 
$-4915_{-1320}^{+1205}$ km s$^{-1}$ and an intrinsic column density of the absorbing gas of $N_\textrm{H} \sim 3 \times 10^{24}$ cm$^{-2}$ (for Solar abundances). The accretion-disk wind in NGC 1365 is persistent, as it regularly appears in the less obscured nuclear states, yet it is also rather variable. Both the outflow velocity and the column density derived above lie close to the upper end of the respective ranges, as observed over the years (e.g., \citealt{Risaliti:2005aa, Brenneman:2013aa, Braito:2014aa}). While these values can certainly provide an upper limit to the mass outflow rate at subpc scales, in order to obtain a more sensible estimate to compare to the mass outflow rate in the optical cones we adopt the average values of $v = -3000$ km s$^{-1}$ and $N_\textrm{H} = 10^{23}$ cm$^{-2}$. While the latter might seem too conservative, it also takes into account the possible overabundance of iron by a factor of a few with respect to Solar standards (e.g., \citealt{Risaliti:2013aa}), which proportionally reduces the equivalent hydrogen column density. 

We compute the mass outflow rate using equation (2) in \cite{Nardini:2018aa}, which assumes that the wind covers a solid angle of 2$\pi$ (comparable to the opening angle of the optical cones in a biconical geometry) and that the launch radius corresponds to the escape radius for the adopted outflow velocity. We get $\dot{M}_\textrm{out,X} \sim 0.15$ $M_7$ $M_{\odot}$ yr$^{-1}$, where $M_7$ is the mass of the central SMBH (in units of $10^7$ $M_\odot$). As no reverberation mapping measure exists for NGC 1365, there is large uncertainty on the SMBH mass, but this is most likely few times 10$^6$ $M_{\odot}$ (e.g., \citealt{Risaliti:2009aa}, \citealt{Davis:2014aa}, \citealt{Onori:2017aa}). 
For $M_7 = 0.45$ (\citealt{Onori:2017aa}), the mass outflow rate of the X-ray wind is $\dot{M}_\textrm{out,X}$ $\sim$ 0.067 $M_\odot$ yr$^{-1}$ and the kinetic energy rate is $\dot{E}_\textrm{k,out,X}$ $\sim$ 1.9 $\times$ 10$^{41}$ erg s$^{-1}$. This is about 0.9$\%$ of the bolometric luminosity of the AGN ($L_\textrm{bol}$ $\sim$ 2.2 $\times$ 10$^{43}$ erg s$^{-1}$), which we calculated from the average de-absorbed 2-10 keV luminosity from our six observations $L_\textrm{X, 2-10 keV}$ $\sim$ $1.7 \times 10^{42}$ erg s$^{-1}$ using the median bolometric correction for Type 2 AGN $\langle k_\textrm{bol} \rangle \sim$ 13 from \cite{Lusso:2011aa}.

Moreover, from $L_\textrm{X, 2-10 keV}$ we can calculate the expected [O\,\textsc{iii}] luminosity in the NLR and compare it with the one we measure from MUSE. Using $L_\textrm{X, 2-10 keV}$--$L_\textrm{[O\,\textsc{iii}]}^\textrm{cor}$ relation from \cite{Ueda:2015aa} for Type 2 AGN, we obtain an extinction-corrected [O\,\textsc{iii}] luminosity of $L_\textrm{[O\,\textsc{iii}]}^\textrm{cor} \sim 2.6 \times 10^{40}$ erg s$^{-1}$, which is about a factor of three smaller than $L_\textrm{[O\,\textsc{iii}]}^\textrm{cor} \sim 8 \times 10^{40}$ erg s$^{-1}$ that we obtain from the two [O\,\textsc{iii}] cones from MUSE. However, this discrepancy is well within the scatter on the $L_\textrm{X, 2-10 keV}$--$L_\textrm{[O\,\textsc{iii}]}^\textrm{cor}$ relation of \cite{Ueda:2015aa} and also consistent with the fact that in their sample all the Type 2 AGN at low $L_\textrm{X, 2-10 keV}$, on the order of  $10^{42}$ erg s$^{-1}$, reside above their relation.
We obtain the extinction-corrected [O\,\textsc{iii}] luminosity of the cones by extracting and fitting two integrated spectra (one per cone) produced by summing up all the spaxels of each of the two cones. The spatial region covered by each cone is defined by the angles used in Fig. \ref{fig:moutrate} extended to the margins of the FOV. We use two Gaussian components per emission line, one to reproduce the disk and one for the outflow. The obtained line flux of the outflow component is corrected for extinction using the average Balmer decrement from the fit of the radial shells in Sect. \ref{ssec:moutrate}. We do not adopt the H$\alpha$/H$\beta$ from these two fits because, as we integrate over the whole cones, we are taking parts of the disk and of the outflow with different velocities, which can overlap together in some areas,
 giving a fitted outflow component of H$\alpha$ and H$\beta$ that may be in part affected by the emission from the disk and consequently an unreliable extinction. 
We take the total [O\,\textsc{iii}] luminosity (outflow + disk) for this comparison with \cite{Ueda:2015aa}, instead of the luminosity of the outflow component only, because the [O\,\textsc{iii}] luminosities used in their work (and similar works involving $L_\textrm{[O\,\textsc{iii}]}$--$L_\textrm{X}$ relation) are the total ones, as the imaging data used did not allow them to distinguish between outflow and disk component. 

Having the bolometric luminosity $L_\textrm{bol}$, we can also estimate the mass accretion rate as follows:
\begin{equation}
\dot{M}_\textrm{accr} = \dfrac{L_\textrm{bol}}{\eta c^2},
\label{eq:Maccr}
\end{equation}
being $\eta$ the radiative efficiency and $c$ the speed of light. 
Assuming $\eta = 0.1$, which is the typical value for a Shakura-Sunyaev accretion disk (\citealt{Shakura:1973aa}), we get $\dot{M}_\textrm{accr}$ $\sim$ $4 \times 10^{-3}$ $M_\odot$ yr$^{-1}$. This is a factor of 17 smaller than the nuclear mass outflow rate we obtained from the blueshifted Fe\,\textsc{xxv} and Fe\,\textsc{xxvi} absorption lines. 
This suggests that the X-ray outflow might have already experienced mass loading since its launch from the innermost regions close to the BH. Indeed, the X-ray outflow in NGC 1365 is an anomalous case, as its velocity of $\sim$3000 km s$^{-1}$ is about one order of magnitude lower than the typical velocities of ultra fast outflows (UFOs) and at the upper limit of the velocity of warm absorbers (\citealt{Tombesi:2013aa}), although having an ionization typical of UFOs. UFOs are associated to a very initial stage of the outflow close to the BH (e.g., \citealt{Nardini:2015aa}), but they might not be observable in absorption if the gas participating in them is completely ionized by the strong AGN radiation field close to the BH (\citealt{Pinto:2018aa}). A UFO might then be present in NGC 1365 but not detectable, becoming observable only on a larger scale after it has done mass loading and slowed down. Another possibility to explain the UFO-like ionization of the outflow and its velocity a factor of at least 10 smaller than UFO typical velocities might be that the outflow is actually a UFO but only a little fraction of its velocity is directed in our line of sight. However, this would imply an inclination of the nuclear outflow axis close to 90$^{\circ}$ with respect to the line of sight, which is highly unlikely, either considering an accretion disk coplanar with the galaxy disk (which is inclined by 35$^{\circ}$ relative to the line of sight; \citealt{Hjelm:1996aa}) or the estimate of 63$^{\circ}$ for the inclination of the accretion disk from \cite{Risaliti:2013aa}.

\section{MUSE--Chandra matching}
In this section we present the spatial matching between MUSE optical data showed in Sect. \ref{sec:muse_maps} and {\it Chandra} ACIS-S X-ray data presented in Sect. \ref{sec:x_rays}. With the aim of comparing the distribution of the X-ray emitting ionized gas and of the optically-emitting one from MUSE, we extracted different images in selected X-ray energy bands. We also compare the properties of the nuclear X-ray wind with those of the extended ionized outflow to obtain a more comprehensive picture of the outflow in NGC 1365.

\subsection{Diffuse emission}

\begin{figure*}
\centering
\begin{subfigure}[t]{0.01\textwidth}
\textbf{a}
\end{subfigure}
\begin{subfigure}[t]{0.45\textwidth}
\centering
\includegraphics[trim={0 0.5cm 0 0},clip,width=0.895\textwidth,valign=b]{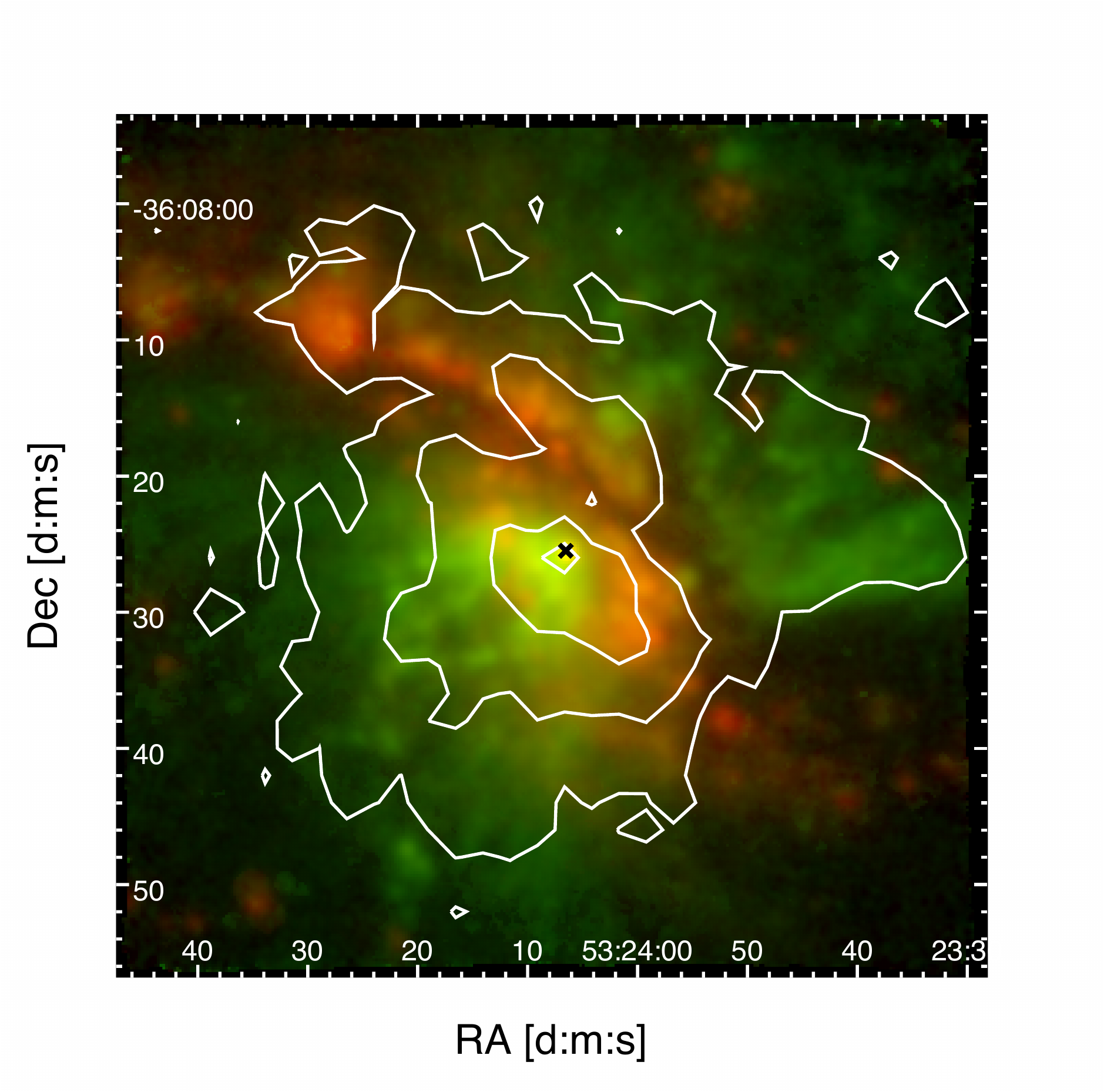}
\end{subfigure}
\quad
\begin{subfigure}[t]{0.01\textwidth}
\textbf{b}
\end{subfigure}
\begin{subfigure}[t]{0.45\textwidth}
\includegraphics[trim={0 0 0 -1cm},clip,width=\textwidth,valign=b]{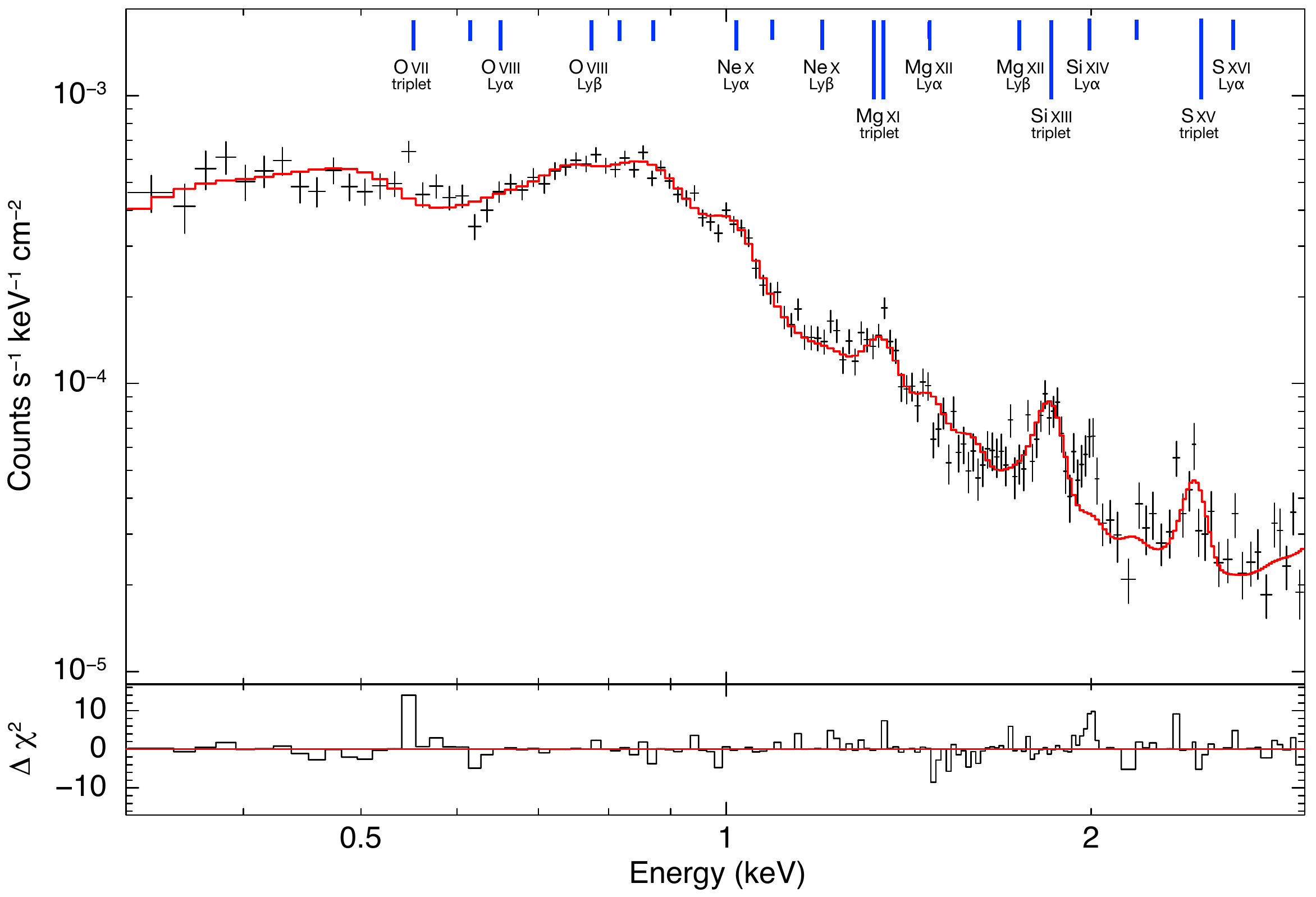}
\end{subfigure}\\
\vspace{0.5cm}
\begin{subfigure}[t]{0.01\textwidth}
\textbf{c}
\end{subfigure}
\begin{subfigure}[t]{0.45\textwidth}
\centering
\includegraphics[trim={0cm 0 0cm 0},clip,width=0.725\textwidth,valign=b]{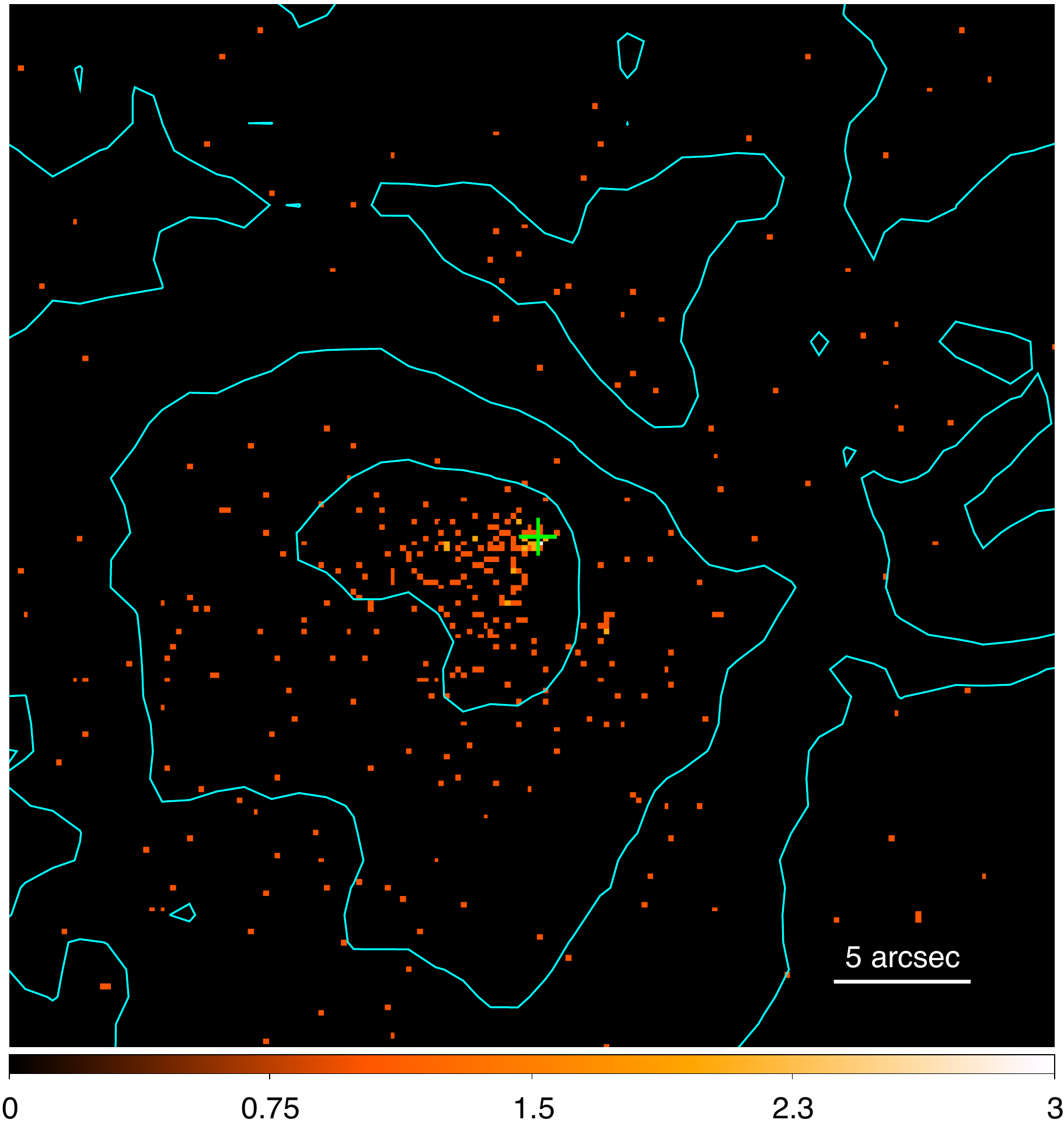}
\end{subfigure}
\quad
\begin{subfigure}[t]{0.01\textwidth}
\textbf{d}
\end{subfigure}
\begin{subfigure}[t]{0.45\textwidth}
\includegraphics[width=\textwidth,valign=b]{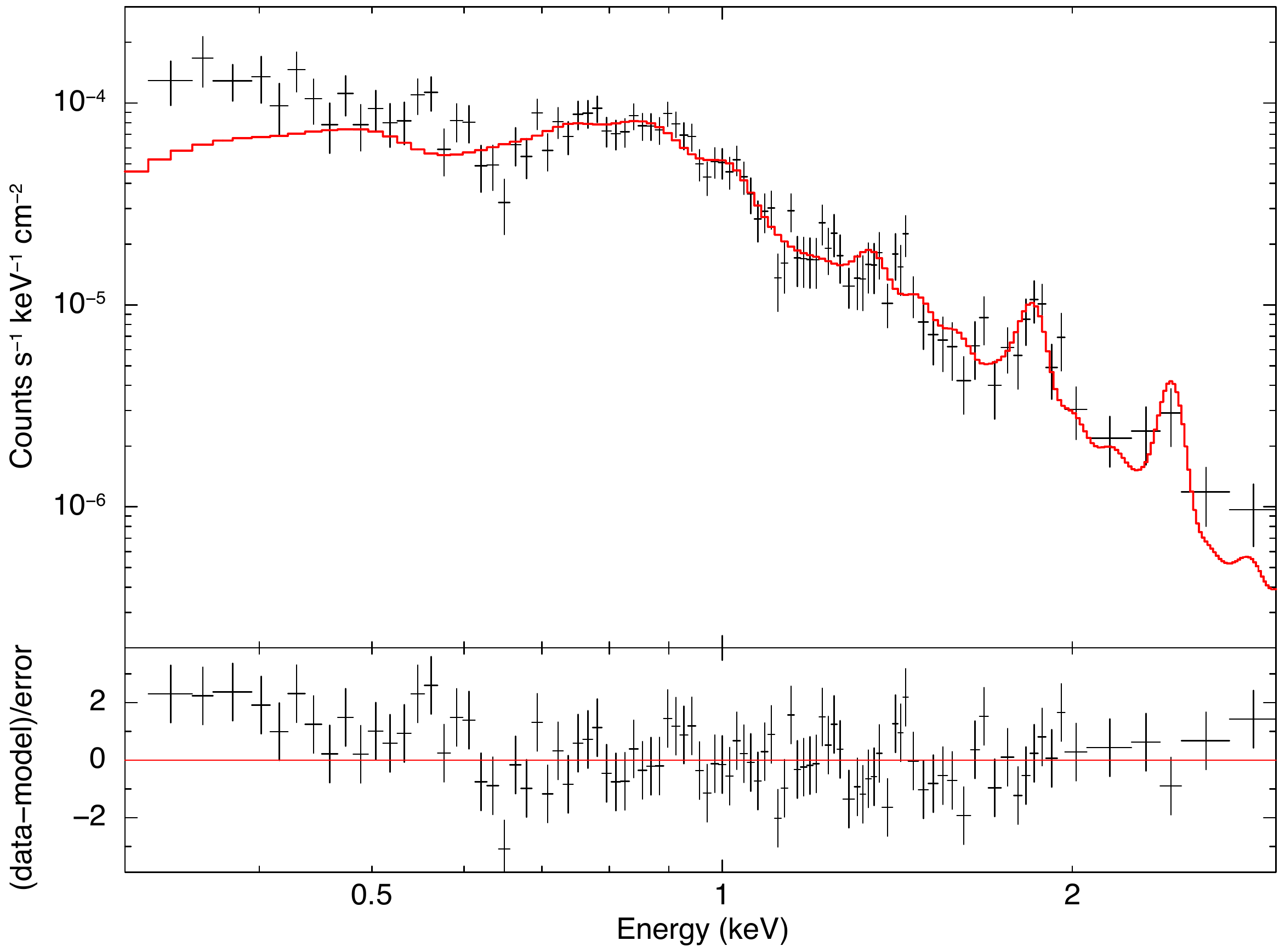}
\end{subfigure}
\caption{{\bf(a)} 0.3--1.2 keV {\it Chandra} ACIS-S soft X-ray emission (contours) superimposed on the [O\,\textsc{iii}] (green) and H$\alpha$ (red) emission fitted from the star-subtracted smoothed MUSE data cube (same as in Fig. \ref{fig:maps_redd_ratio}a). {\bf(b)} Soft X-ray emission extracted from the yellow polygonal region in Fig. \ref{fig:x_merged}. The data are in black, resulting from the merging of the six extracted spectra. Red is the best fit, having used a thermal emission model. The residuals are reported on the bottom. The labeled blue lines on top indicate the statistically significant (at 95$\%$ level) emission lines in the {\it Chandra} HETG spectra, after accounting for the collisional component, from \cite{Nardini:2015ab}. {\bf(c)} Unsmoothed 0.3--0.4 keV emission and [O\,\textsc{iii}] contours from MUSE, zooming in the central regions of MUSE FOV. {\bf(d)} Spectrum extracted from the inner contour of [\textsc{O\,iii}] in panel c (excluding the nucleus).}
\label{fig:muse_chandra_match}
\end{figure*}

\begin{figure*}[t]
\centering
\begin{subfigure}[t]{0.01\textwidth}
\textbf{a}
\end{subfigure}
\begin{subfigure}[t]{0.45\textwidth}
\includegraphics[trim={0 0.5cm 0 0},clip,width=0.895\textwidth,valign=b]{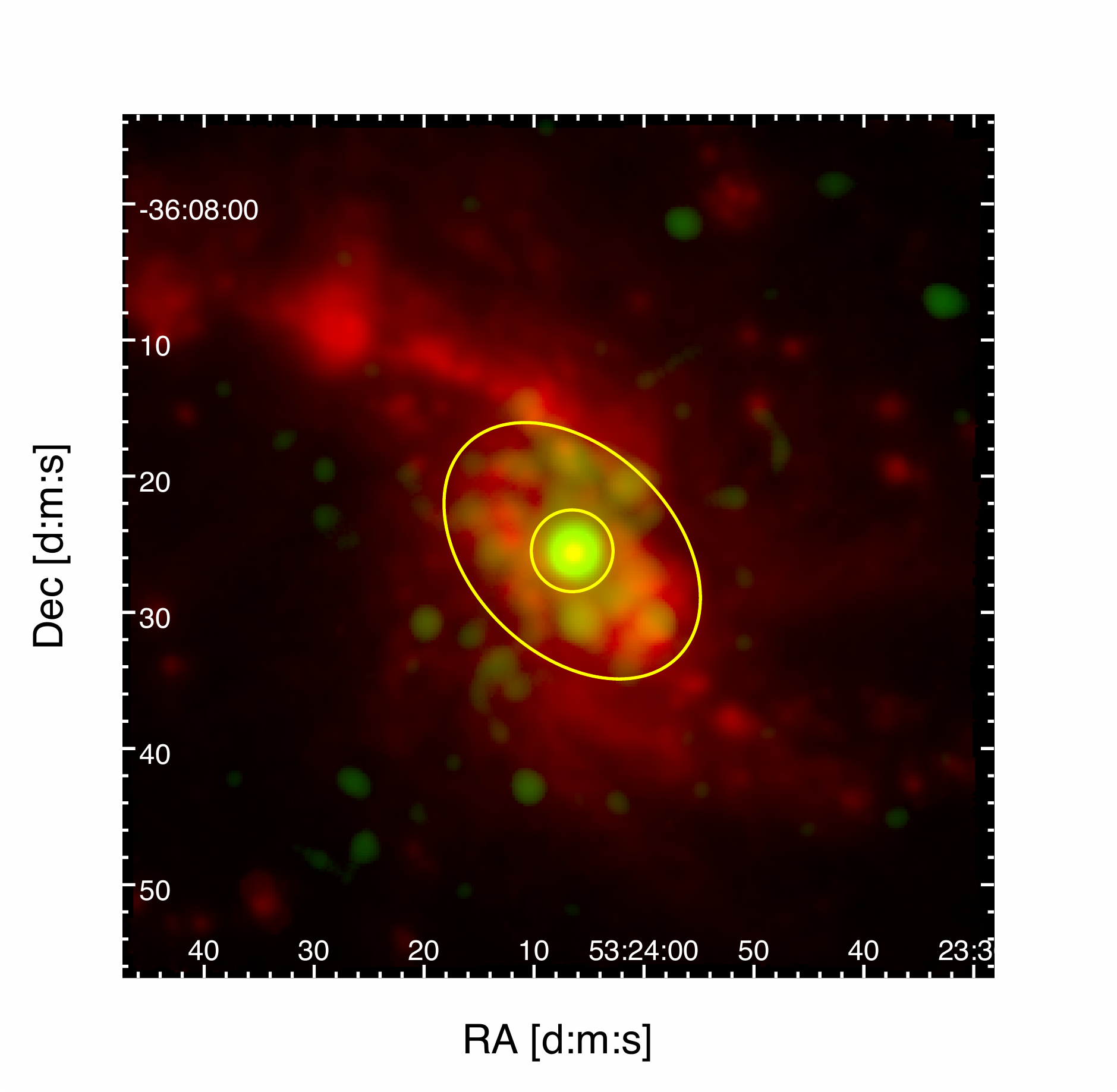}
\end{subfigure}
\begin{subfigure}[t]{0.01\textwidth}
\textbf{b}
\end{subfigure}
\begin{subfigure}[t]{0.45\textwidth}
\includegraphics[trim={0 0 0 -0.5cm},clip,width=0.95\textwidth,valign=b]{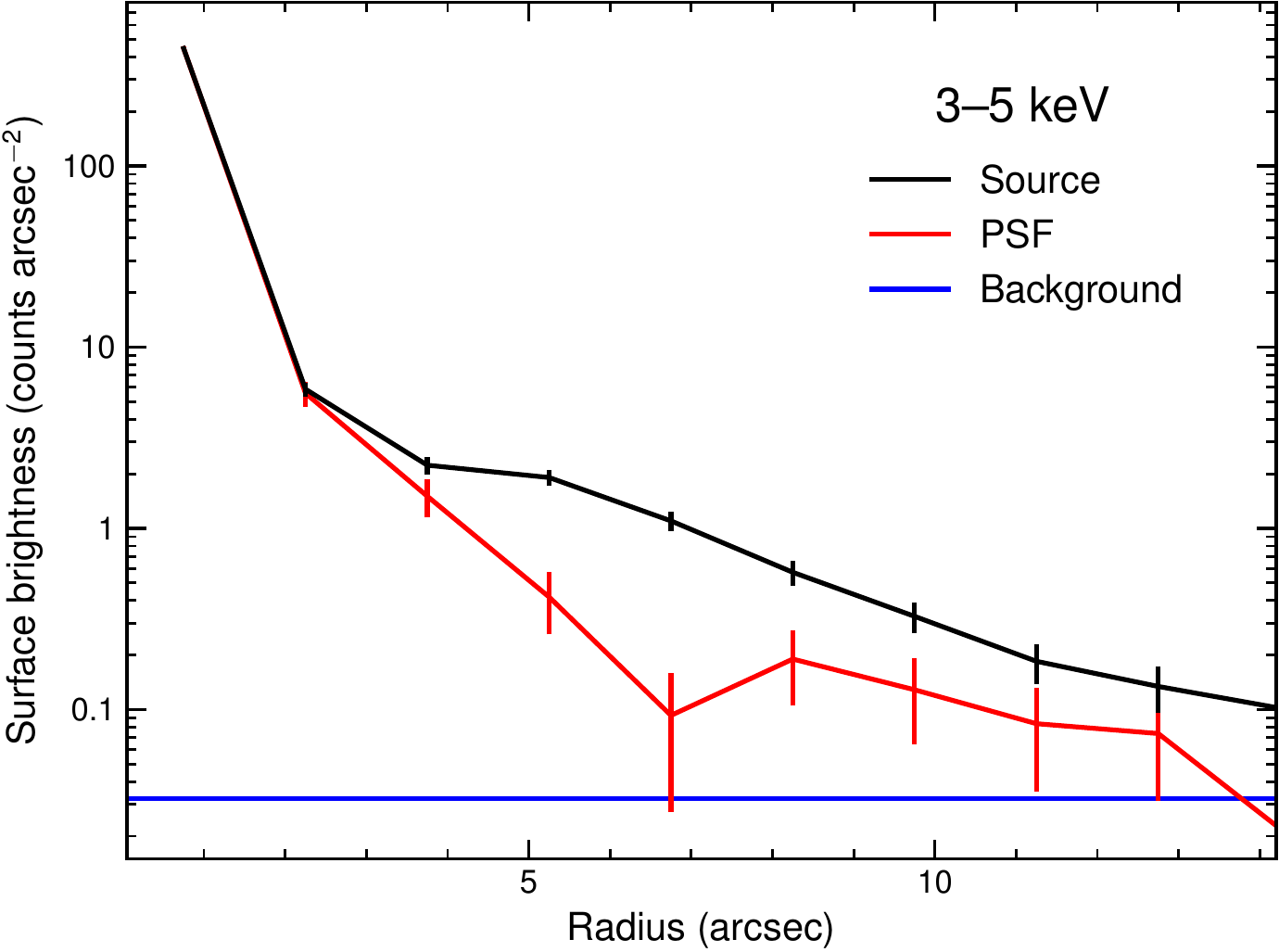}
\end{subfigure}
\caption{{\bf(a)} 3--5 keV emission (green) together with H$\alpha$ emission (red) from MUSE. {\bf(b)} Radial profile of the surface brightness in the energy range 3--5 keV (black) versus the simulated PSF of the nucleus (red) from the brightest, less obscured state of the AGN (ObsID 6870); the level of the background is also reported (blue).}
\label{fig:sb_ring}
\end{figure*}

The smoothed contours of the 0.3--1.2 keV soft X-ray emission (compare with Fig. \ref{fig:fit_x}) are superimposed in Fig. \ref{fig:muse_chandra_match}a on the [O\,\textsc{iii}] (green) and H$\alpha$ (red) maps obtained from the fit of the star-subtracted smoothed MUSE data cube. The 0.3--1.2 keV contours trace very well both the [O\,\textsc{iii}]-dominated regions, corresponding to the SE and NW cones, and the H$\alpha$ ones, following the shape of the circumnuclear star-forming ring and of the long lane at the edge of the dust lane. The soft X-ray emission then seems to be associated in part with the AGN-ionized cone, in part with the star formation dominating in the circumnuclear ring and, in general, in the elongated diagonal region in the direction NE--SW (see the spatially resolved BPT maps in Fig. \ref{fig:bpt}). 

In order to analyze in more detail the diffuse soft X-ray emission, we merged the six spectra extracted from the polygonal region in Fig. \ref{fig:x_merged}, having constrained the nuclear X-ray emission with the full-band spectral fitting in Sect. \ref{ssec:fit_Xfull}. The combined spectrum was then binned to a S/N of 5 per channel. We thus performed a fit in the energy range 0.3--3 keV, fixing the parameters of the nuclear components to the values found in Sect. \ref{ssec:fit_Xfull} (see Table \ref{table:x}), only allowing the normalization of the primary continuum and the column density of the neutral absorber to slightly re-adjust to fit the average spectral shape, as they both vary among the single observations. We note that the transmitted nuclear contribution becomes rapidly negligible below 3 keV, and only the faint and virtually featureless reflected component survives.

We fitted the soft X-ray emission with a single \texttt{vapec} thermal component. The best fit gave fully consistent results with the previous analysis, but the individual parameters are now better constrained, with $kT$ = 0.66$_{-0.03}^{+0.01}$ keV and $N_\textrm{H}$ = 0.065$_{-0.015}^{+0.016} \times 10^{22}$ cm$^{-2}$. The same holds for the elemental abundances. The fit statistics, however, is rather poor ($\chi^2/\nu = 255/135$). The merged spectrum together with the best-fit model and the relative residuals is shown in Fig. \ref{fig:muse_chandra_match}b. Following previous works (e.g., \citealt{Wang:2009aa}, \citealt{Guainazzi:2009aa}), we tried to add another collisionally-ionized component with the same abundances, but, although slightly improving the goodness of fit ($\chi^2/\nu = 212/132$), this does not return a statistically acceptable solution, indicating that a thermal model is unable to account for the main residuals. Indeed, as reported by \cite{Nardini:2015ab} and \cite{Whewell:2016aa}, the soft X-ray spectrum of NGC 1365 also contains a multiphase component photoionized by the AGN, producing a wealth of emission lines. The energies of the lines detected at the 95\% confidence level after accounting for the collisional component in the high-resolution {\it Chandra} HETG spectrum by \cite{Nardini:2015ab} are reported in the figure. The most prominent residuals in our low resolution spectrum, at $\sim$0.55 and 2 keV, correspond to the O\,\textsc{vii} triplet and to the Si\,\textsc{xiii} Ly$\alpha$, respectively, which both have a photoionization origin. Despite the limited data quality, tentative positive residuals can be seen at the energy of several other photoionized lines. 

\subsection{Inner ionization cone}
Thanks to a very deep analysis of all the available high-resolution X-ray data of NGC 1365, \cite{Whewell:2016aa} identified two collisionally ionized and three photoionized gas phases in the soft X-ray emission-line spectrum. In their spectral decomposition, the photoionized components start to prevail below $\sim$0.5 keV. Unfortunately, there are only few strong lines at these energies. Neglecting the N\,\textsc{vi} triplet at $\sim$0.43 keV, which requires nitrogen abundance of 4.5 times Solar to be properly accounted for, we focused on the C\,\textsc{vi} Ly$\alpha$ line at $\sim$0.37 keV, and extracted a narrowband {\it Chandra} image in the 0.3--0.4 keV range. Fig. \ref{fig:muse_chandra_match}c shows a zoom on the central regions of the FOV, with the unsmoothed 0.3--0.4 keV emission and the [O\,\textsc{iii}] contours from MUSE superimposed on it. The figure reveals that the very soft X-ray emission, dominated by photoionized gas, perfectly traces the inner and most intense part of the [O\,\textsc{iii}]-emitting AGN-dominated SE cone. 
We thus extracted the spectrum (one for each observation) of this region to further inspect the soft X-ray emission in the cone. 
The merged spectrum can be seen in Fig. \ref{fig:muse_chandra_match}d, together with the best-fit thermal model (red line) previously obtained for the extended soft X-ray emission (Fig. \ref{fig:muse_chandra_match}b), simply rescaled to the flux of this spectrum.
An excess with respect to the scaled model emerges below $\sim$0.5 keV, indicating that the emission in the inner part of the cone has an additional contribution compared to the extended thermal emission. We also tried to refit the model (an absorbed \texttt{vapec}) to the data, but it was still not able to reproduce the observed excess, which can then be attributed to a blend of emission lines (due to low resolution and limited statistics) photoionized by the AGN.

In order to obtain a rough guess of the luminosity of this excess component, we fitted it with a simple blackbody model, getting a luminosity of $\sim$ 2 $\times$ 10$^{39}$ erg s$^{-1}$. We get an [O\,\textsc{iii}] luminosity over the same region of $\sim$ 5 $\times$ 10$^{39}$ erg s$^{-1}$ from MUSE data, which is of the same order of magnitude, thus supporting the scenario of a common energy input for the two emissions, that is, the AGN photoionization. 
This is expected for Sy2 galaxies (see e.g., \citealt{Bianchi:2006aa}), but it is not straightforward for NGC 1365, being the soft (0.3--2 keV) X-ray band dominated by thermal emission due to the intense star formation. Nevertheless, we have shown that using the appropriate strategy the photoionized component can be isolated. While \cite{Nardini:2015ab} were able to reveal the diluted photoionized emission by adopting a narrow extraction region close to the nucleus, here we have reached the same results by choosing the proper energy band, which allowed us to find the cospatiality of the X-ray photoionized component with the [O\,\textsc{iii}]. NGC 1365 is then a standard obscured AGN, whose peculiar spectral properties (such as thermal emission dominating over photoionization in the soft X-rays) are most likely due to the much higher star-formation compared to normal Sy2s.

\subsection{Star-forming ring}
While the AGN radiation field clearly dominates the photoionization in the cones, the circumnuclear starburst ring can, in principle, contribute to the driving of the outflow. Soft X-ray emission from collisionally-ionized gas is typical of shock heating associated with starburst winds (e.g., \citealt{Veilleux:2005aa}, and references therein). In general, one should expect that the larger the gas temperature, the faster the shocks and the larger the injection of mechanical energy (e.g., \citealt{McKee:1980aa}). Even if we were able to obtain a good fit of the soft X-ray continuum with a single \texttt{vapec} component of $kT \simeq 0.66$ keV, this is just a phenomenological approximation of the emission from a multitemperature gas known to be present in the circumnuclear environment. 
In Fig. \ref{fig:sb_ring}a the 3--5 keV smoothed hard X-ray emission (green) can be compared with H$\alpha$ emission (red). 
Besides the nuclear point source, there is an extended component that is cospatial with the elongated star-forming ring traced by the strong H$\alpha$ emission, suggesting that the diffuse 3--5 keV emission is associated with the star-forming processes occurring in the ring. This can be due to hotter diffuse gas and/or X-ray binaries. 

The extended nature of the 3--5 keV emission is confirmed by the radial profile of the surface brightness in this band reported in Fig. \ref{fig:sb_ring}b (black curve), compared to the simulated PSF of the nucleus (red) generated with the {\it Chandra} Ray Tracer\footnote{\url{http://cxc.harvard.edu/chart/}.} and MARX\footnote{\url{http://space.mit.edu/cxc/marx/}.} suites. The brightest, less obscured state of the AGN (ObsID 6870) was considered to be as much conservative as possible. Still, there is a significant excess in the surface brightness with respect to the wings of nuclear PSF at the scale of the ring, confirming that there is extended local emission in addition to the unresolved source. In order to evaluate the spectral properties of this component, we selected an elliptical region encompassing the ring, and excising the central 3 arcsec (in radius) around the nucleus (Fig. \ref{fig:sb_ring}a). This ensures that any contribution from the wings of the nuclear PSF is minimized. The combined spectrum of the ring from the six observations was fitted with the same model of Sect.~\ref{ssec:fit_Xfull}, including an absorbed \texttt{pshock} component to account for any emission from hot, shocked gas (following \citealt{Wang:2014aa}). The collisionally-ionized and nuclear components were fixed to the best fit values of Table~\ref{table:x}, simply rescaled through constant factors. From the former, we get that about 40\% of the thermal emission arises in the ring, while the latter implies that the nuclear PSF leakage amounts to only 2\%, in excellent agreement with the simulated PSF. The shock component, absorbed by a column of $(5 \pm 2) \times 10^{21}$ cm$^{-2}$, improves the $\chi^2/\nu$ from 341/182 to 213/179, but its temperature is poorly constrained. Taking the nominal best-fit value of 4.8 keV as the post-shock temperature in a fully ionized, monoatomic gas, we would derive a shock velocity of $\sim$2000 km s$^{-1}$. The intrinsic luminosity of this component is $\approx$ $1.5 \times 10^{40}$ erg s$^{-1}$ (this is most likely an upper limit, as a similar contribution is expected from the high-mass X-ray binary population of the ring, based on the SFR of the ring of $\sim$5.6 $M_\odot$ yr$^{-1}$ from \citealt{Alonso-Herrero:2012aa} and the predictions from \citealt{Mineo:2012aa}), to be compared to the expected rate of energy injection from SNe of $\sim 2 \times 10^{41}$ erg s$^{-1}$, assuming the standard 10$^{51}$ erg per SN and 10\% efficiency, and a SN rate in the ring of 0.07 yr$^{-1}$ (\citealt{Wang:2009aa}). 
In order to achieve a kinetic energy rate comparable to the one previously estimated for the AGN wind ($\sim$ 2 $\times$ 10$^{41}$ erg s$^{-1}$), 2 $M_{\odot}$ per SN would be required to be entrained in such a fast outflow from the SF ring.

From an energetic point of view this SNe shock might then be capable of driving the extended optical outflow. Moreover, the highest values of the mass outflow rate are observed on the scale of the SF ring (Fig. \ref{fig:moutrate_plot}a), which might suggest a correlation between the ring and the outflow, even though the mass outflow rate remains still high at larger distances compared to the ring scale, before dropping.

So, the SNe shock in the ring, if actually present, might contribute in principle to the kpc-scale wind we observe from MUSE, but it is not clear to which extent or how (if by directly pushing out material to large distances or only by adding additional mass to the passing AGN outflow). 
With the present X-ray data quality, it is therefore impossible to 
quantify the additional contribution of the starburst to the AGN outflow.

\subsection{Multiphase outflow properties}\label{ssec:outflow}
\begin{figure*}
\centering
\begin{subfigure}[t]{0.01\textwidth}
\textbf{a}
\end{subfigure}
\begin{subfigure}[t]{0.475\textwidth}
\includegraphics[trim={0.5cm 0 2cm 1cm},clip,width=\textwidth,valign=b]{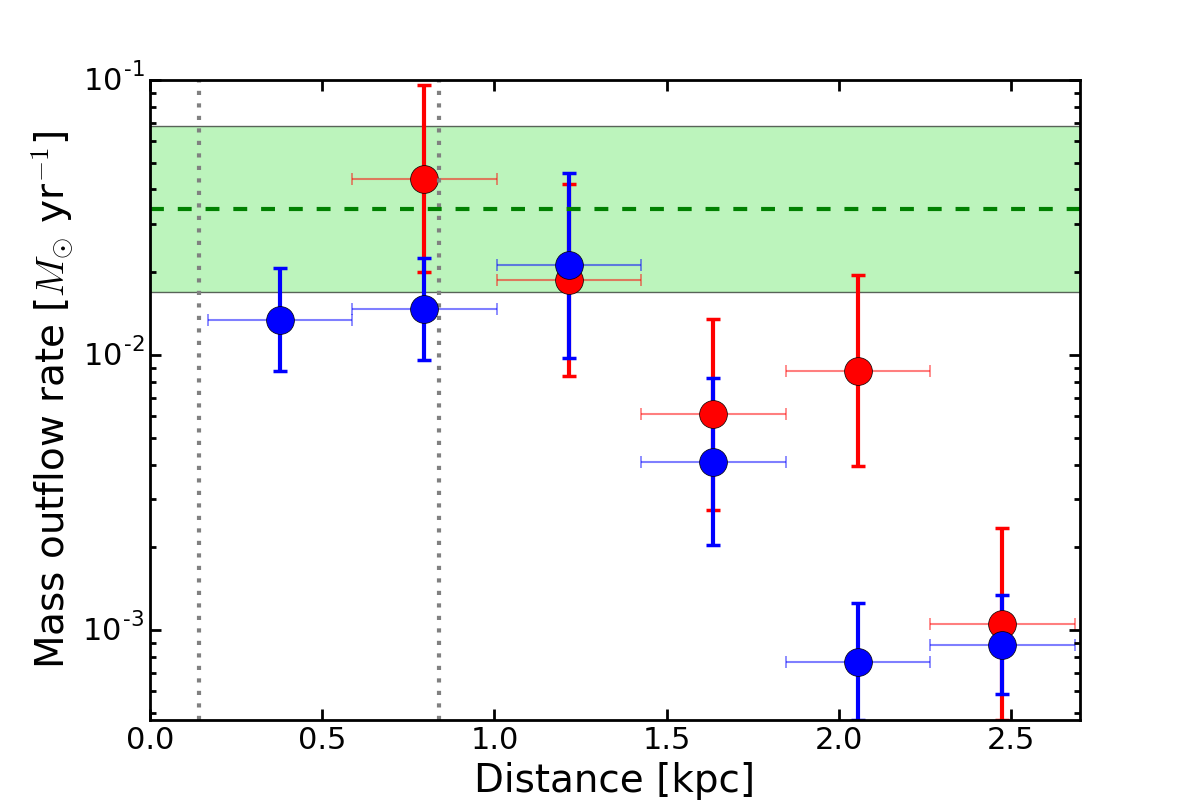}
\end{subfigure}
\hfill
\begin{subfigure}[t]{0.01\textwidth}
\textbf{b}
\end{subfigure}
\begin{subfigure}[t]{0.475\textwidth}
\includegraphics[trim={0.5cm 0 2cm 1cm},clip,width=\textwidth,valign=b]{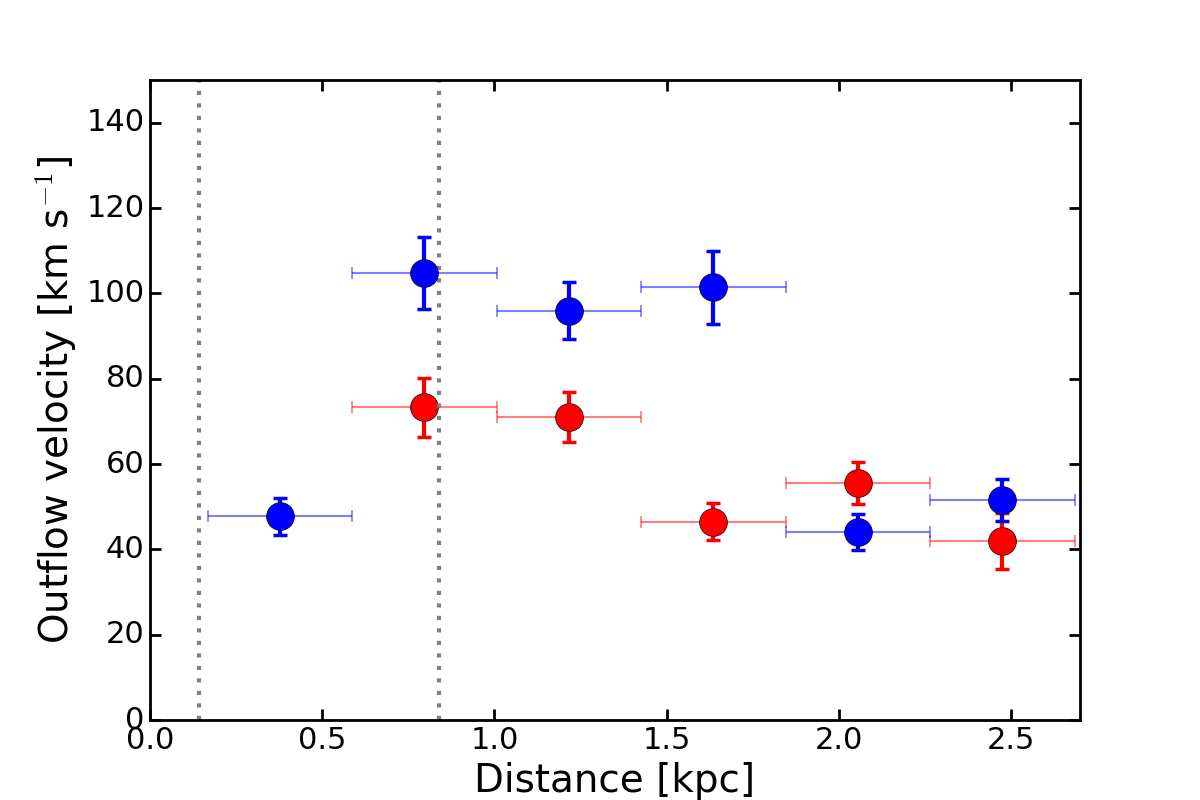}
\end{subfigure}\\
\vspace{0.2cm}
\begin{subfigure}[t]{0.01\textwidth}
\textbf{c}
\end{subfigure}
\begin{subfigure}[t]{0.475\textwidth}
\includegraphics[trim={0.5cm 0 2cm 1cm},clip,width=\textwidth,valign=b]{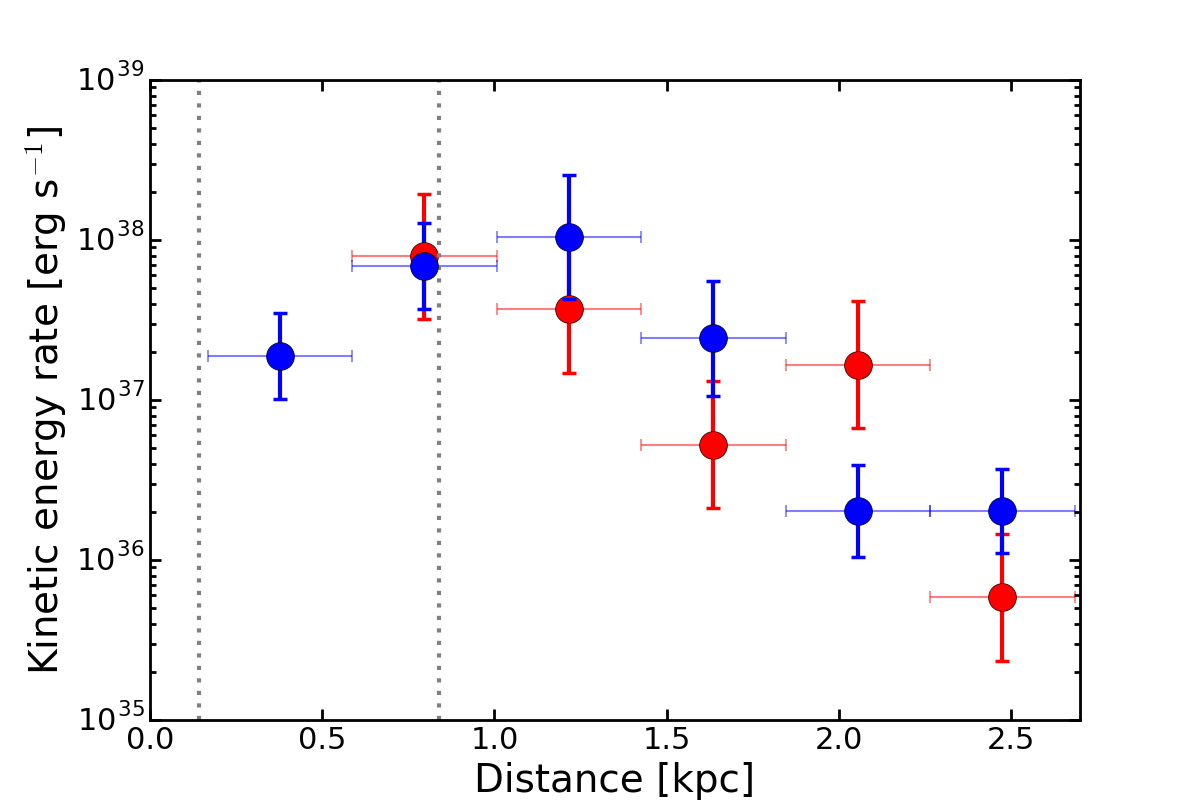}
\end{subfigure}
\hfill
\begin{subfigure}[t]{0.01\textwidth}
\textbf{d}
\end{subfigure}
\begin{subfigure}[t]{0.475\textwidth}
\includegraphics[trim={0.5cm 0 2cm 1cm},clip,width=\textwidth,valign=b]{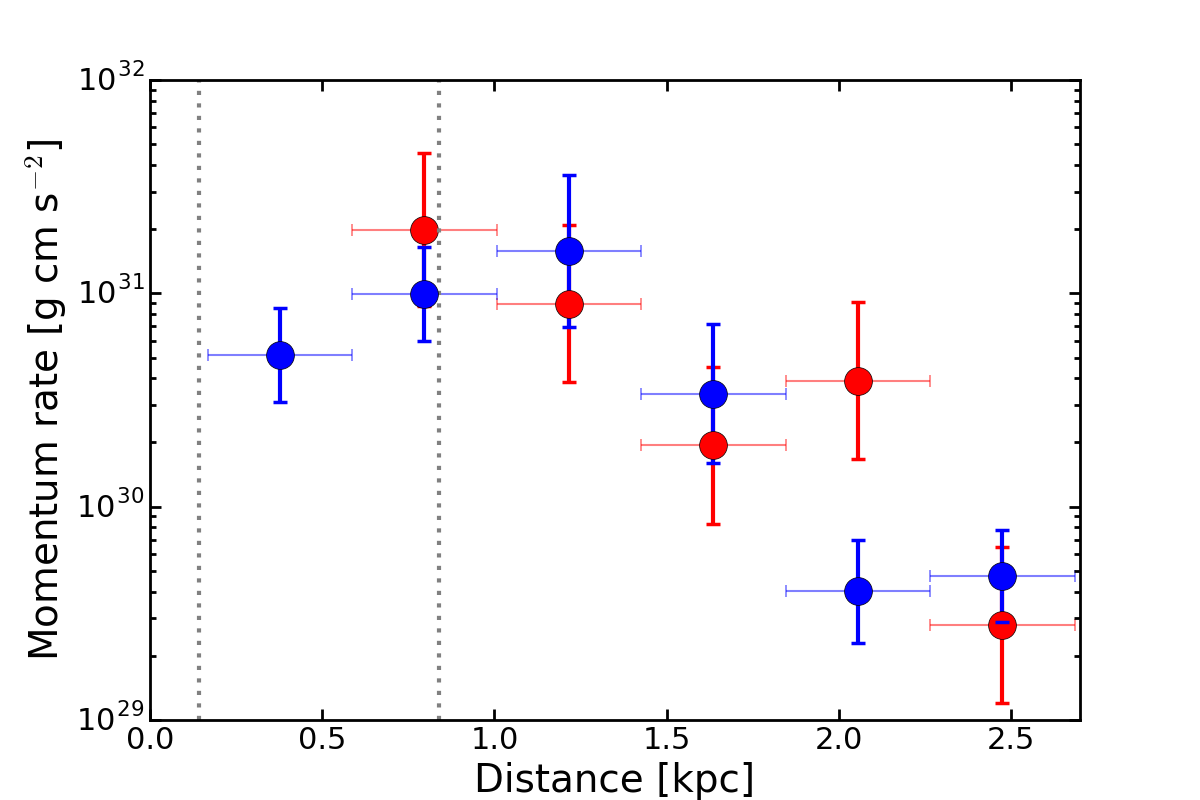}
\end{subfigure}
\caption{{\bf(a)} Mass outflow rate of the ionized gas as a function of distance from the active nucleus, obtained from the gridded map in Fig. \ref{fig:moutrate}; blue and red points indicate the sum of the mass outflow rates of all the grid elements at a certain distance, for the SE (blueshifted) cone and for the NW (redshifted) one, respectively. The error bars on the distance indicate the width of the radial shells. 
The green dashed line is the mass outflow rate obtained for the nuclear outflow from Fe\,\textsc{xxv} and Fe\,\textsc{xxvi} absorption lines from {\it Chandra} X-ray data (Sect. \ref{sec:x_rays}), together with its typical uncertainty in shaded green; the two dotted gray lines mark the scale of the inner minor semi-axis and of the outer major semi-axis of the circumnuclear SF ring, respectively. {\bf(b)} Absolute value of the outflow velocity; the points indicate the average velocity at the given distance. {\bf(c)} Kinetic energy rate radial profile. {\bf(d)} Momentum rate radial profile.}
\label{fig:moutrate_plot}
\end{figure*}

\begin{table*}[tp]
\centering
\caption{Mass outflow rate ($\dot{M}_\textrm{out}$), outflow velocity ($v_\textrm{out}$), kinetic energy rate ($\dot{E}_\textrm{k,out}$), and momentum rate ($\dot{p}_\textrm{out}$) of the ionized gas as a function of distance from the center for both the SE (blueshifted) and the NW (redshifted) cone, as reported in Fig. \ref{fig:moutrate_plot}.}
\setlength{\extrarowheight}{6pt}
\begin{tabular*}{\textwidth}{@{} @{\extracolsep{\stretch{1}}}*{2}{l} @{\extracolsep{\stretch{1}}}*{7}{c}  @{}}
\hline\hline
\hspace{10pt} Distance [kpc] & & 0.38 & 0.80 & 1.22 & 1.64 & 2.06 & 2.48 \rule[-1.2ex]{0pt}{0pt}\\
\hline
\multirow{2}{*}{\hspace{10pt} $\dot{M}_\textrm{out}$ [10$^{-2}$ $M_\odot$ yr$^{-1}$]} & SE cone & $1.3_{-0.5}^{+0.7}$ & $1.5_{-0.5}^{+0.8}$ & $2.1_{-1.1}^{+2.5}$ & $0.4_{-0.2}^{+0.4}$ & $0.08_{-0.03}^{+0.05}$ & $0.09_{-0.03}^{+0.05}$ \\
 & NW cone & $-$ & $4_{-2}^{+5}$ & $1.9_{-1.0}^{+2.3}$ & $0.6_{-0.3}^{+0.7}$ & $0.9_{-0.5}^{+1.1}$ & $0.11_{-0.06}^{+0.13}$ \rule[-1.5ex]{0pt}{0pt}\\
 \hline
\multirow{2}{*}{\hspace{10pt} $v_\textrm{out}$ [km s$^{-1}$]} & SE cone & $-48\pm4$ & $-105\pm8$ & $-96\pm7$ & $-101\pm9$ & $-44\pm4$ & $-52\pm5$ \\
& NW cone & $-$ & $+73\pm7$ & $+71\pm6$ & $+47\pm4$ & $+56\pm5$ & $+42\pm7$ \rule[-1.2ex]{0pt}{0pt}\\
\hline
\multirow{2}{*}{\hspace{10pt} $\dot{E}_\textrm{k,out}$ [10$^{37}$ erg s$^{-1}$]} & SE cone & $1.9_{-0.9}^{+1.6}$ & $7_{-3}^{+6}$ & $10_{-6}^{+15}$ & $2.4_{-1.4}^{+3.1}$ & $0.20_{-0.10}^{+0.19}$ & $0.20_{-0.09}^{+0.17}$ \\
 & NW cone & $-$ & $8_{-5}^{+12}$ & $4_{-2}^{+6}$ & $0.5_{-0.3}^{+0.8}$ & $1.7_{-1.0}^{+2.5}$ & $0.006_{-0.004}^{+0.009}$ \rule[-1.5ex]{0pt}{0pt}\\
\hline 
\multirow{2}{*}{\hspace{10pt} $\dot{p}_\textrm{out}$ [10$^{30}$ g cm s$^{-2}$]} & SE cone & $5_{-2}^{+3}$ & $10_{-4}^{+7}$ & $16_{-9}^{+20}$ & $3.4_{-1.8}^{+3.8}$ & $0.40_{-0.17}^{+0.30}$ & $0.47_{-0.18}^{+0.30}$ \\
 & NW cone & $-$ & $20_{-11}^{+26}$ & $9_{-5}^{+12}$ & $1.9_{-1.1}^{+2.6}$ & $4_{-2}^{+5}$ & $0.28_{-0.16}^{+0.37}$ \rule[-1.5ex]{0pt}{0pt}\\
 \hline
\end{tabular*}
\label{table:moutrate_tab} 
\end{table*}

Here we present an extensive analysis of the outflow in NGC 1365, by comparing the properties and energetics of the extended ionized phase with those of the nuclear X-ray phase.

In Fig. \ref{fig:moutrate_plot} we show the mass outflow rate, the outflow velocity, the kinetic energy rate, and the momentum rate as a function of distance from the active nucleus. These are obtained from 
the gridded map of the mass outflow rate (Fig. \ref{fig:moutrate}) that we have produced in Sect. \ref{ssec:moutrate}. 
The kinetic energy rate and the momentum rate are defined as $\dot{E}_\textrm{k,out} = 1/2 \, \dot{M}_\textrm{out} v^2_\textrm{out}$ and $\dot{p}_\textrm{out} = \dot{M}_\textrm{out} v_\textrm{out}$, respectively.
The plotted points of mass outflow rate, kinetic energy rate, and momentum rate are the sum of the values of these quantities in the grid elements at the same distance, while in the case of the outflow velocity they are the average of the velocities at the given distance. The values reported in the plots are tabulated in Table \ref{table:moutrate_tab}. 
The uncertainties have been estimated through a Monte Carlo simulation\footnote{We have generated mock spectra from the observed data by perturbing the flux value of each spectral channel from its real value according to its statistical variance. We have then performed a spectral fitting of each mock spectrum, giving the estimates of the free parameters. This simulation results in a distribution for all free parameters, from which we have derived the various uncertainties, that we propagated.}.

By inspecting the mass outflow rate in the two cones as a function of distance (Fig. \ref{fig:moutrate_plot}a) we see that, especially in the SE blueshifted cone, there is an initial almost constant or slightly increasing trend up to $\sim$1 kpc, followed by a sharp drop of up to an order of magnitude at larger distances, from $>$10$^{-2}$ $M_\odot$ yr$^{-1}$ to $\sim$10$^{-3}$ $M_\odot$ yr$^{-1}$. The decreasing trend is more irregular in the NW redshifted cone but still clearly present. The constant or slightly increasing values at smaller distances seen in the SE cone are not evident in the NW one, but the shell at the smallest distance is absent in the latter. Moreover, the second shell has the highest values of reddening, which could have been slightly overestimated, resulting in an overestimated mass outflow rate. At any rate, the red points (NW cone) are generally consistent with the blue ones (SE cone) within the uncertainties.

The green dashed line in Fig. \ref{fig:moutrate_plot}a indicates the estimate of the nuclear mass outflow rate from the {\it Chandra} X-ray data from the Fe\,\textsc{xxv} and Fe\,\textsc{xxvi} absorption lines (Sect. \ref{sec:x_rays}), divided by 2 to be compared to the optical mass outflow rates of the single cones, giving 0.034 $M_\odot$ yr$^{-1}$ for $M_7 = 0.45$ (\citealt{Onori:2017aa}). The nuclear X-ray mass outflow rate is consistent with the optical one in the inner three radial shells (after which the mass outflow rate in the cones drops). 

However, we stress that, while the X-ray mass outflow rate is an instantaneous quantity, the optical mass outflow rate of the extended gas traces the time-averaged past activity, and this aspect has to be kept in mind when comparing them (although the mass outflow rate calculated in Eq. \ref{eq:moutrate} is defined as instantaneous according to the usual conventions for extended outflows; e.g., \citealt{Gonzalez-Alfonso:2017aa}). In addition, the velocities we measure for the two outflows may also be subject to different projection effects, due to the different nature of the two signatures (absorption lines for the X-ray outflow, emission lines for the optical one).
Moreover, the mass outflow rate measured from the X-rays is the total nuclear mass outflow rate, as all the gas belonging to the nuclear outflow is expected to be in a highly ionized phase, while the optical outflow we observe from MUSE only samples the ionized gas, thus missing the contribution to the outflow mass from the neutral atomic and molecular phases. Therefore, the total mass outflow rate of the kpc-scale outflow is expected to be larger than the sky-projected one we measure only for the ionized gas and, consequently, larger than the nuclear one. 

As just emphasized, we are only measuring the ionized phase of the gas, whose fraction at a given distance from the ionizing source (which in case of the outflow in NGC 1365 is the AGN) depends on the fraction of ionizing photons intercepting the gas clouds at that distance, which is, in turn, determined by the covering factor of the clouds and by the shading of the AGN flux by the gas from smaller distances. The putative case that the mass outflow rate is constant with distance (when considering all the gas phases) might then result in an ionized mass outflow rate which radially decreases if the gas at larger distances receives less ionizing photons than the gas at smaller distances. Assuming that the total gas mass in each radial shell is conserved, this would imply that velocity would also remain constant with distance.
Overall, this explanation does not seem to apply to NGC 1365, since the outflow velocity clearly varies with distance over the radial extent of the outflow (Fig. \ref{fig:moutrate_plot}b). However, we cannot rule out that this occurs locally, as the velocity stays roughly constant over two to three consecutive radial shells both in the SE and in the NW cone. 
The radial profile of the velocity reflects in general that of the mass outflow rate, as there is a decrease of the velocity in the outer two-three points compared to the inner ones. However, there are some differences between the two, as the innermost shell of the SE cone has almost half the velocity of the three following ones and the NW cone has lower maximum velocities (of about 20-30 km s$^{-1}$) compared to those of the SE cone. Moreover, the velocity of the NW cone has already declined to about 40-50 km s$^{-1}$ at $\sim$1.5 kpc, where the velocity of the SE cone is instead still as high as in the inner shells ($\sim$100 km s$^{-1}$).

The radial profile of kinetic energy rate and momentum rate reflects that of mass outflow rate and velocity (being a by-product of those quantities), though the rising and dropping trend with distance is more regular in this case. 
Decreasing trends with distance of either outflow velocity, mass outflow rate, and outflow energetics are observed in AGN also for example by \cite{Cicone:2015aa}, \cite{Karouzos:2016aa}, \cite{Karouzos:2016ab}, \cite{Bae:2017aa}. \cite{Crenshaw:2015aa} and \cite{Revalski:2018aa} performed a grid- or shell-like analysis of the outflow similar to ours on two single galaxies (NGC 4151 and Mrk 573, respectively), both finding a rising and dropping trend of mass outflow rate, kinetic rate, and momentum rate as we do, but on scales smaller than ours. They cover in fact a radial distance smaller than $\sim$140 pc and $\sim$600 pc, respectively, against the $\sim$2.5 kpc of our case, and the above outflow quantities peak at $\sim$70-90 pc and $\sim$210 pc in their respective cases, while they peak at $\sim$1 kpc in our work.

To compare the energetics of nuclear and extended outflow in NGC 1365, 
we consider their kinetic energy rate and momentum rate, which are $\dot{E}_\textrm{k,out,X}$ $\sim$ 1.9 $\times$ 10$^{41}$ erg s$^{-1}$ and $\dot{p}_\textrm{out,X}$ $\sim$ $1.3\times10^{33}$ g cm s$^{-2}$ for the X-ray outflow and $\dot{E}_\textrm{k,out}$ $\lesssim$ $2 \times 10^{38}$ erg s$^{-1}$ and $\dot{p}_\textrm{out}$ $\lesssim$ $3 \times 10^{31}$ g cm s$^{-2}$ for the optical outflow (these are the peak values of the optical outflow, corresponding to the second and third radial shell; Fig. \ref{fig:moutrate_plot}c and \ref{fig:moutrate_plot}d). 
In order to have momentum conservation between the nuclear X-ray outflow and the large scale outflow in all its phases (ionized, neutral atomic, and molecular), a mass of neutral atomic and molecular gas a factor of at least $\sim$40 larger than the ionized one would be required, supposing that all the three gas phases have the same velocity (which is found for ionized and molecular outflows e.g., by \citealt{Carniani:2015aa}, \citealt{Fiore:2017aa}). This threshold is consistent with the molecular versus ionized mass outflow rate ratios found by \cite{Carniani:2015aa}, \cite{Fiore:2017aa}, \cite{Fluetsch:2018aa}.
Energy conservation would instead require a fraction of neutral atomic and molecular gas a factor of at least $\sim$10$^3$ larger than the ionized gas for the large scale kinetic rate to match the nuclear one, which is unlikely.
Therefore, a momentum-conserving outflow seems to be the favored scenario for NGC 1365. 
However, momentum-driven outflows are predicted to occur on scales $\lesssim$kpc (e.g., \citealt{King:2015aa}), while we are dealing with distances $\gtrsim$kpc in our case. 
Direct AGN radiation pressure on dusty clouds could be potentially capable of driving the outflow, as the AGN photon momentum $L_\textrm{bol}/c$ $\sim$ $7 \times 10^{32}$ g cm s$^{-2}$ is a factor $\sim$20 greater than the peak value of the optical extended outflow ($\sim$ $3 \times 10^{31}$ g cm s$^{-2}$). Momentum rates ranging from $\sim$ 1 up to 5 $L_\textrm{bol}/c$ are expected on $\sim$kpc scales in this regime (\citealt{Thompson:2015aa}, \citealt{Ishibashi:2018aa}), implying that a fraction of neutral atomic and molecular gas a factor at least $\sim$ 20 up to 100 larger than the optical one would be required, still broadly compatible with what \cite{Carniani:2015aa}, \cite{Fiore:2017aa}, \cite{Fluetsch:2018aa} find.

\begin{figure*}[t]
\centering
\begin{subfigure}[t]{0.005\textwidth}
\textbf{a}
\end{subfigure}
\begin{subfigure}[t]{0.32\textwidth}
\includegraphics[trim={0.43cm 0 0 0},clip,width=\textwidth,valign=b]{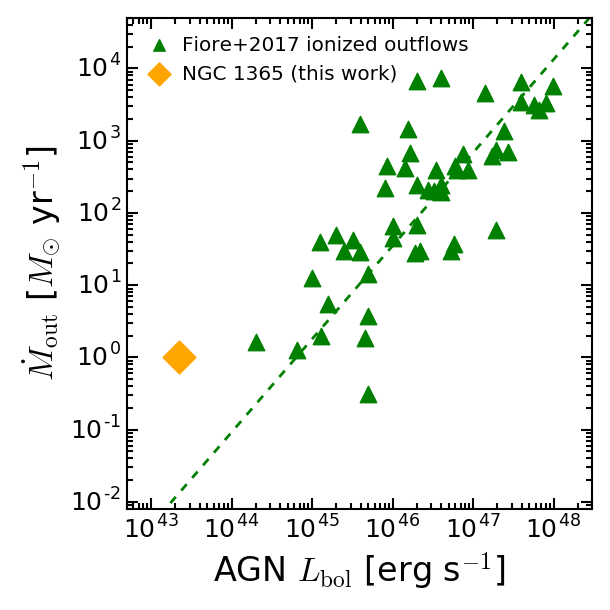}
\end{subfigure}
\begin{subfigure}[t]{0.005\textwidth}
\textbf{b}
\end{subfigure}
\begin{subfigure}[t]{0.32\textwidth}
\includegraphics[trim={0.43cm 0 0 0},clip,width=\textwidth,valign=b]{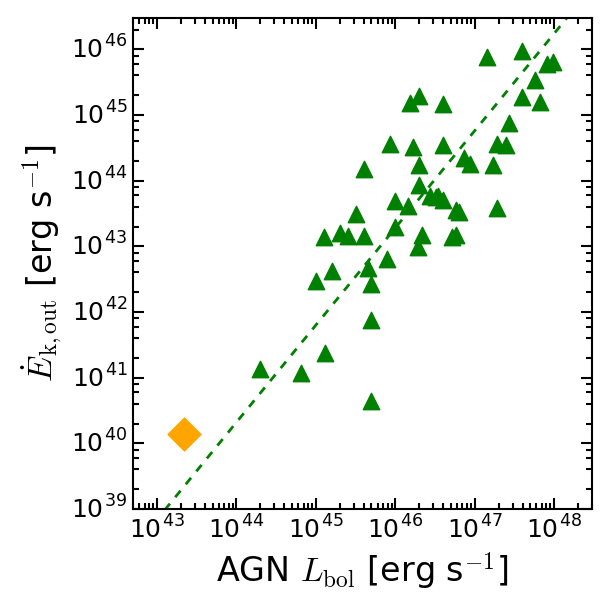}
\end{subfigure}
\begin{subfigure}[t]{0.005\textwidth}
\textbf{c}
\end{subfigure}
\begin{subfigure}[t]{0.32\textwidth}
\includegraphics[trim={0.43cm 0 0 0},clip,width=\textwidth,valign=b]{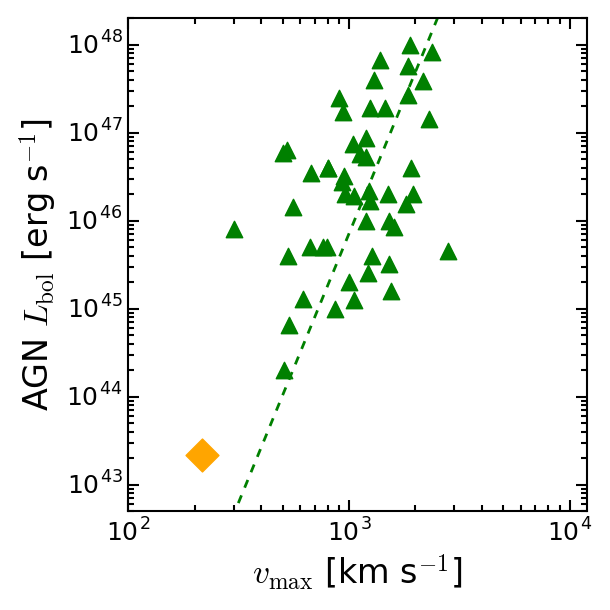}
\end{subfigure}
\caption{{\bf(a)} Mass outflow rate as a function of the AGN bolometric luminosity. The green symbols mark the ionized outflow measurements from \cite{Fiore:2017aa}, while the orange diamond is our estimate for the ionized outflow in NGC 1365, having calculated the mass outflow rate and the maximum outflow velocity consistently with \cite{Fiore:2017aa} from the integrated spectra of the two outflowing cones (see text). The green dashed line is the best-fit correlation from \cite{Fiore:2017aa} for their ionized outflow sample. {\bf(b)} Kinetic energy rate as a function of the AGN bolometric luminosity. {\bf(c)} AGN bolometric luminosity as a function of the maximum outflow velocity.}
\label{fig:fiore}
\end{figure*}

In order to derive integrated estimates of the outflow quantities, to be compared with our spatially resolved measurements, we followed the same approach adopted in Sect. \ref{ssec:fit_Xfull} to obtain the total [O\,\textsc{iii}] luminosity of the two cones, by fitting the two integrated spectra of the cones with two Gaussians per line, one for the outflow and one for the disk. We then obtain a total outflow mass from the luminosity of the H$\alpha$ outflow component of $M_\textrm{out}$ $\sim$ 4 $\times$ 10$^6$ $M_\odot$ and a total kinetic energy of the outflow of $E_\textrm{k,out}$ = 1/2 $M_\textrm{out} \, v_\textrm{out}^2$ $\sim$ 1.2 $\times$ 10$^{53}$ erg.  
They are, respectively, a factor of $\sim$ 4.3 and $\sim$ 2.7 larger than the total outflow mass $M_\textrm{out}$ $\sim$ 9.2 $\times$ 10$^5$ $M_\odot$ and kinetic energy $E_\textrm{k,out}$ $\sim$ 4.5 $\times$ 10$^{52}$ erg we previously measured by summing the single outflow masses and kinetic energies of the grid elements.
This can be partly due to the fact that we are now integrating over a wider area (a factor $\sim$1.7 larger than that covered by the grid elements) and to a contamination by H$\alpha$ emission from the disk. Therefore, the values just obtained have to be taken as rough estimates of the total kinetic energy and of the total outflow mass.

We can estimate the dynamical timescale of the optical outflow by dividing the distance of the most external shell ($\sim$2.5 kpc) by its velocity ($\sim$47 km s$^{-1}$, average between the blueshifted and the redshifted cone; see Fig. \ref{fig:moutrate_plot}b). We then get $t_\textrm{dyn}$ $\sim$ $5.2 \times 10^7$ yr. However, this is an upper limit, as the outflow might have been faster at smaller distances in the past. A constant deceleration is a more reasonable assumption, and, considering as initial velocity of the outflow that of the nuclear X-ray wind ($\sim$3000 km s$^{-1}$), we obtain $t_\textrm{dyn}$ $\sim$ $1.6 \times 10^6$ yr.
We can also calculate the time that would be needed by the AGN at its present-day rate to inject in the system the total amount of kinetic energy observed from the two optical outflowing cones, as $t_\textrm{out}$ = 1/2 $M_\textrm{out} \, v_\textrm{out}^2$ / $\dot{E}_\textrm{k,out,X}$.
With $\dot{E}_\textrm{k,out,X} \sim 1.9 \times 10^{41}$ erg s$^{-1}$ (from Sect. \ref{sec:x_rays}), we obtain $t_\textrm{out} \sim 1.9 \times 10^4$ yr or $t_\textrm{out} \sim 7.5 \times 10^3$ yr, by adopting the kinetic energy from the two fits of the integrated spectra of the cones in the first case and the sum of kinetic energies of the grid elements in the second case. Anyway, the two values have the same order of magnitude of $\sim$10$^4$ yr. This is quite a short timescale for AGN activity and also compared to $t_\textrm{dyn}$ calculated above. 
It is worth noting that, even in the energy-conserving case, simulations predict a coupling efficiency of 10-20\% between the nuclear wind and the extended outflow (e.g., \citealt{Richings:2017aa}), so we should not expect $t_\textrm{out} \sim t_\textrm{dyn}$  (e.g., \citealt{Nardini:2018aa}). However, even considering our most conservative estimation of $t_\textrm{dyn}$ given above, we obtain $t_\textrm{out}/t_\textrm{dyn} < 10^{-2}$, which is far too small. 
This might indicate that the AGN is now in a phase of more intense activity compared to the time-averaged past one from the optical outflow. The fact that both the outflow velocity and the mass outflow rate are lower at larger distances seems to enforce this possibility. 
In principle, the decrease of the mass outflow rate at larger distances could be also ascribed to the radial density profile. We do not find, however, any clear trend in this sense from our [\textsc{S\,ii}]-based estimates, and the same clumpy nature of the outflow seems to rule out such a simple explanation. 
Anyway, the total kinetic energy of all the gas phases rather than of the ionized gas only would be needed for a more significant estimation of $t_\textrm{out}$.

\cite{Fiore:2017aa} studied the dependence of the outflow characteristic quantities (velocity, mass outflow rate, kinetic energy rate) on AGN bolometric luminosity. Their sample of ionized outflows covers bolometric luminosities ranging from ${\sim}2 \times 10^{44}$ erg s$^{-1}$ to $\sim$10$^{48}$ erg s$^{-1}$. As we estimate a bolometric luminosity of $L_\textrm{bol}$ $\sim$ $2.2 \times 10^{43}$ erg s$^{-1}$ for NGC 1365 (see Sect. \ref{ssec:fit_Xfull}), we are then able to extend the work by \cite{Fiore:2017aa} down by an order of magnitude in bolometric luminosity. As the results in their work come from integrated spectra, we adopt $M_\textrm{out}$ and $v_\textrm{out}$ from our fit of the entire cones previously mentioned. 
Moreover, while in our work we measure the outflow velocity $v_\textrm{out}$ from the position of the centroid of the outflow component, in \cite{Fiore:2017aa} the outflow velocity is defined as $v_\textrm{max} = v_\textrm{out} + 2\sigma$ (as e.g., in \citealt{Rupke:2013aa}, \citealt{Feruglio:2015aa}, \citealt{Bischetti:2017aa}), being $\sigma$ the velocity dispersion of the outflow component. We thus calculate the velocity in the above way to make the comparison. The mass outflow rate is obtained from Eq. \ref{eq:moutrate}, where $v_\textrm{max}$ is used instead of $v_\textrm{out}$ and $\Delta R$ is now the extension of the entire outflow, namely $\sim$30$''$ ($\sim$2.7 kpc) per cone as the outflow reaches the margins of MUSE FOV. Moreover, the mass outflow rate is multiplied by 3 to be consistent with the relation adopted in \cite{Fiore:2017aa} for a spherical or multiconical outflow uniformly filled with clumps.

The resulting ionized mass outflow rate ($\dot{M}_\textrm{out}$ $\sim$ 1 $M_\odot$ yr$^{-1}$) is two orders of magnitude larger compared to the the extrapolation at $L_\textrm{bol}$ $\sim$ $2.2 \times 10^{43}$ erg s$^{-1}$ of \cite{Fiore:2017aa} best fit for ionized outflows ($\sim$0.01 $M_\odot$ yr$^{-1}$; Fig. \ref{fig:fiore}a). However, the scatter in their points is large, spanning up to almost three orders of magnitude at high bolometric luminosities and thus the extrapolation at low luminosities is quite uncertain. 
We obtain a kinetic energy rate $\dot{E}_\textrm{k,out}$ $\sim$ $1.4 \times 10^{40}$ erg s$^{-1}$, which is one order of magnitude larger than the extrapolated value from \cite{Fiore:2017aa} ($1.4 \times 10^{39}$ erg s$^{-1}$), but within the scatter of their points (Fig. \ref{fig:fiore}b). The velocity that we get ($v_\textrm{max}$ $\sim$ 215 km s$^{-1}$) is also consistent with the extrapolation at low luminosities from their best fit ($\sim$370 km s$^{-1}$) within the scatter (Fig. \ref{fig:fiore}c).  

We emphasize that the integrated mass outflow rate obtained following \cite{Fiore:2017aa} prescriptions is much higher (by a factor >20) than the radial values we mapped with our own method (Fig. \ref{fig:moutrate_plot}a). 
This highlights that different methods for calculating the outflow quantities give a large scatter in the results and underlines the importance of having spatially resolved information to properly disentangle the outflows from the disk motions.

\section{Conclusions}
We have presented a study of the ionized gas in the central $\sim$5.3 kpc of the nearby Seyfert galaxy NGC 1365, both in its warm and hot phase, using spatially resolved optical VLT/MUSE observations and {\it Chandra} ACIS-S X-ray data. Thanks to the capabilities of MUSE, we could map the spatial distribution of the optical emission lines to an unprecedented detail, and obtained unique insights by comparing them with the hot gas from {\it Chandra}.

The main results of our analysis are summarized below:
\begin{itemize}
\item We detect a biconical outflow prominent in [O\,\textsc{iii}] and extended out to about 2.5 kpc in both directions, with line-of-sight velocities up to $\pm$170 km s$^{-1}$. We resolve the clumpy structure of the outflow down to a scale of $\sim$70 pc. The cone to the SE of the nucleus has approaching velocities, while the NW one recedes and is located behind the disk, being partly obscured by the dust in the disk, especially closer to the center. The SE cone, despite residing above the disk, has a typical extinction of $A_V$ $\simeq$ 0.7, indicating that the outflow contains dust, in agreement with the polar extra-disk dust features observed on smaller scales in some AGN. Moreover, 
 the dust might be coupled with cold molecular gas which would be potentially capable of forming stars within the outflow.
\item The AGN ionizes the outflowing cones as inferred from the spatially resolved BPT diagnostic diagrams. The AGN origin of the ionization in the two cones is further supported by the fact that the extinction-corrected [O\,\textsc{iii}] emission measured in the cones is compatible with the value that we get from the 2-10 keV X-ray emission using the latest $L_\textrm{X, 2-10 keV}$--$L_\textrm{[O\,\textsc{iii}]}^\textrm{cor}$ correlation for Type 2 AGN.
\item Star formation dominates the disk and the bar of the galaxy (which extends in the direction SW-NE, perpendicular to the biconical outflow), where a strong H$\alpha$ emitting circumnuclear star-forming ring is observed, residing between the outer and the inner Inner Lindblad Resonances. The ring is also traced by the density map (from the [S\,\textsc{ii}] line doublet ratio), where values up to 10$^3$ cm$^{-3}$ are found in correspondence with it. The H$\alpha$ emission also nicely follows the leading edges of two dust lanes in the bar in an S-shaped pattern wrapping the circumnuclear ring.
\item The soft X-rays in NGC 1365 are dominated by thermal emission due to star-forming processes, with about 40$\%$ of the soft X-ray emission originating from the SF ring. However, by selecting appropriate X-ray energy intervals we are able to recover the AGN-photoionized component, which spatially matches the [O\,\textsc{iii}] optical AGN-photoionized emission from MUSE. 
\item The mass outflow rate of the kpc-scale ionized outflow from the MUSE optical emission lines is of the same order of magnitude as the one obtained for the nuclear highly-ionized outflow from the Fe\,\textsc{xxv} and Fe\,\textsc{xxvi} absorption lines from {\it Chandra} X-ray data ($\approx$0.067 $M_\odot$ yr$^{-1}$). This, together with the fact that the AGN ionizes the double-conical outflow, suggests that the AGN is the driver of the wind. We observe a decrease of the mass outflow rate with distance, which might either imply that the outflow slows down at larger distances ($\sim$1.5 kpc) or that the AGN pushing the wind has become more powerful recently. An energy-driven scenario for the extended outflow is unlikely, as a fraction of neutral atomic and molecular gas $\gtrsim$10$^3$ than the ionized one would be required to have energy conservation, which is too large compared to what is observed in other AGN. A lower fraction ($\gtrsim$40) would be required for momentum conservation, but models predict that the momentum-driven mode operates on scales $\lesssim$1 kpc, while the outflow in NGC 1365 extends on few kpc. We note that the constraints of $\gtrsim$10$^3$ and $\gtrsim$40 on the gas fraction are partially relaxed in the hypothesis that the AGN (and so the X-ray wind) is more powerful at present time than in the past. 
Finally, direct acceleration by the AGN radiation pressure on dusty clouds could be a possible driver of the outflow. 
\item The integrated mass outflow rate, kinetic energy rate, and outflow velocity are broadly consistent with the extrapolation at low AGN bolometric luminosities of \cite{Fiore:2017aa} relations for more luminous AGN ($10^{44}$ $\lesssim$ $L_\textrm{bol}$ $\lesssim$ $10^{48}$ erg s$^{-1}$), being the AGN bolometric luminosity of NGC 1365 ${\sim} 2.2 \times 10^{43}$ erg s$^{-1}$. However, we note that the integrated mass outflow rate, calculated according to \cite{Fiore:2017aa} prescriptions, is 20 times larger than the values that we obtain from our radial analysis of the outflowing shells.
This indicates that different methods may lead to a large discrepancy in estimating the outflow quantities and stresses the importance of having spatially resolved information to properly disentangle outflow and disk kinematics.
\item We observe an excess of hard X-ray (3-5 keV) extended emission cospatial with the SF ring in H$\alpha$. This could be tentatively associated with a hotter gas component due to SNe in the ring driving a shocked wind, which might contribute in principle to the kpc-scale outflow. However, higher quality X-ray data are needed to discern if star formation actually contributes to the observed optical outflow and, if it does, to which extent.
\end{itemize}

\section{Acknowledgments}
E.N., C.C., and C.F. acknowledge funding from the European Union’s Horizon 2020 research and innovation program under the Marie Sk\l{}odowska-Curie grant agreement no. 664931. 
G.C. acknowledges the support by INAF/Frontiera through the ``Progetti 
Premiali'' funding scheme of the Italian Ministry of Education, 
University, and Research. G.C. has been supported by the INAF PRIN-SKA 
2017 program 1.05.01.88.04.
S.Z. has been supported by the EU Marie Curie Career Integration Grant ``SteMaGE'' Nr. PCIG12-GA-2012-326466  (Call Identifier: FP7-PEOPLE-2012 CIG) and by the INAF PRIN-SKA 2017 program 1.05.01.88.04.
F.F. acknowledges support from PRIN INAF 2016 FORECAST.
This research has made use of NASA's Astrophysics Data System Service. This research has made use of the NASA/IPAC Extragalactic Database (NED) which is operated by the Jet Propulsion Laboratory, California Institute of Technology, under contract with the National Aeronautics and Space Administration.
We acknowledge the usage of the HyperLeda database (\url{http://leda.univ-lyon1.fr}; \citealt{Makarov:2014aa}).

\newpage
\bibliographystyle{aa}
\bibliography{bibliography_1}

\begin{thebibliography}{110}
\expandafter\ifx\csname natexlab\endcsname\relax\def\natexlab#1{#1}\fi

\bibitem[{{Alonso-Herrero} {et~al.}(2012){Alonso-Herrero},
  {S{\'a}nchez-Portal}, {Ramos Almeida}, {Pereira-Santaella}, {Esquej},
  {Garc{\'{\i}}a-Burillo}, {Castillo}, {Gonz{\'a}lez-Mart{\'{\i}}n},
  {Levenson}, {Hatziminaoglou}, {Acosta-Pulido}, {Gonz{\'a}lez-Serrano},
  {Povi{\'c}}, {Packham}, \& {P{\'e}rez-Garc{\'{\i}}a}}]{Alonso-Herrero:2012aa}
{Alonso-Herrero}, A., {S{\'a}nchez-Portal}, M., {Ramos Almeida}, C., {et~al.}
  2012, \mnras, 425, 311

\bibitem[{{Arnaud}(1996)}]{Arnaud:1996aa}
{Arnaud}, K.~A. 1996, in Astronomical Society of the Pacific Conference Series,
  Vol. 101, Astronomical Data Analysis Software and Systems V, ed. G.~H.
  {Jacoby} \& J.~{Barnes}, 17

\bibitem[{{Asmus} {et~al.}(2016){Asmus}, {H{\"o}nig}, \&
  {Gandhi}}]{Asmus:2016aa}
{Asmus}, D., {H{\"o}nig}, S.~F., \& {Gandhi}, P. 2016, \apj, 822, 109

\bibitem[{{Bacon} {et~al.}(2010){Bacon}, {Accardo}, {Adjali}, {Anwand},
  {Bauer}, {Biswas}, {Blaizot}, {Boudon}, {Brau-Nogue}, {Brinchmann},
  {Caillier}, {Capoani}, {Carollo}, {Contini}, {Couderc}, {Daguis{\'e}},
  {Deiries}, {Delabre}, {Dreizler}, {Dubois}, {Dupieux}, {Dupuy}, {Emsellem},
  {Fechner}, {Fleischmann}, {Fran{\c c}ois}, {Gallou}, {Gharsa}, {Glindemann},
  {Gojak}, {Guiderdoni}, {Hansali}, {Hahn}, {Jarno}, {Kelz}, {Koehler},
  {Kosmalski}, {Laurent}, {Le Floch}, {Lilly}, {Lizon}, {Loupias}, {Manescau},
  {Monstein}, {Nicklas}, {Olaya}, {Pares}, {Pasquini}, {P{\'e}contal-Rousset},
  {Pell{\'o}}, {Petit}, {Popow}, {Reiss}, {Remillieux}, {Renault}, {Roth},
  {Rupprecht}, {Serre}, {Schaye}, {Soucail}, {Steinmetz}, {Streicher}, {Stuik},
  {Valentin}, {Vernet}, {Weilbacher}, {Wisotzki}, \& {Yerle}}]{Bacon:2010aa}
{Bacon}, R., {Accardo}, M., {Adjali}, L., {et~al.} 2010, in Society of
  Photo-Optical Instrumentation Engineers (SPIE) Conference Series, Vol. 7735,
  Society of Photo-Optical Instrumentation Engineers (SPIE) Conference Series,
  773508

\bibitem[{{Bae} {et~al.}(2017){Bae}, {Woo}, {Karouzos}, {Gallo}, {Flohic},
  {Shen}, \& {Yoon}}]{Bae:2017aa}
{Bae}, H.-J., {Woo}, J.-H., {Karouzos}, M., {et~al.} 2017, \apj, 837, 91

\bibitem[{{Baldwin} {et~al.}(1981){Baldwin}, {Phillips}, \&
  {Terlevich}}]{Baldwin:1981aa}
{Baldwin}, J.~A., {Phillips}, M.~M., \& {Terlevich}, R. 1981, \pasp, 93, 5

\bibitem[{{Belfiore} {et~al.}(2016){Belfiore}, {Maiolino}, {Maraston},
  {Emsellem}, {Bershady}, {Masters}, {Yan}, {Bizyaev}, {Boquien}, {Brownstein},
  {Bundy}, {Drory}, {Heckman}, {Law}, {Roman-Lopes}, {Pan}, {Stanghellini},
  {Thomas}, {Weijmans}, \& {Westfall}}]{Belfiore:2016aa}
{Belfiore}, F., {Maiolino}, R., {Maraston}, C., {et~al.} 2016, \mnras, 461,
  3111

\bibitem[{{Bianchi} {et~al.}(2006){Bianchi}, {Guainazzi}, \&
  {Chiaberge}}]{Bianchi:2006aa}
{Bianchi}, S., {Guainazzi}, M., \& {Chiaberge}, M. 2006, \aap, 448, 499

\bibitem[{{Binney} \& {Tremaine}(2008)}]{Binney:2008aa}
{Binney}, J. \& {Tremaine}, S. 2008, {Galactic Dynamics: Second Edition}
  (Princeton University Press)

\bibitem[{{Bischetti} {et~al.}(2017){Bischetti}, {Piconcelli}, {Vietri},
  {Bongiorno}, {Fiore}, {Sani}, {Marconi}, {Duras}, {Zappacosta}, {Brusa},
  {Comastri}, {Cresci}, {Feruglio}, {Giallongo}, {La Franca}, {Mainieri},
  {Mannucci}, {Martocchia}, {Ricci}, {Schneider}, {Testa}, \&
  {Vignali}}]{Bischetti:2017aa}
{Bischetti}, M., {Piconcelli}, E., {Vietri}, G., {et~al.} 2017, \aap, 598, A122

\bibitem[{{Braito} {et~al.}(2014){Braito}, {Reeves}, {Gofford}, {Nardini},
  {Porquet}, \& {Risaliti}}]{Braito:2014aa}
{Braito}, V., {Reeves}, J.~N., {Gofford}, J., {et~al.} 2014, \apj, 795, 87

\bibitem[{{Brenneman} {et~al.}(2013){Brenneman}, {Risaliti}, {Elvis}, \&
  {Nardini}}]{Brenneman:2013aa}
{Brenneman}, L.~W., {Risaliti}, G., {Elvis}, M., \& {Nardini}, E. 2013, \mnras,
  429, 2662

\bibitem[{{Bureau} {et~al.}(1996){Bureau}, {Mould}, \&
  {Staveley-Smith}}]{Bureau:1996aa}
{Bureau}, M., {Mould}, J.~R., \& {Staveley-Smith}, L. 1996, \apj, 463, 60

\bibitem[{{Calzetti} {et~al.}(2000){Calzetti}, {Armus}, {Bohlin}, {Kinney},
  {Koornneef}, \& {Storchi-Bergmann}}]{Calzetti:2000aa}
{Calzetti}, D., {Armus}, L., {Bohlin}, R.~C., {et~al.} 2000, \apj, 533, 682

\bibitem[{{Cano-D{\'{\i}}az} {et~al.}(2012){Cano-D{\'{\i}}az}, {Maiolino},
  {Marconi}, {Netzer}, {Shemmer}, \& {Cresci}}]{Cano-Diaz:2012aa}
{Cano-D{\'{\i}}az}, M., {Maiolino}, R., {Marconi}, A., {et~al.} 2012, \aap,
  537, L8

\bibitem[{{Cappellari} \& {Copin}(2003)}]{Cappellari:2003aa}
{Cappellari}, M. \& {Copin}, Y. 2003, \mnras, 342, 345

\bibitem[{{Cappellari} \& {Emsellem}(2004)}]{Cappellari:2004aa}
{Cappellari}, M. \& {Emsellem}, E. 2004, \pasp, 116, 138

\bibitem[{{Carniani} {et~al.}(2016){Carniani}, {Marconi}, {Maiolino},
  {Balmaverde}, {Brusa}, {Cano-D{\'{\i}}az}, {Cicone}, {Comastri}, {Cresci},
  {Fiore}, {Feruglio}, {La Franca}, {Mainieri}, {Mannucci}, {Nagao}, {Netzer},
  {Piconcelli}, {Risaliti}, {Schneider}, \& {Shemmer}}]{Carniani:2016aa}
{Carniani}, S., {Marconi}, A., {Maiolino}, R., {et~al.} 2016, \aap, 591, A28

\bibitem[{{Carniani} {et~al.}(2015){Carniani}, {Marconi}, {Maiolino},
  {Balmaverde}, {Brusa}, {Cano-D{\'{\i}}az}, {Cicone}, {Comastri}, {Cresci},
  {Fiore}, {Feruglio}, {La Franca}, {Mainieri}, {Mannucci}, {Nagao}, {Netzer},
  {Piconcelli}, {Risaliti}, {Schneider}, \& {Shemmer}}]{Carniani:2015aa}
{Carniani}, S., {Marconi}, A., {Maiolino}, R., {et~al.} 2015, \aap, 580, A102

\bibitem[{{Cicone} {et~al.}(2015){Cicone}, {Maiolino}, {Gallerani}, {Neri},
  {Ferrara}, {Sturm}, {Fiore}, {Piconcelli}, \& {Feruglio}}]{Cicone:2015aa}
{Cicone}, C., {Maiolino}, R., {Gallerani}, S., {et~al.} 2015, \aap, 574, A14

\bibitem[{{Crenshaw} {et~al.}(2015){Crenshaw}, {Fischer}, {Kraemer}, \&
  {Schmitt}}]{Crenshaw:2015aa}
{Crenshaw}, D.~M., {Fischer}, T.~C., {Kraemer}, S.~B., \& {Schmitt}, H.~R.
  2015, \apj, 799, 83

\bibitem[{{Cresci} {et~al.}(2015{\natexlab{a}}){Cresci}, {Mainieri}, {Brusa},
  {Marconi}, {Perna}, {Mannucci}, {Piconcelli}, {Maiolino}, {Feruglio},
  {Fiore}, {Bongiorno}, {Lanzuisi}, {Merloni}, {Schramm}, {Silverman}, \&
  {Civano}}]{Cresci:2015ab}
{Cresci}, G., {Mainieri}, V., {Brusa}, M., {et~al.} 2015{\natexlab{a}}, \apj,
  799, 82

\bibitem[{{Cresci} \& {Maiolino}(2018)}]{Cresci:2018aa}
{Cresci}, G. \& {Maiolino}, R. 2018, Nature Astronomy, 2, 179

\bibitem[{{Cresci} {et~al.}(2015{\natexlab{b}}){Cresci}, {Marconi}, {Zibetti},
  {Risaliti}, {Carniani}, {Mannucci}, {Gallazzi}, {Maiolino}, {Balmaverde},
  {Brusa}, {Capetti}, {Cicone}, {Feruglio}, {Bland-Hawthorn}, {Nagao}, {Oliva},
  {Salvato}, {Sani}, {Tozzi}, {Urrutia}, \& {Venturi}}]{Cresci:2015aa}
{Cresci}, G., {Marconi}, A., {Zibetti}, S., {et~al.} 2015{\natexlab{b}}, \aap,
  582, A63

\bibitem[{{Cresci} {et~al.}(2017){Cresci}, {Vanzi}, {Telles}, {Lanzuisi},
  {Brusa}, {Mingozzi}, {Sauvage}, \& {Johnson}}]{Cresci:2017aa}
{Cresci}, G., {Vanzi}, L., {Telles}, E., {et~al.} 2017, \aap, 604, A101

\bibitem[{{Davis} {et~al.}(2014){Davis}, {Berrier}, {Johns}, {Shields},
  {Hartley}, {Kennefick}, {Kennefick}, {Seigar}, \& {Lacy}}]{Davis:2014aa}
{Davis}, B.~L., {Berrier}, J.~C., {Johns}, L., {et~al.} 2014, \apj, 789, 124

\bibitem[{{de Vaucouleurs} {et~al.}(1991){de Vaucouleurs}, {de Vaucouleurs},
  {Corwin}, {Buta}, {Paturel}, \& {Fouqu{\'e}}}]{de-Vaucouleurs:1991aa}
{de Vaucouleurs}, G., {de Vaucouleurs}, A., {Corwin}, Jr., H.~G., {et~al.}
  1991, {Third Reference Catalogue of Bright Galaxies. Volume I: Explanations
  and references. Volume II: Data for galaxies between 0$^{h}$ and 12$^{h}$.
  Volume III: Data for galaxies between 12$^{h}$ and 24$^{h}$.}

\bibitem[{{Dopita} \& {Sutherland}(1995)}]{Dopita:1995aa}
{Dopita}, M.~A. \& {Sutherland}, R.~S. 1995, \apj, 455, 468

\bibitem[{{Edmunds} {et~al.}(1988){Edmunds}, {Taylor}, \&
  {Turtle}}]{Edmunds:1988aa}
{Edmunds}, M.~G., {Taylor}, K., \& {Turtle}, A.~J. 1988, \mnras, 234, 155

\bibitem[{{Fabian}(2012)}]{Fabian:2012aa}
{Fabian}, A.~C. 2012, \araa, 50, 455

\bibitem[{{Feruglio} {et~al.}(2015){Feruglio}, {Fiore}, {Carniani},
  {Piconcelli}, {Zappacosta}, {Bongiorno}, {Cicone}, {Maiolino}, {Marconi},
  {Menci}, {Puccetti}, \& {Veilleux}}]{Feruglio:2015aa}
{Feruglio}, C., {Fiore}, F., {Carniani}, S., {et~al.} 2015, \aap, 583, A99

\bibitem[{{Fiore} {et~al.}(2017){Fiore}, {Feruglio}, {Shankar}, {Bischetti},
  {Bongiorno}, {Brusa}, {Carniani}, {Cicone}, {Duras}, {Lamastra}, {Mainieri},
  {Marconi}, {Menci}, {Maiolino}, {Piconcelli}, {Vietri}, \&
  {Zappacosta}}]{Fiore:2017aa}
{Fiore}, F., {Feruglio}, C., {Shankar}, F., {et~al.} 2017, \aap, 601, A143

\bibitem[{{Fluetsch} {et~al.}(2018){Fluetsch}, {Maiolino}, {Carniani},
  {Marconi}, {Cicone}, {Bourne}, {Costa}, {Fabian}, {Ishibashi}, \&
  {Venturi}}]{Fluetsch:2018aa}
{Fluetsch}, A., {Maiolino}, R., {Carniani}, S., {et~al.} 2018, ArXiv e-prints
  [\eprint[arXiv]{1805.05352}]

\bibitem[{{Forbes} \& {Norris}(1998)}]{Forbes:1998aa}
{Forbes}, D.~A. \& {Norris}, R.~P. 1998, \mnras, 300, 757

\bibitem[{{Freudling} {et~al.}(2013){Freudling}, {Romaniello}, {Bramich},
  {Ballester}, {Forchi}, {Garc{\'{\i}}a-Dabl{\'o}}, {Moehler}, \&
  {Neeser}}]{Freudling:2013aa}
{Freudling}, W., {Romaniello}, M., {Bramich}, D.~M., {et~al.} 2013, \aap, 559,
  A96

\bibitem[{{Fruscione} {et~al.}(2006){Fruscione}, {McDowell}, {Allen},
  {Brickhouse}, {Burke}, {Davis}, {Durham}, {Elvis}, {Galle}, {Harris},
  {Huenemoerder}, {Houck}, {Ishibashi}, {Karovska}, {Nicastro}, {Noble},
  {Nowak}, {Primini}, {Siemiginowska}, {Smith}, \& {Wise}}]{Fruscione:2006aa}
{Fruscione}, A., {McDowell}, J.~C., {Allen}, G.~E., {et~al.} 2006, in Society
  of Photo-Optical Instrumentation Engineers (SPIE) Conference Series, Vol.
  6270, Society of Photo-Optical Instrumentation Engineers (SPIE) Conference
  Series, 62701V

\bibitem[{{Garmire} {et~al.}(2003){Garmire}, {Bautz}, {Ford}, {Nousek}, \&
  {Ricker}}]{Garmire:2003aa}
{Garmire}, G.~P., {Bautz}, M.~W., {Ford}, P.~G., {Nousek}, J.~A., \& {Ricker},
  Jr., G.~R. 2003, in Society of Photo-Optical Instrumentation Engineers (SPIE)
  Conference Series, Vol. 4851, X-Ray and Gamma-Ray Telescopes and Instruments
  for Astronomy., ed. J.~E. {Truemper} \& H.~D. {Tananbaum}, 28--44

\bibitem[{{Gonz{\'a}lez-Alfonso} {et~al.}(2017){Gonz{\'a}lez-Alfonso},
  {Fischer}, {Spoon}, {Stewart}, {Ashby}, {Veilleux}, {Smith}, {Sturm},
  {Farrah}, {Falstad}, {Mel{\'e}ndez}, {Graci{\'a}-Carpio}, {Janssen}, \&
  {Lebouteiller}}]{Gonzalez-Alfonso:2017aa}
{Gonz{\'a}lez-Alfonso}, E., {Fischer}, J., {Spoon}, H.~W.~W., {et~al.} 2017,
  \apj, 836, 11

\bibitem[{{Guainazzi} {et~al.}(2009){Guainazzi}, {Risaliti}, {Nucita}, {Wang},
  {Bianchi}, {Soria}, \& {Zezas}}]{Guainazzi:2009aa}
{Guainazzi}, M., {Risaliti}, G., {Nucita}, A., {et~al.} 2009, \aap, 505, 589

\bibitem[{{Harrison} {et~al.}(2012){Harrison}, {Alexander}, {Swinbank},
  {Smail}, {Alaghband-Zadeh}, {Bauer}, {Chapman}, {Del Moro}, {Hickox},
  {Ivison}, {Men{\'e}ndez-Delmestre}, {Mullaney}, \&
  {Nesvadba}}]{Harrison:2012aa}
{Harrison}, C.~M., {Alexander}, D.~M., {Swinbank}, A.~M., {et~al.} 2012,
  \mnras, 426, 1073

\bibitem[{{Hjelm} \& {Lindblad}(1996)}]{Hjelm:1996aa}
{Hjelm}, M. \& {Lindblad}, P.~O. 1996, \aap, 305, 727

\bibitem[{{Ishibashi} {et~al.}(2018){Ishibashi}, {Fabian}, \&
  {Maiolino}}]{Ishibashi:2018aa}
{Ishibashi}, W., {Fabian}, A.~C., \& {Maiolino}, R. 2018, \mnras, 476, 512

\bibitem[{{Iyomoto} {et~al.}(1997){Iyomoto}, {Makishima}, {Fukazawa},
  {Tashiro}, \& {Ishisaki}}]{Iyomoto:1997aa}
{Iyomoto}, N., {Makishima}, K., {Fukazawa}, Y., {Tashiro}, M., \& {Ishisaki},
  Y. 1997, \pasj, 49, 425

\bibitem[{{Jones} \& {Jones}(1980)}]{Jones:1980aa}
{Jones}, J.~E. \& {Jones}, B.~J.~T. 1980, \mnras, 191, 685

\bibitem[{{Jorsater} \& {van Moorsel}(1995)}]{Jorsater:1995aa}
{Jorsater}, S. \& {van Moorsel}, G.~A. 1995, \aj, 110, 2037

\bibitem[{{Kalberla} {et~al.}(2005){Kalberla}, {Burton}, {Hartmann}, {Arnal},
  {Bajaja}, {Morras}, \& {P{\"o}ppel}}]{Kalberla:2005aa}
{Kalberla}, P.~M.~W., {Burton}, W.~B., {Hartmann}, D., {et~al.} 2005, \aap,
  440, 775

\bibitem[{{Kallman} \& {Bautista}(2001)}]{Kallman:2001aa}
{Kallman}, T. \& {Bautista}, M. 2001, \apjs, 133, 221

\bibitem[{{Karouzos} {et~al.}(2016{\natexlab{a}}){Karouzos}, {Woo}, \&
  {Bae}}]{Karouzos:2016aa}
{Karouzos}, M., {Woo}, J.-H., \& {Bae}, H.-J. 2016{\natexlab{a}}, \apj, 819,
  148

\bibitem[{{Karouzos} {et~al.}(2016{\natexlab{b}}){Karouzos}, {Woo}, \&
  {Bae}}]{Karouzos:2016ab}
{Karouzos}, M., {Woo}, J.-H., \& {Bae}, H.-J. 2016{\natexlab{b}}, \apj, 833,
  171

\bibitem[{{Kauffmann} {et~al.}(2003){Kauffmann}, {Heckman}, {Tremonti},
  {Brinchmann}, {Charlot}, {White}, {Ridgway}, {Brinkmann}, {Fukugita}, {Hall},
  {Ivezi{\'c}}, {Richards}, \& {Schneider}}]{Kauffmann:2003aa}
{Kauffmann}, G., {Heckman}, T.~M., {Tremonti}, C., {et~al.} 2003, \mnras, 346,
  1055

\bibitem[{{Kewley} {et~al.}(2001){Kewley}, {Dopita}, {Sutherland}, {Heisler},
  \& {Trevena}}]{Kewley:2001ab}
{Kewley}, L.~J., {Dopita}, M.~A., {Sutherland}, R.~S., {Heisler}, C.~A., \&
  {Trevena}, J. 2001, \apj, 556, 121

\bibitem[{{Kewley} {et~al.}(2006){Kewley}, {Groves}, {Kauffmann}, \&
  {Heckman}}]{Kewley:2006aa}
{Kewley}, L.~J., {Groves}, B., {Kauffmann}, G., \& {Heckman}, T. 2006, \mnras,
  372, 961

\bibitem[{{King} \& {Pounds}(2015)}]{King:2015aa}
{King}, A. \& {Pounds}, K. 2015, \araa, 53, 115

\bibitem[{{Komossa} \& {Schulz}(1998)}]{Komossa:1998aa}
{Komossa}, S. \& {Schulz}, H. 1998, \aap, 339, 345

\bibitem[{{Kristen} {et~al.}(1997){Kristen}, {Jorsater}, {Lindblad}, \&
  {Boksenberg}}]{Kristen:1997aa}
{Kristen}, H., {Jorsater}, S., {Lindblad}, P.~O., \& {Boksenberg}, A. 1997,
  \aap, 328, 483

\bibitem[{{Lena} {et~al.}(2016){Lena}, {Robinson}, {Storchi-Bergmann}, {Couto},
  {Schnorr-M{\"u}ller}, \& {Riffel}}]{Lena:2016aa}
{Lena}, D., {Robinson}, A., {Storchi-Bergmann}, T., {et~al.} 2016, \mnras, 459,
  4485

\bibitem[{{Lindblad} {et~al.}(1996{\natexlab{a}}){Lindblad}, {Lindblad}, \&
  {Athanassoula}}]{Lindblad:1996ab}
{Lindblad}, P.~A.~B., {Lindblad}, P.~O., \& {Athanassoula}, E.
  1996{\natexlab{a}}, \aap, 313, 65

\bibitem[{{Lindblad}(1999)}]{Lindblad:1999aa}
{Lindblad}, P.~O. 1999, \aapr, 9, 221

\bibitem[{{Lindblad} {et~al.}(1996{\natexlab{b}}){Lindblad}, {Hjelm},
  {Hoegbom}, {Joersaeter}, {Lindblad}, \& {Santos-Lleo}}]{Lindblad:1996aa}
{Lindblad}, P.~O., {Hjelm}, M., {Hoegbom}, J., {et~al.} 1996{\natexlab{b}},
  \aaps, 120, 403

\bibitem[{{L{\'o}pez-Gonzaga} {et~al.}(2016){L{\'o}pez-Gonzaga}, {Burtscher},
  {Tristram}, {Meisenheimer}, \& {Schartmann}}]{Lopez-Gonzaga:2016aa}
{L{\'o}pez-Gonzaga}, N., {Burtscher}, L., {Tristram}, K.~R.~W., {Meisenheimer},
  K., \& {Schartmann}, M. 2016, \aap, 591, A47

\bibitem[{{Lusso} {et~al.}(2011){Lusso}, {Comastri}, {Vignali}, {Zamorani},
  {Treister}, {Sanders}, {Bolzonella}, {Bongiorno}, {Brusa}, {Civano}, {Gilli},
  {Mainieri}, {Nair}, {Aller}, {Carollo}, {Koekemoer}, {Merloni}, \&
  {Trump}}]{Lusso:2011aa}
{Lusso}, E., {Comastri}, A., {Vignali}, C., {et~al.} 2011, \aap, 534, A110

\bibitem[{{Magdziarz} \& {Zdziarski}(1995)}]{Magdziarz:1995aa}
{Magdziarz}, P. \& {Zdziarski}, A.~A. 1995, \mnras, 273, 837

\bibitem[{{Maiolino} {et~al.}(2017){Maiolino}, {Russell}, {Fabian}, {Carniani},
  {Gallagher}, {Cazzoli}, {Arribas}, {Belfiore}, {Bellocchi}, {Colina},
  {Cresci}, {Ishibashi}, {Marconi}, {Mannucci}, {Oliva}, \&
  {Sturm}}]{Maiolino:2017aa}
{Maiolino}, R., {Russell}, H.~R., {Fabian}, A.~C., {et~al.} 2017, \nat, 544,
  202

\bibitem[{{Makarov} {et~al.}(2014){Makarov}, {Prugniel}, {Terekhova},
  {Courtois}, \& {Vauglin}}]{Makarov:2014aa}
{Makarov}, D., {Prugniel}, P., {Terekhova}, N., {Courtois}, H., \& {Vauglin},
  I. 2014, \aap, 570, A13

\bibitem[{{Markwardt}(2009)}]{Markwardt:2009aa}
{Markwardt}, C.~B. 2009, in Astronomical Society of the Pacific Conference
  Series, Vol. 411, Astronomical Data Analysis Software and Systems XVIII, ed.
  D.~A. {Bohlender}, D.~{Durand}, \& P.~{Dowler}, 251

\bibitem[{{McKee} \& {Hollenbach}(1980)}]{McKee:1980aa}
{McKee}, C.~F. \& {Hollenbach}, D.~J. 1980, \araa, 18, 219

\bibitem[{{Mineo} {et~al.}(2012){Mineo}, {Gilfanov}, \&
  {Sunyaev}}]{Mineo:2012aa}
{Mineo}, S., {Gilfanov}, M., \& {Sunyaev}, R. 2012, \mnras, 419, 2095

\bibitem[{{Nardini} {et~al.}(2015{\natexlab{a}}){Nardini}, {Gofford}, {Reeves},
  {Braito}, {Risaliti}, \& {Costa}}]{Nardini:2015ab}
{Nardini}, E., {Gofford}, J., {Reeves}, J.~N., {et~al.} 2015{\natexlab{a}},
  \mnras, 453, 2558

\bibitem[{{Nardini} {et~al.}(2015{\natexlab{b}}){Nardini}, {Reeves}, {Gofford},
  {Harrison}, {Risaliti}, {Braito}, {Costa}, {Matzeu}, {Walton}, {Behar},
  {Boggs}, {Christensen}, {Craig}, {Hailey}, {Matt}, {Miller}, {O'Brien},
  {Stern}, {Turner}, \& {Ward}}]{Nardini:2015aa}
{Nardini}, E., {Reeves}, J.~N., {Gofford}, J., {et~al.} 2015{\natexlab{b}},
  Science, 347, 860

\bibitem[{{Nardini} \& {Zubovas}(2018)}]{Nardini:2018aa}
{Nardini}, E. \& {Zubovas}, K. 2018, ArXiv e-prints
  [\eprint[arXiv]{1805.00040}]

\bibitem[{{Onori} {et~al.}(2017){Onori}, {Ricci}, {La Franca}, {Bianchi},
  {Bongiorno}, {Brusa}, {Fiore}, {Maiolino}, {Marconi}, {Sani}, \&
  {Vignali}}]{Onori:2017aa}
{Onori}, F., {Ricci}, F., {La Franca}, F., {et~al.} 2017, \mnras, 468, L97

\bibitem[{{Osterbrock} \& {Ferland}(2006)}]{Osterbrock:2006aa}
{Osterbrock}, D.~E. \& {Ferland}, G.~J. 2006, {Astrophysics of gaseous nebulae
  and active galactic nuclei}

\bibitem[{{Perna} {et~al.}(2015){Perna}, {Brusa}, {Salvato}, {Cresci},
  {Lanzuisi}, {Berta}, {Delvecchio}, {Fiore}, {Lutz}, {Le Floc'h}, {Mainieri},
  \& {Riguccini}}]{Perna:2015aa}
{Perna}, M., {Brusa}, M., {Salvato}, M., {et~al.} 2015, \aap, 583, A72

\bibitem[{{Phillips} {et~al.}(1983){Phillips}, {Edmunds}, {Pagel}, \&
  {Turtle}}]{Phillips:1983aa}
{Phillips}, M.~M., {Edmunds}, M.~G., {Pagel}, B.~E.~J., \& {Turtle}, A.~J.
  1983, \mnras, 203, 759

\bibitem[{{Pinto} {et~al.}(2018){Pinto}, {Alston}, {Parker}, {Fabian}, {Gallo},
  {Buisson}, {Walton}, {Kara}, {Jiang}, {Lohfink}, \&
  {Reynolds}}]{Pinto:2018aa}
{Pinto}, C., {Alston}, W., {Parker}, M.~L., {et~al.} 2018, \mnras, 476, 1021

\bibitem[{{Revalski} {et~al.}(2018){Revalski}, {Crenshaw}, {Kraemer},
  {Fischer}, {Schmitt}, \& {Machuca}}]{Revalski:2018aa}
{Revalski}, M., {Crenshaw}, D.~M., {Kraemer}, S.~B., {et~al.} 2018, \apj, 856,
  46

\bibitem[{{Richings} \& {Faucher-Giguere}(2017)}]{Richings:2017aa}
{Richings}, A.~J. \& {Faucher-Giguere}, C.-A. 2017, ArXiv e-prints
  [\eprint[arXiv]{1710.09433}]

\bibitem[{{Risaliti} {et~al.}(2005){Risaliti}, {Bianchi}, {Matt}, {Baldi},
  {Elvis}, {Fabbiano}, \& {Zezas}}]{Risaliti:2005aa}
{Risaliti}, G., {Bianchi}, S., {Matt}, G., {et~al.} 2005, \apjl, 630, L129

\bibitem[{{Risaliti} {et~al.}(2007){Risaliti}, {Elvis}, {Fabbiano}, {Baldi},
  {Zezas}, \& {Salvati}}]{Risaliti:2007aa}
{Risaliti}, G., {Elvis}, M., {Fabbiano}, G., {et~al.} 2007, \apjl, 659, L111

\bibitem[{{Risaliti} {et~al.}(2013){Risaliti}, {Harrison}, {Madsen}, {Walton},
  {Boggs}, {Christensen}, {Craig}, {Grefenstette}, {Hailey}, {Nardini},
  {Stern}, \& {Zhang}}]{Risaliti:2013aa}
{Risaliti}, G., {Harrison}, F.~A., {Madsen}, K.~K., {et~al.} 2013, \nat, 494,
  449

\bibitem[{{Risaliti} {et~al.}(2009){Risaliti}, {Miniutti}, {Elvis}, {Fabbiano},
  {Salvati}, {Baldi}, {Braito}, {Bianchi}, {Matt}, {Reeves}, {Soria}, \&
  {Zezas}}]{Risaliti:2009aa}
{Risaliti}, G., {Miniutti}, G., {Elvis}, M., {et~al.} 2009, \apj, 696, 160

\bibitem[{{Rupke} \& {Veilleux}(2013)}]{Rupke:2013aa}
{Rupke}, D.~S.~N. \& {Veilleux}, S. 2013, \apj, 768, 75

\bibitem[{{Rupke} \& {Veilleux}(2015)}]{Rupke:2015aa}
{Rupke}, D.~S.~N. \& {Veilleux}, S. 2015, \apj, 801, 126

\bibitem[{{Sandqvist} {et~al.}(1995){Sandqvist}, {Joersaeter}, \&
  {Lindblad}}]{Sandqvist:1995aa}
{Sandqvist}, A., {Joersaeter}, S., \& {Lindblad}, P.~O. 1995, \aap, 295, 585

\bibitem[{{Schulz} {et~al.}(1999){Schulz}, {Komossa}, {Schmitz}, \&
  {M{\"u}cke}}]{Schulz:1999aa}
{Schulz}, H., {Komossa}, S., {Schmitz}, C., \& {M{\"u}cke}, A. 1999, \aap, 346,
  764

\bibitem[{{Shakura} \& {Sunyaev}(1973)}]{Shakura:1973aa}
{Shakura}, N.~I. \& {Sunyaev}, R.~A. 1973, \aap, 24, 337

\bibitem[{{Sharp} \& {Bland-Hawthorn}(2010)}]{Sharp:2010aa}
{Sharp}, R.~G. \& {Bland-Hawthorn}, J. 2010, \apj, 711, 818

\bibitem[{{Shlosman}(1996)}]{Shlosman:1996aa}
{Shlosman}, I. 1996, in Lecture Notes in Physics, Berlin Springer Verlag, Vol.
  474, Lecture Notes in Physics, Berlin Springer Verlag, ed. A.~{Sandqvist} \&
  P.~O. {Lindblad}, 141

\bibitem[{{Singh} {et~al.}(2013){Singh}, {van de Ven}, {Jahnke}, {Lyubenova},
  {Falc{\'o}n-Barroso}, {Alves}, {Cid Fernandes}, {Galbany},
  {Garc{\'{\i}}a-Benito}, {Husemann}, {Kennicutt}, {Marino}, {M{\'a}rquez},
  {Masegosa}, {Mast}, {Pasquali}, {S{\'a}nchez}, {Walcher}, {Wild}, {Wisotzki},
  \& {Ziegler}}]{Singh:2013aa}
{Singh}, R., {van de Ven}, G., {Jahnke}, K., {et~al.} 2013, \aap, 558, A43

\bibitem[{{Speights} \& {Rooke}(2016)}]{Speights:2016aa}
{Speights}, J.~C. \& {Rooke}, P.~C. 2016, \apj, 826, 2

\bibitem[{{Stalevski} {et~al.}(2017){Stalevski}, {Asmus}, \&
  {Tristram}}]{Stalevski:2017aa}
{Stalevski}, M., {Asmus}, D., \& {Tristram}, K.~R.~W. 2017, \mnras, 472, 3854

\bibitem[{{Stevens} {et~al.}(1999){Stevens}, {Forbes}, \&
  {Norris}}]{Stevens:1999aa}
{Stevens}, I.~R., {Forbes}, D.~A., \& {Norris}, R.~P. 1999, \mnras, 306, 479

\bibitem[{{Storchi-Bergmann} \& {Bonatto}(1991)}]{Storchi-Bergmann:1991aa}
{Storchi-Bergmann}, T. \& {Bonatto}, C.~J. 1991, \mnras, 250, 138

\bibitem[{{Teuben} {et~al.}(1986){Teuben}, {Sanders}, {Atherton}, \& {van
  Albada}}]{Teuben:1986aa}
{Teuben}, P.~J., {Sanders}, R.~H., {Atherton}, P.~D., \& {van Albada}, G.~D.
  1986, \mnras, 221, 1

\bibitem[{{Thompson} {et~al.}(2015){Thompson}, {Fabian}, {Quataert}, \&
  {Murray}}]{Thompson:2015aa}
{Thompson}, T.~A., {Fabian}, A.~C., {Quataert}, E., \& {Murray}, N. 2015,
  \mnras, 449, 147

\bibitem[{{Tombesi} {et~al.}(2013){Tombesi}, {Cappi}, {Reeves}, {Nemmen},
  {Braito}, {Gaspari}, \& {Reynolds}}]{Tombesi:2013aa}
{Tombesi}, F., {Cappi}, M., {Reeves}, J.~N., {et~al.} 2013, \mnras, 430, 1102

\bibitem[{{Ueda} {et~al.}(2015){Ueda}, {Hashimoto}, {Ichikawa}, {Ishino},
  {Kniazev}, {V{\"a}is{\"a}nen}, {Ricci}, {Berney}, {Gandhi}, {Koss},
  {Mushotzky}, {Terashima}, {Trakhtenbrot}, \& {Crenshaw}}]{Ueda:2015aa}
{Ueda}, Y., {Hashimoto}, Y., {Ichikawa}, K., {et~al.} 2015, \apj, 815, 1

\bibitem[{{Vayner} {et~al.}(2017){Vayner}, {Wright}, {Murray}, {Armus},
  {Larkin}, \& {Mieda}}]{Vayner:2017aa}
{Vayner}, A., {Wright}, S.~A., {Murray}, N., {et~al.} 2017, \apj, 851, 126

\bibitem[{{Vazdekis} {et~al.}(2010){Vazdekis}, {S{\'a}nchez-Bl{\'a}zquez},
  {Falc{\'o}n-Barroso}, {Cenarro}, {Beasley}, {Cardiel}, {Gorgas}, \&
  {Peletier}}]{Vazdekis:2010aa}
{Vazdekis}, A., {S{\'a}nchez-Bl{\'a}zquez}, P., {Falc{\'o}n-Barroso}, J.,
  {et~al.} 2010, \mnras, 404, 1639

\bibitem[{{Veilleux} {et~al.}(2005){Veilleux}, {Cecil}, \&
  {Bland-Hawthorn}}]{Veilleux:2005aa}
{Veilleux}, S., {Cecil}, G., \& {Bland-Hawthorn}, J. 2005, \araa, 43, 769

\bibitem[{{Veilleux} \& {Osterbrock}(1987)}]{Veilleux:1987aa}
{Veilleux}, S. \& {Osterbrock}, D.~E. 1987, \apjs, 63, 295

\bibitem[{{Veilleux} {et~al.}(2003){Veilleux}, {Shopbell}, {Rupke},
  {Bland-Hawthorn}, \& {Cecil}}]{Veilleux:2003aa}
{Veilleux}, S., {Shopbell}, P.~L., {Rupke}, D.~S., {Bland-Hawthorn}, J., \&
  {Cecil}, G. 2003, \aj, 126, 2185

\bibitem[{{Venturi} {et~al.}(2017){Venturi}, {Marconi}, {Mingozzi}, {Carniani},
  {Cresci}, {Risaliti}, \& {Mannucci}}]{Venturi:2017aa}
{Venturi}, G., {Marconi}, A., {Mingozzi}, M., {et~al.} 2017, Frontiers in
  Astronomy and Space Sciences, 4, 46

\bibitem[{{Venturi} {et~al.}(2018)}]{Venturi:2018aa}
{Venturi}, G. {et~al.} 2018, in prep.

\bibitem[{{V{\'e}ron-Cetty} \& {V{\'e}ron}(2006)}]{Veron-Cetty:2006aa}
{V{\'e}ron-Cetty}, M.-P. \& {V{\'e}ron}, P. 2006, \aap, 455, 773

\bibitem[{{Wang} {et~al.}(2009){Wang}, {Fabbiano}, {Elvis}, {Risaliti},
  {Mazzarella}, {Howell}, \& {Lord}}]{Wang:2009aa}
{Wang}, J., {Fabbiano}, G., {Elvis}, M., {et~al.} 2009, \apj, 694, 718

\bibitem[{{Wang} {et~al.}(2014){Wang}, {Nardini}, {Fabbiano}, {Karovska},
  {Elvis}, {Pellegrini}, {Max}, {Risaliti}, {U}, \& {Zezas}}]{Wang:2014aa}
{Wang}, J., {Nardini}, E., {Fabbiano}, G., {et~al.} 2014, \apj, 781, 55

\bibitem[{{Whewell} {et~al.}(2016){Whewell}, {Branduardi-Raymont}, \&
  {Page}}]{Whewell:2016aa}
{Whewell}, M., {Branduardi-Raymont}, G., \& {Page}, M.~J. 2016, \aap, 595, A85

\bibitem[{{Williams} {et~al.}(2017){Williams}, {Maiolino}, {Krongold},
  {Carniani}, {Cresci}, {Mannucci}, \& {Marconi}}]{Williams:2017aa}
{Williams}, R.~J., {Maiolino}, R., {Krongold}, Y., {et~al.} 2017, \mnras, 467,
  3399

\bibitem[{{Z{\'a}nmar S{\'a}nchez} {et~al.}(2008){Z{\'a}nmar S{\'a}nchez},
  {Sellwood}, {Weiner}, \& {Williams}}]{Zanmar-Sanchez:2008aa}
{Z{\'a}nmar S{\'a}nchez}, R., {Sellwood}, J.~A., {Weiner}, B.~J., \&
  {Williams}, T.~B. 2008, \apj, 674, 797

\end{thebibliography}

\end{document}